\documentclass[twocolumn]{aastex631}

\shorttitle{Dwarf diversity in $\Lambda$CDM with baryons}
\shortauthors{Cruz et al.}

\begin{document}

\title{Galaxy size and rotation curve diversity in $\Lambda$CDM with baryons}

\author[0000-0001-7831-4892]{Akaxia Cruz}
\affiliation{Center for Computational Astrophysics, Flatiron Institute, New York, NY 10010, USA}

\affiliation{Department of Physics, Princeton University, Princeton, NJ 08544, USA}

\affiliation{Department of Astrophysical Sciences, Princeton University, Princeton, NJ 08544, USA}
\author[0000-0002-0372-3736]{Alyson M. Brooks}
\affiliation{Center for Computational Astrophysics, Flatiron Institute, New York, NY 10010, USA}
\affiliation{Department of Physics \& Astronomy, Rutgers, the State University of New Jersey, Piscataway, NJ 08854, USA}

\author[0000-0002-8495-8659]{Mariangela Lisanti}
\affiliation{Center for Computational Astrophysics, Flatiron Institute, New York, NY 10010, USA}

\affiliation{Department of Physics, Princeton University, Princeton, NJ 08544, USA}

\author[0000-0002-8040-6785]{Annika H. G. Peter}
\affiliation{Department of Physics, Ohio State University, Columbus, OH 43210, USA}
\affiliation{Department of Astronomy, Ohio State University, Columbus, OH 43210, USA}
\affiliation{Center for Cosmology and AstroParticle Physics (CCAPP), The Ohio State University, Columbus, OH 43210, USA}

\author[0000-0003-1509-9966]{Robel Geda}
\affiliation{Department of Astrophysical Sciences, Princeton University, Princeton, NJ 08544, USA}

\author[0000-0003-1282-7825]{Thomas Quinn}
\affiliation{Department of Astronomy, University of Washington, Seattle, WA 98195, USA}

\author[0000-0002-4353-0306]{Michael Tremmel}
\affiliation{School of Physics, University College Cork, Ireland}

\author[0000-0002-9581-0297]{Ferah Munshi}
\affiliation{Department of Physics $\&$ Astronomy, George Mason University,  Fairfax, VA 22030, USA}

\author[0000-0002-9642-7193]{Ben Keller}
\affiliation{Department of Physics and Material Science, The University of Memphis, Memphis, TN 38152, USA}

\author[0000-0001-8745-0263]{James Wadsley}
\affiliation{Department of Physics and Astronomy, McMaster University, Ontario, Canada}

\begin{abstract}

Observed rotation curves of dwarf galaxies exhibit significant diversity at fixed halo mass, challenging galaxy formation within the Cold Dark Matter (CDM) model. Previous cosmological galaxy formation simulations with baryonic physics fail to reproduce the full diversity of rotation curves, suggesting either that there is a flaw in baryonic feedback models, or that an alternative to CDM must be invoked.  In this work, we use the Marvelous Massive Dwarf zoom-in simulations, a suite of high-resolution dwarf simulations with $M_{200}~\sim 10^{10}$--$10^{11}$~M$_{\odot}$ and $M_{*}\sim 10^{7}$--$10^{9}$~M$_{\odot}$, designed to target the mass range where galaxy rotation curve diversity is maximized, i.e., between $V_{\rm max} \sim 70$--$100~ {\rm km/s}$. We add to this a set of low-mass galaxies from the Marvel Dwarf Zoom Volumes to extend the galaxy mass range to lower values. Our fiducial star formation and feedback models produce simulated dwarfs with a broader range of rotation curve shapes, similar to observations. These are the first simulations that can both create dark matter cores via baryonic feedback, reproducing the slower rising rotation curves, while also allowing for compact galaxies and steeply rising rotation curves. Our simulated dwarfs also reproduce the observed size--$M_*$ relation, including scatter, producing both extended and compact dwarfs for the first time in simulated field dwarfs.  We explore star formation and feedback models and conclude that previous simulations may have had feedback that was too strong to produce compact dwarfs.
 
\end{abstract}

\section{Introduction} \label{sec:intro}
Observations showing that rotation curves~(RCs) are flat at large radii provided some of the first evidence for more matter in galaxies than can be accounted for by gas and stars \citep{Rubin1980, Bosma1981}.  Since then, it has been determined that dark matter comprises $\sim$85$\%$ of all matter in the Universe \citep{Planck2020}.  The Cold Dark Matter~(CDM) model describes the Universe well on large scales \citep[e.g.,][]{Springel2006}, but has encountered trouble explaining the diversity of RC shapes, particularly in dwarf galaxies \citep{Sales2022}.  Dark matter-only (DMO) simulations show universal, self-similar halo density profiles \citep[NFW profiles,][]{NFW1996, NFW1997} that are steeper than some inferred from the RCs of dwarf galaxies \citep{Moore1994, Flores1994, deBlok2001, deBlok2010, Oh2011}.  Hydrodynamic simulations have shown that the inclusion of energetic feedback from baryons can fluctuate the gravitational potential at the centers of dwarf galaxies, transforming an initially steep density profile into a more shallow profile, thought to reconcile simulations with slowly rising RCs \citep{Read2005, Mashchenko2008, Governato2012, Pontzen2012, Teyssier2013, Dicinto2014a, Tollet2015, Read2016, Llambay2019, Dutton2019}.  

However, it has more recently been recognized that the full range of RCs, from quickly to slowly rising, has not been fully reproduced by galaxy simulations \citep{Oman2015, SantosSantos2018, SantosSantos2020}. In particular, \cite{Oman2015} highlights the diversity of observed RCs, %with two-dimensional velocity fields to accurately connect the data with underlying gravitational potentials 
%to demonstrate the observed diversity in the centers of galaxies.
showing that diversity is maximized at $V_{\rm max}$ (the maximum observed RC velocity) $ \sim 70$--$80$~km/s, with a factor $\sim$ 4 spread within the central 2 kpc ($V_{\rm 2 kpc}$) at this fixed $V_{\rm max}$. A similar central velocity spread relative to $V_{\rm max}$ is found in Spitzer Photometry and Accurate RCs \citep[SPARC,][]{Lelli2016}, as well as more recent $\rm{H}\alpha$ RCs \citep{Relatores2019a, Relatores2019b}.  

Previous simulations with CDM and baryons have struggled to reproduce the full range of observed diversity of RCs. In particular, \cite{Oman2015} found a lack of diversity in the Evolution and Assembly of GaLaxies and their Environments \citep[EAGLE,][]{Schaye2015, Crain2015} simulations, though they explicitly used simulations that do not transform dark matter cusps into cores.~\cite{SantosSantos2018} examined diversity in the Numerical Investigation of a Hundred Astrophysical Objects~\citep[NIHAO,][]{Wang2015} simulations, which do create dark matter cores~\citep[e.g.,][]{Dicinto2014a, Dicintio2014b}.  They demonstrated that using the gravitational midplane potential to calculate RCs better modeled dwarf galaxies that were out of dynamical equilibrium due to bursty feedback, as opposed to using shells of enclosed mass that assume spherical symmetry.  No matter the RC calculation method, the NIHAO simulations 
fail to capture the RCs that rise more quickly than predicted by NFW \citep[see also][]{Frosst2022}. Likewise, \cite{SantosSantos2020} showed that EAGLE simulations with a high density threshold for star formation (the gas density above which stars form) can produce cored dark matter profiles with slowing rising RCs, but struggle to simultaneously produce the most quickly rising RCs.  In contrast, simulations with lower density thresholds for star formation are able to produce the more quickly rising RCs but not the more slowly rising ones. Similarly, \cite{Roper2023} showed a correspondence between high (low) star formation density threshold and slowly (quickly) rising RCs.  Further, \cite{Roper2023} showed that simulations that do not create dark matter cores can reproduce the full range of observed diversity due to non-circular motions introduced in mock HI observations.

SPARC suggests an additional observed diversity in the galaxy size--stellar mass plane. Observations show the presence of both compact and extended galaxies at a given stellar mass~$(M_*)$. Previous hydrodynamical simulations within CDM struggle not only to produce diversity in RCs, but also diversity in the size--$M_*$ plane~\citep{Sales2022}. \cite{Jiang2019} showed that NIHAO, and Feedback in Realistic Environments~2~\citep[FIRE-2,][]{Hopkins2018} and simulations presented in \cite{Lupi2017} produce galaxies that are more diffuse than observed galaxies and struggle to produce the most compact galaxies at and below $M_* \sim 10^9$~M$_{\odot}$.

In this work, we use high-resolution hydrodynamical, cosmological simulations of dwarf galaxies produced using the {\sc ChaNGa} code with multiple feedback models to examine the diversity of RCs and galaxy sizes compared to observations. We show that a new suite of dwarf galaxies, the Marvelous Massive Dwarf zooms, produce a range of both quickly and slowly rising RCs at a given halo mass when run with our fiducial model.  
This suite is the first to reproduce a broad range of RC diversity in CDM with baryons, and without creating mock HI observations.  In addition, the galaxies also produce a more realistic range of sizes at stellar masses $M_* \sim 10^7$--$10^9$~M$_{\odot}$, consistent with the scatter showing both extended and compact galaxies in observations.

This paper is organized as follows. In Section~\ref{sec:sims}, we detail the galaxy formation models used in our simulated sample. In Section~\ref{sec:data}, we briefly describe the observational data with which we compare our simulated results. We detail our analysis techniques used throughout this work in Section~\ref{sec:analysis}. We present our RC diversity shapes in Section~\ref{sec:RC_shapes} and explore the impact of baryons on RC shapes in Section~\ref{subsec:baryons_RC}. In Section~\ref{sec:size_mass}, we then discuss diversity in the size--$M_*$ plane, and in Section~\ref{subsec:subgrid_phys}, we consider how subgrid physics can prevent the formation of compact galaxies. We discuss how our results compare to previous works in CDM with baryons and in alternative dark matter models in Section \ref{sec:discussion}, and finally summarize our results in Section \ref{sec:conclusions}. In Appendix~\ref{appen:RC_methods}, we detail how the method used to obtain RCs influences their shapes. In Appendix~\ref{appen:feedback_RC}, we briefly discuss feedback differences on RC shape. In Appendix~\ref{appen:den_profs}, we detail our density profiles and fitting procedure and show RC shape trends with dark matter density slope.

\section{Simulations} \label{sec:sims}

In this study, we use two suites of simulated dwarf galaxies: one collection of ``zoom volumes" and a complementary set of more massive individual zoom-in galaxy simulations.  Both suites are produced using the N-body$+$Smooth Particle Hydrodynamics~(SPH) code Charm Nbody Gravity solver~ \citep[{\sc ChaNGa},][]{Menon2015} which adopts hydrodynamic physics from {\sc Gasoline2} \citep{Wadsley2004, Wadsley2017}, but modified to use the {\sc charm}$++$ runtime code for load balancing and communication. All simulations are run in a fully cosmological context. We run with a fiducial model on which most of the results are based, but also rerun two galaxies with varying subgrid physics in Section \ref{subsec:subgrid_phys}. First, we detail the physics in common between the two simulation suites, then discuss their differences in Sections~\ref{subsec:Marvel} and \ref{sec:MassiveDwarfs}.  In Section~\ref{sec:nihao}, we describe the additional variations used for the tests in Section~\ref{subsec:subgrid_phys}. The final sample selection compared to the SPARC data \citep[][see Section \ref{sec:data}]{Lelli2016} is defined in Section~\ref{sec:analysis}.

The fiducial star formation model for both simulation suites utilizes the metal-line-cooling and star formation scheme as presented in \citet{Christensen2012}. This model includes non-equilibrium formation and destruction of H$_2$ and a uniform time-dependent UV field~\citep{Haardt2012} to model photoionization and heating.  Lyman-Werner radiation from young stars is additionally tracked and can further heat and photo-dissociate H$_2$ and suppress star formation, mimicking early stellar feedback. Star particles represent stellar populations with a \cite{Kroupa2001} initial mass function. The star particle initial mass is 30$\%$ of the original gas particle mass. Mass and metals from stars are returned to surrounding gas particles in winds and in Type I and Type II supernovae~(SNe)~\citep{stinson2006}.

 Star formation is restricted to occur stochastically in the presence of H$_2$ when gas particles become cold ($T_{*} \leq10^3$~K) and dense ($n_{*}\geq0.1$ cm$^{-3}$). The probability of creating a star particle (with mass $m_{\rm star}$) from a gas particle (with mass $m_{\rm gas}$) with local dynamical time, $t_{\rm dyn}$, in time $\Delta t = 10^6~{\rm yr}$ is:

\begin{equation} \label{eqn:DCstarform}
p = \frac{m_{\mathrm{gas}}}{m_{\mathrm{star}}} (1-\mathrm{e}^{-c_* X_{H_2} \Delta t/t_{\mathrm{dyn}}}) \, ,
\end{equation}
\noindent where the value adopted for star formation efficiency is $c_{*} = 0.1$. The star formation efficiency, $c_*$, times the fraction of non-ionized molecular hydrogen, $X_{H_2}$, produces the appropriate normalization of the Kennicutt-Schmidt relation \citep{Christensen2014}.  Although the density threshold is low compared to other bursty star formation prescriptions \citep[e.g.,][]{Governato2010, Llambay2019, Dutton2019, Dutton2020}, the actual density of star-forming gas tends to be closer to 100~cm$^{-3}$ due to the additional requirement that H$_2$ be present. 

All of the simulations in our zoom volume and fiducial zoom-in models include black hole~(BH) formation and growth, with BH accretion and feedback models from \citet{Tremmel2015} and \citet{Tremmel2017}.  We do not make assumptions about BH halo occupation fractions. Instead, BH seed formation is related to the physical state of the gas (i.e., temperature, density, metallicity, molecular fraction) at high redshift. 

The BHs are subject to a prescription of dynamical friction which follows the orbital evolution of BHs, as introduced by \citet{Tremmel2015}. Correctly accounting for this effect can lead to pairs of supermassive BHs at kpc-scale separations that survive for several Gyrs \citep{Tremmel2018a}. BHs in our simulation also obey a modified Bondi accretion \citep{Tremmel2017}. Parameters for seeding and accretion are detailed for each simulation suite below.

Here, we have described the physics that the simulations have in common.  Below, we describe the differences between the two suites.  In summary, they use slightly different cosmologies and resolutions, they have slightly different seeding mechanisms for BHs, and their feedback implementation (for Type II SNe and BHs) is different.  Despite that, previous works have used both sets of simulations simultaneously to increase the mass range of the simulated sample size.  They were unable to find differences in galaxy properties between the two runs, e.g., in stellar-to-halo mass, HI content and gas fractions, sizes, or metallicity~\citep[see][]{Piacitelli2025, Ruan2025}.  We again combine them in this work to increase the range of galaxy masses and explore potential differences (see Appendix~\ref{appen:feedback_RC}.)

\subsection{Marvel Dwarf \label{subsec:Marvel} Zoom Volumes}
The Marvel dwarf suite (hereafter Marvel)
consists of a population of dwarfs from ``zoom volumes'' \citep{Bellovary2019, Munshi2021, Christensen2024}. In contrast to typical zoom simulations that focus their Lagrangian regions on individual halos, in these simulations, the Lagrangian regions are larger volumes that contain dozens of dwarf galaxies. The volumes for the Marvel zooms are selected to be between $\sim 1.5$--$7$~Mpc away from Milky Way-mass galaxies \citep{Christensen2024}. These volumes represent isolated local-volume galaxies. Four such simulations are run, named CptMarvel, Elektra, Rogue, and Storm, with a WMAP3 cosmology \citep[$\Omega_{\Lambda} = 0.736 $, $\Omega_{\rm m} = 0.264$, $\Omega_b = 0.0441$, $\sigma_8 = 0.805$, $n_s = 0.967$, $h = 0.712$,][]{Spergel2007}. Each zoom volume has a spline force softening of $\epsilon = 60$~pc, an initial gas mass resolution of 1410~M$_{\odot}$, an initial stellar mass resolution of 420~M$_{\odot}$, and a dark matter mass resolution of 6650~M$_{\odot}$. 

The Marvel simulations include a model for stellar feedback from Type~II SNe, which adopts a ``blastwave" implementation as detailed in \cite{stinson2006}. This model deposits mass, thermal energy, and metals into nearby gas when a massive star of $8$--40~M$_{\odot}$ becomes a SN. Each SN releases 1.5 $\times 10^{51}$ ergs of energy into 32 neighboring gas particles, and a star particle deposits 1.5~$\times 10^{49}$ ergs/M$_{\odot}$ over the lifetime of SNe~II. After each SN, gas cooling is turned off for a time determined by Equation~11 in \cite{stinson2006}, the timescale over which the gas particle does not radiatively cool. The deposited energy, combined with the turning off of cooling, is used to mimic how inefficient cooling  
influences the local interstellar medium~(ISM). The Storm zoom volume is additionally run with superbubble feedback, which is introduced in the next subsection.  

BH seeds form in the early universe if the gas particle has already met the star formation threshold. Additionally, BH particles form in extremely dense ($n >1.5 \times10^4$~cm$^{-3}$) and cool ($T < 2 \times 10^4$~K) gas regions. To prevent fragmentation into stars, gas particles that can form BHs must also have low metallicity ($Z < 10^{-4}$) and have low molecular fraction ($X_{H_2} < 10^{-4}$).  The mass of a BH seed is $5\times10^4$~M$_{\odot}$. BHs accrete according to a modified Bondi-Hoyle accretion, which accounts for the rotational support of the surrounding gas \citep{Tremmel2017}. The energy of accretion onto the BHs is isotropically transferred to nearby gas particles as thermal energy using the blastwave feedback model of \cite{stinson2006} for the Marvel zoom volumes. However, \citet{Bellovary2019} found that BHs in Marvel zoom volumes barely grow after their formation and thus impart very little energy.

\subsection{The Marvelous Massive and Extended Dwarf Zooms} \label{sec:MassiveDwarfs}

We analyze the Marvelous Massive Dwarf zooms, a collection of 40 zoom-in simulations with $M_{200}\sim (0.1$--$2) \times 10^{11}$~M$_{\odot}$ and stellar masses $M_{*} \sim (0.06$--$3)\times 10^{9
}$~M$_{\odot}$, as well as the Extended Marvelous Dwarf zooms that have stellar masses down to $M_{*} \sim 6 \times 10^6$~M$_{\odot}$. As suggested by the name, the Marvelous Massive Dwarfs are designed to have higher masses than the Marvel set. They are specifically run to investigate the mass range where the diversity of RCs is maximized and selected based on their stellar mass, aiming for 10$^8$--10$^9$~M$_{\odot}$. Both the Marvelous Massive and Extended Marvelous dwarf zooms are selected from the (25 Mpc)$^3$ box, Romulus25 \citep{Tremmel2017}. We choose halos that are at least 1.5~Mpc away from any galaxy with $M_* > 10^{10}$~M$_{\odot}$ at redshift $z = 0.1$.  At a fixed stellar mass, they are chosen to span a range in halo mass. No other selection criteria are used. Since they are run with the same physics, we refer to the Marvelous Massive Dwarf zooms and the Extended Marvelous Dwarf zooms collectively as the Marvelous Massive Dwarf zooms or just the Massive Dwarfs to not confuse results with previous work using the Marvel zoom-volume galaxies.

These 55 simulations are run at ``Mint" resolution  with spline force softening of $\epsilon = 87$~pc, an initial gas mass resolution of 3300~M$_{\odot}$, an initial stellar mass resolution of 990~M$_{\odot}$, and a dark matter mass resolution of $1.8 
\times 10^4$~M$_{\odot}$.  All of our Marvelous Massive Dwarf zooms are run with a Planck Cosmology \citep[$\Omega_{\Lambda} = 0.6914$, $\Omega_{\mathrm{m}} = 0.3086$, $\Omega_{\mathrm{b}} = 0.04825$,  $\sigma_8 = 0.8288$, $n_s = 0.9611$, $h = 0.6777$,][]{Plank2014} and used the ``zoom-in" methods described by \citet{Pontzen2008}. 

We run the 40 Marvelous Massive Dwarf zooms and the 15 Extended Marvelous Dwarf zooms with a fiducial model that includes H$_2$ cooling, star formation, and superbubble feedback. Stars preferentially form in clusters \citep{Pfalzner2012, Krause2020}. Feedback from clustered stars produces superbubbles, which are distinct from isolated SNe, as modeled by the blastwave model. The superbubble feedback model used here~\citep{Keller2014} introduces a multiphase treatment of gas, evaporation, and thermal conduction. Unlike the blastwave model, this feedback model does not explicitly shut off cooling. Instead, the evaporation from thermal conduction is designed to regulate the amount of hot gas. As in \cite{Keller2014}, a star particle deposits $10^{49}$~erg/M$_{\odot}$ of thermal energy due to SNe\,II. This model deposits feedback into resolution elements in a brief multi-phase state: the particles have different specific energies, masses, and densities in the hot bubble phase and the cold ISM phase. Then, for each phase, separate cooling rates can be derived using their respective densities and temperatures. Gas particles become single phase again when the hot phase either cools or evaporates the cool phase.

BH particles form in gas particles with high density ($n > 1.5~\times 10^4$~cm$^{-3}$), cool temperature ($T < 5 \times 10^3$~K), low metallicity ($Z < 10^{-5}$), low molecular fraction ($X_{H_2} < 2 \times 10^{-3}$), and large Jeans mass~ ($M_J > 10^5$~M$_{\odot}$). These criteria are motivated by \cite{Wise2019, Regan2020, Regan2023}. The mass of a BH seed is $10^5$~M$_{\odot}$. BHs have a density-boosted Bondi accretion motivated by \cite{Booth2009} where the boost density threshold is 100 cm$^{-3}$. The energy of accretion onto the BHs is isotropically transferred to nearby gas particles as thermal energy using the superbubble feedback model of \cite{Keller2014} for the Massive Dwarf zooms. \cite{Bellovary2019} found little accretion in the Marvel galaxies, and we explored a subset of the Massive Dwarf zoom galaxies run with and without BHs and found no preferential shift in their stellar sizes or masses.

\begin{figure*}
\epsscale{0.86}
\plotone{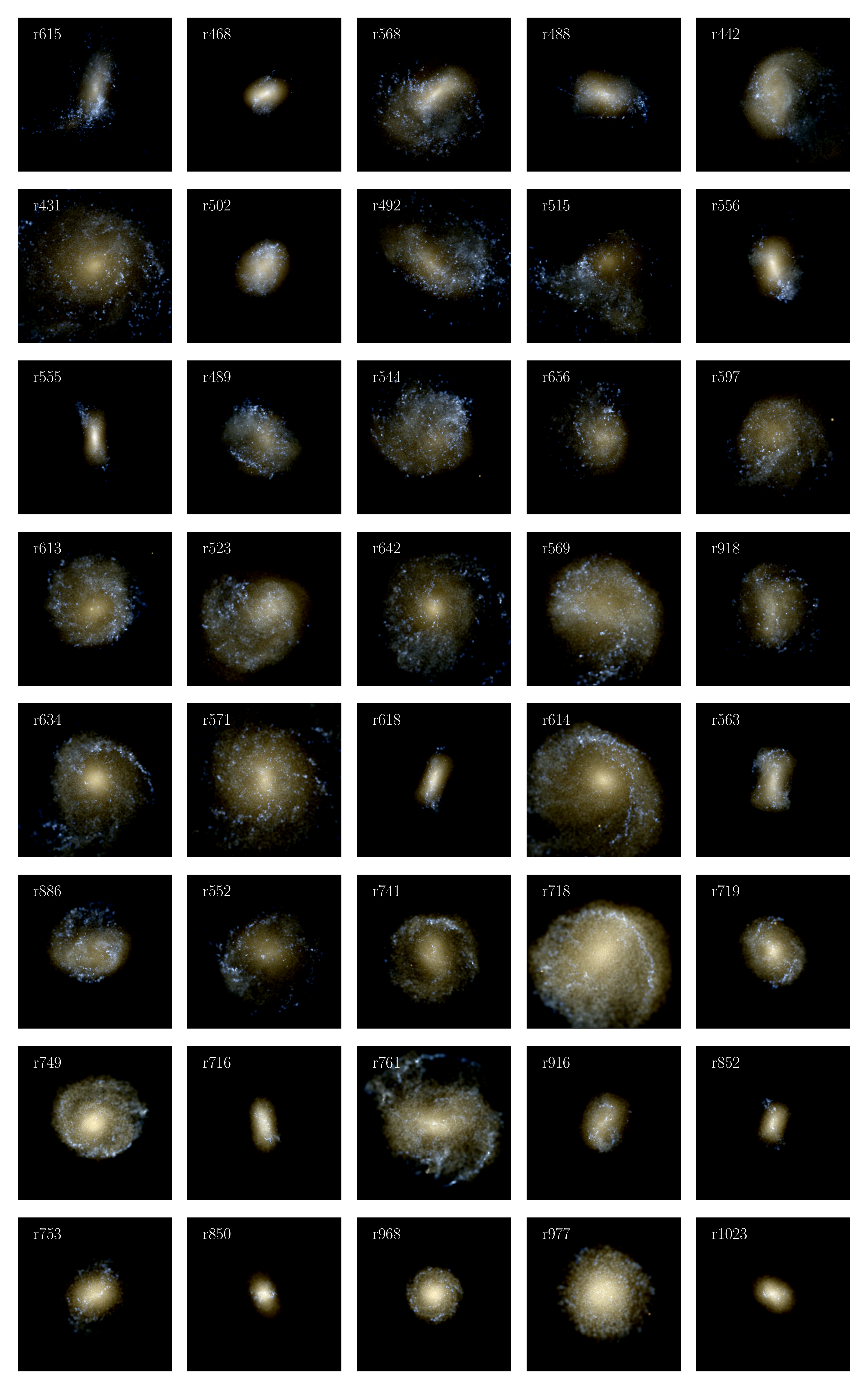}
\caption{{{\sc Mock UVI images of 40 of the Massive Dwarf zoom-in galaxies at $z = 0$}, ordered by decreasing $M_{*}$ from top left to bottom right. The zoom-in name is annotated in the upper left of each subpanel. Each galaxy is shown face-on (i.e., with the angular momentum vector out of the page), and each square panel has a side length of 15~kpc. Visually, it is clear that the Massive Dwarf zooms have diverse morphologies.} 
\label{fig:uvi_grid}}
\end{figure*}

We visually examine the morphology of 40 of the most massive galaxies in the Massive Dwarf zooms, shown as mock-UVI images in  Figure~\ref{fig:uvi_grid}. The images are ``face-on" in the x--y plane, so that the angular momentum of each galaxy is rotated into the z-direction, which is out of the page. The zooms are ordered by stellar mass, $M_*$, from top to bottom and left to right in a given row. The zoom-in name is annotated in the upper left of each subpanel. 
The details of how $M_{200}$ and angular momentum are calculated are described in Section~\ref{subsec:amiga}. From visual inspection, it can be seen that the galaxies have diverse morphologies and sizes, as a given row represents roughly the same stellar mass, and there is a diversity in size. 

Beyond these galaxies having diverse morphologies, the simulations also agree well with other observables. %and have offered additional predictions to further test the ChaNGa models used in this paper. 
In forthcoming work, we will demonstrate the ability to produce stellar dwarf galaxy disks \citep{Geda2025} with stellar masses down to $10^7$~M$_{\odot}$, which are observed~\citep{Lelli2016}, but have evaded previous simulations~\citep{ElBadry2018a, Benavides2025, Celiz2025}. Additionally, the ChaNGa models make galaxies that are in better agreement with observed stellar age gradients in dwarfs \citep{Riggs2024}, have HI sizes and gas fractions that match observed galaxies \citep{Ruan2025}, and match the mass--metallicity relation, the sSFR-mass trends, and reasonably reproduce dwarf CGM column densities \citep{Piacitelli2025}. We have additionally made predictions for the turn down in the baryonic Tully-Fisher relationship~\citep{Ruan2025} and in stellar shapes \citep{Keith2025}, which can further test the ChaNGa models used in this work.

\subsection{NIHAO Reruns}\label{sec:nihao}
Because ChaNGa adopts hydrodynamic physics from {\sc Gasoline}, the code used for the NIHAO simulations, we can rerun initial conditions with some of the subgrid physics used in \cite{Wang2015}. In Section~\ref{subsec:subgrid_phys}, we rerun two of the Massive Dwarf zoom-in galaxies that have 
quiescent merger histories after cosmic time of $\sim 2$~Gyrs to better understand the differences between the NIHAO results and our simulations. Starting from the Mint resolution, fiducial star formation, and superbubble feedback model, we make incremental changes, progressing towards the NIHAO model and resolution. 
In the NIHAO star formation model, gas is eligible to form stars once it has reached specified density and temperature thresholds in agreement with the Kennicutt-Schmidt Law. In the model adopted in \cite{Wang2015}, which we mimic here, stars form in cold ($T_* < 1.5 \times 10^4~$K) and dense ($n_{*} > 10.3$~cm$^{-3}$) gas.  The NIHAO star formation model deviates from our fiducial model significantly, with stars forming in much less dense and much warmer gas. 

NIHAO uses the blastwave SN feedback model of \cite{stinson2006}, with star particles releasing a total of 10$^{49}$~erg/M$_{\odot}$ thermal energy due to SNe\,II. 
as well as early stellar feedback from massive stars \citep{Stinson2013}. To limit star formation to agree with the abundance matching value at the Milky Way scale, the early feedback efficiency, $\epsilon_{\scriptscriptstyle{{\rm ESF}}} =13\%$, of the total stellar flux, $2 \times 10^{50}$~erg/M$_{\odot}$ is ejected into the surrounding gas over the 4~Myr between a star's formation and when Type~II SNe begin to occur. For our runs, we adopt a slightly lower $\epsilon_{\scriptscriptstyle{{\rm ESF}}} =10\%$, consistent with \cite{Stinson2013}.

The NIHAO simulations are run with constant relative resolution between the gravitational softening of the dark matter $\epsilon_{\scriptscriptstyle{\rm DM}}$ and halo virial radius across $\sim 3$~dex of halo mass and have $\sim 10^6$ dark matter particles per halo. This means that each halo mass scale is run at a different resolution. We approach the mass resolution of the NIHAO runs for the $M_{200}~\sim~10^{10}$--$10^{11}$~M$_{\odot}$ halos\footnote{In NIHAO, $M_{200} = (2.63$--$7.12)\times10^{10}$~M$_\odot$ halos have a dark matter mass resolution of $6.426 \times 10^4$~M$_{\odot}$, baryon (gas and stars) mass resolution of $1.173 \times 10^4$~M$_{\odot}$, and a gravitational softening of $\epsilon_{\rm gas} = 132.6$~pc for gas and $\epsilon_{\scriptscriptstyle{\rm DM}} = 310.5$~pc for dark matter.} by stepping from our ``Mint'' resolution to ``Near Mint" that has half the force resolution (174 pc) and 8$\times$ lower mass resolution (initial gas particle masses of 2.64 $\times 10^4$~M$_{\odot}$, dark matter mass resolution of $1.4 
\times 10^5$~M$_{\odot}$). 

We change one subgrid physics prescription at a time to isolate the impact on stellar half-light radius and mass. Independent of resolution, in all of the runs with NIHAO subgrid physics models, we exclude BHs since the NIHAO runs we compare to do not include them~\citep{SantosSantos2018}. The details of these physics variations are discussed in Section~\ref{subsec:subgrid_phys}.

\section{Observational Data} \label{sec:data}
We compare our numerical results to observed galaxies from the SPARC dataset \citep{Lelli2016}. These observations comprise a sample of 175 nearby galaxies with homogeneous photometry at 3.6 $\mu$m, which complements high-quality RC data from previous HI and/or H$\alpha$ observations. Since the velocity dispersion of HI is typically on the order of $\sim10$~km/s, we expect pressure support to be significant only in very low-mass dwarf galaxies with rotation velocities of $\sim
20$~km/s \citep{Lelli2012b}. However, the reported SPARC RCs have been corrected for pressure support, beam smearing, and inclination. Stellar masses are derived by assuming a mass-to-light ratio of $M_*/L_{[3.6~\mu{\rm m}]} = 0.5$. 

The sample of SPARC galaxies considered in this analysis only includes those with inclinations $i > 30^{\circ}$ as in \cite{SantosSantos2020} and those suitable for dynamical studies with quality flag $Q \neq 3$ as in \cite{Zentner2022}, to minimize RCs with substantial uncertainties. We use the RCs from \cite{Lelli2016} to calculate $V_{\rm max}$ values for the observed sample. We take the maximum value of the observed RCs out to the last observed radial bin to be $V_{\rm max}$ for the observed sample, noting that in some cases this value might under-predict the true maximum circular velocity. This is more common in the lower-mass galaxies, as the observational HI-limit is still on the rising part of the RC \citep[see e.g.,][]{Ruan2025} for more details on how our simulated galaxies are impacted.

\section{Analysis} \label{sec:analysis}

\subsection{Halo Finding and Centering}

\label{subsec:amiga}
We identify halos in a given simulation using the AMIGA halo finder \citep[AHF;][]{Knollmann2009}. The virial radius, $R_{\rm vir}$, of a halo is the radius at which the density corresponds to $\Delta_{h} = 200$ times the critical density of the Universe, $\rho_c$. All halos in this work are analyzed at $z=0$. We calculate the virial mass of halos as: 
\begin{equation}\label{eqn:M200}
    M_{\mathrm{vir}} = \frac{4\pi}{3} \, \Delta_{h} \, \rho_c \, R^3_{\mathrm{vir}} \, .
\end{equation}

We center each halo using the shrinking-sphere procedure \citep{Power2003} and Lorentz boost into the rest frame of the center-of-mass of the particles within 1~kpc of the center. We then rotate the galaxy in the x--y plane such that the angular momentum vector of the gas within 5~kpc of the halo center points in the z-direction. We do produce disk galaxies and, in these systems, the z-component of the total angular momentum is well aligned with the z-direction \citep{Geda2025}. However, even for the non-disky galaxies, we use the same orientation methodology. 

\subsection{Rotation Curves} \label{subsec:rotcurves}

This subsection describes the variety of methods used to calculate the RC velocities from our simulations and how we compare them with observations. We calculate full RCs between $r\in [r_{\rm conv}, 15~{\rm kpc}]$ using 100 linearly spaced bins. Below the convergence radius, $r_{\rm conv}$, two-body scattering can alter the mass distribution~\citep{Power2003}. \cite{LSB2019} notes that the convergence radius interior to which we expect artificial heating can be conveniently approximated as $r_{\rm conv} = 0.055\times l$,  where $l = L_{\rm box}/N_{\rm part}^{1/3}$ is the mean interparticle spacing in physical units for a simulation with box length $L_{\rm box}$ and $N_{\rm part}$ total particles.  In the Massive zooms, %$r_{\rm conv}~=~0.055~\times~25~{\rm Mpc}/3072 \sim~0.45~{\rm kpc}$ and in the Marvel simulations, $r_{\rm conv}~=~0.055~\times~25~{\rm Mpc}/4096~\sim 0.34~{\rm kpc}$. 
$r_{\rm conv}\sim0.45~{\rm kpc}$ and in the Marvel simulations, $r_{\rm conv}\sim0.34~{\rm kpc}$.
The upper end of the considered radial range is motivated by the radius that corresponds to the true maximum halo velocity: we find $R_{\rm max}<15~{\rm kpc}$ for all systems considered in this work. The binning scheme is chosen so that each bin contains at least 3000 dark matter particles, even for the smallest mass systems. In the higher-mass systems, each bin contains well over 3000 particles. The limit on particles-per-bin is expected to properly resolve densities~\citep{Power2003}. 

We calculate the RCs using two different methods and then use two separate proxies to determine the maximum and central value.  First, we assume spherical symmetry and calculate 
\begin{equation} \label{eqn:vcirc}
    V^{\rm circ}(r) = \sqrt{\frac{GM(<r)}{r}} \, ,
\end{equation}
where $G$ is Newton's constant and $M(<r)$ is the total enclosed mass at galactocentric radius $r$. The total enclosed mass within a sphere of specified radius is determined using {\sc pynbody}~\citep{pynbody}. The second approach is to use the velocity derived from the gravitational potential, $\Phi(R)$, in the midplane of the galaxy as 
\begin{equation} \label{eqn:vmid}
V^{\rm mid}(R) = \sqrt{R \frac{\partial\Phi(R)}{\partial R} }\Bigg|_{{\rm z} = 0} \, ,
\end{equation}
where $R$ is the cylindrical radius, and ${\rm z} = 0$ denotes the x--y plane, as discussed in Section~\ref{subsec:amiga}. This calculation is also done using {\sc pynbody}.
 
We determine the maximum velocity of a given RC in two ways.  The first is to take the maximum of the total simulated RC. In this case, we define maximum velocities for RCs using spherical symmetry and the midplane potential as $V_{\rm max}^{\rm circ}$ and $V_{\rm max}^{\rm mid}$, respectively.  The second is to instead use the maximum velocity for HI-limited RCs within a radius, $R_1$, that is consistent with observational limits. Observed low-mass dwarfs have RCs that are sometimes still rising at the detection limit in their last radial bin~\citep{Kuzio2006, Catinella2006}, implying that the true maximum of their RCs may not be measured~\citep{Salas2017, Ruan2025}.  To align our simulation results with observational limits, we calculate the HI gas mass surface density in cylindrical shells from the host center. We define $R_{\rm 1}$, the 2D projected size of the HI disk, as the radius where $\Sigma_{\rm HI } = 1$~M$_{\odot}~{\rm pc}^{-2}$, as motivated by, e.g.,  \citet{Broeils1997} and \citet{WangJ2016}.  Again, we find $R_1 < 15$~kpc for all systems considered in this work. We define maximum velocities for HI-limited RCs using spherical symmetry and the midplane potential as $V_{\rm max}^{\rm circ+HI}$ and $V_{\rm max}^{\rm mid+HI}$, respectively. 

\begin{figure*}[]
\epsscale{1}
\plotone{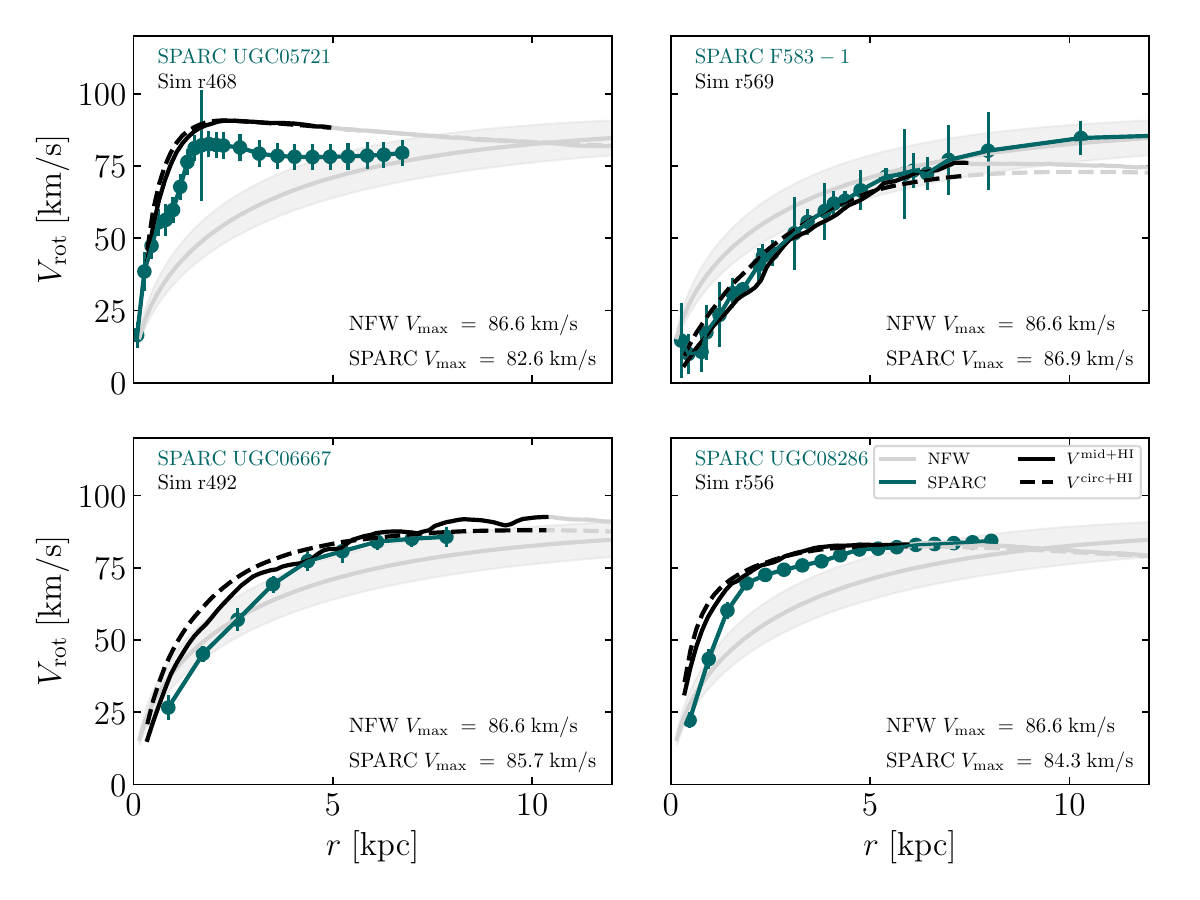}
\caption{{\sc Rotation curves} for sample galaxies from SPARC with $V_{\rm max} \sim 80$--$90$~km/s. Full SPARC RCs are shown in dark teal. In each panel, a simulated Massive Dwarf RC with similar shape and $V_{\rm max}$ to the SPARC RC is shown using spherical symmetry (Equation \ref{eqn:vcirc}) with a dashed line and from the midplane potential (Equation \ref{eqn:vmid}) with a solid line. The full simulated RCs are shown for both cases with dark gray lines ($V^{\rm circ}$ and $V^{\rm mid}$), and the HI-limited RCs are shown in black ($V^{\rm circ+HI}$ and $V^{\rm mid+HI}$). An NFW RC with $V_{\rm max} \sim 87~{\rm km/s}$, intermediate to the observed RCs, is shown with scatter in the NFW mass-concentration relationship by the light-gray shaded region. For the two simulations with disks (r492 and r569), the rotation speeds from the midplane potential peak higher and at larger radii than the RCs from the mass-enclosed method. Both the midplane and mass-enclosed RCs exhibit diversity that mimics the observed RCs; the RCs in the upper-left and lower-right panels are more quickly rising than the NFW case.  In contrast, the upper right is more slowly rising, and the lower left is in rough agreement with the NFW RC. }
\label{fig:vrot_SPARC}
\end{figure*}

Once the maximum velocity $V_{\rm max}$ ($V_{\rm max}^{\rm circ}$ or $V_{\rm max}^{\rm circ+HI}$ for spherical symmetry, or $V_{\rm max}^{\rm mid}$ or $V_{\rm max}^{\rm mid+HI}$ for the midplane potential) is determined, a proxy for the central rotation velocity is taken to be 
\begin{equation} \label{eqn:vfid}
V_{\rm fid} = V_{\rm rot}(r = r_{\rm fid}) \, ,
\end{equation}
where $r_{\rm fid} / {\rm kpc} = V_{\rm max} /(35~{\rm km/s})$ is motivated by \cite{SantosSantos2020}. $V_{\rm rot}$ is the RC obtained using either of the four methods described above (e.g., circ, circ+HI, mid, or mid+HI), which we also indicate in the superscript to $V_{\rm fid}$. $V_{\rm fid}$ is chosen as the proxy for the central velocity so we can consistently explore RC diversity and compare with previous works \citep{SantosSantos2020, Roper2023, Downing2023}. While \cite{Oman2015} used 2~kpc, $r_{\rm fid}$ carries the benefit of probing the central parts of a RC, independent of halo mass. $V_{\rm fid}$ is selected so that simulations and most high-resolution HI observations of dwarfs are able to resolve this radius, allowing for comparison between the two. We explicitly explore the difference between the definitions of maximum and fiducial velocities in Section~\ref{sec:RC_shapes}. 

For the observed RCs from \cite{Lelli2016}, we calculate $V_{\rm fid}$ using the same method as for the simulated curves.  In particular, the maximum velocity, $V_{\rm max}$ is determined for the each galaxy in the SPARC sample identified in Section~\ref{sec:data} out to the last observed radial bin. Then, for each SPARC RC we calculate $r_{\rm fid}$ and take $V_{\rm fid}$ to be the rotational velocity in the radial bin closest to that radius.

We limit our analysis to the simulated and observed dwarf galaxies with stellar masses in the range ${\rm M}_* \sim (5\times10^6)$--$(1\times10^9)~{\rm M}_{\odot}$.  We restrict the SPARC sample to galaxies with $V_{\rm max} < 100$~km/s since that is the range for most of the simulated galaxies.\footnote{One simulated galaxy with $V_{\rm max} > 100$~km/s at $z=0$ has $V_{\rm max} < 100$~km/s at $z = 0.1$, after which it underwent a major merger.} By including resolved galaxies outside the virial radius of the main host, we can increase the Massive Dwarf set by seven more galaxies. 
This results in 47 galaxies from the Marvelous Massive Dwarf zoom suite, 15 galaxies from the Extended Marvelous Massive Dwarf zooms, and 23 from the Marvel suite, for a total of 85 simulated galaxies.

The Massive Dwarf zoom-in simulations include quickly and slowly rising RCs consistent with observations, as demonstrated in Figure~\ref{fig:vrot_SPARC}. In particular, we show examples of four SPARC RCs that exhibit similar behavior at a roughly fixed $V_{\rm max} \sim 80$--$90~{\rm km/s}$. In each panel, we also show the RCs $V^{\rm mid}$~(solid line) and $V^{\rm circ}$~(dashed line) for a selected simulation. HI-limited RCs ($V^{\rm mid+HI}$ and $V^{\rm circ+HI}$) are plotted in black, whereas RCs beyond $R_1$ are shown in dark gray, suggesting that the RC would not be detected in resolved HI observations at these large radii. In the lowest-mass systems ($V_{\rm max} \lesssim 60$~km/s), the true $V_{\rm max}$ can exist at $R_{\rm max} > R_1$ \citep[see, e.g.,][and the right panel of Figure \ref{fig:RCcomp_vfid_vmax} for examples of such systems]{Ruan2025}. Further, the fiducial radius, $r_{\rm fid}$, can also be greater than $R_{1}$. In these cases, the ratio $\eta_{\rm rot} = V_{\rm fid} / V_{\rm max}$, a proxy for how fast or slow the RCs rise in the central region, may be larger than the true value if the RC is still rising. 

In Figure \ref{fig:vrot_SPARC}, the RCs derived using the midplane potential peak at larger velocities and at larger radii than the mass-enclosed case, specifically for the systems with disks (r492 and r569). This general shift is consistent with the analytic prediction for RCs of the same mass enclosed in a disk compared to distributed spherically~\citep{Binney2008}. The full RCs from SPARC~\citep{Lelli2016} are plotted in dark teal. The observed diversity contrasts an NFW RC with $V_{\rm max} \sim 87$~km/s shown in gray, with the shaded band representing the scatter in the mass-concentration relation from~\citet{Ludlow2016}. The NFW $V_{\max}$ is selected to be intermediate to those of the observed RCs. The self-similar NFW halo density profile is parameterized by the dark matter halo mass~\citep{NFW1996, NFW1997} and thus predicts identical RCs for fixed $V_{\rm max}$, which is a proxy for the mass of the system. The SPARC RCs in the upper-left and lower-right panels of Figure~\ref{fig:vrot_SPARC} are more quickly rising than the NFW prediction with $V_{\rm max} = 86.6$ km/s, but have lower $V_{\rm max}$ = 82.6~km/s and $V_{\rm max}$ = 84.3~km/s, respectively. The most slowly rising SPARC RC in the upper-right panel instead has a very similar $V_{\rm max} = 86.9$~km/s. The SPARC RC with the shape closest to NFW has a $V_{\rm max}$ that is 0.9~km/s lower than the NFW RC.

\subsection{Galaxy Sizes} \label{subsec:rh}

To determine the galaxy sizes, we use {\sc pynbody} to measure the half-light radius of its stars, $r_{1/2}$, in the Johnson $K$-band using cylindrical coordinates. The $K$-band is used to mimic SPARC~\citep{Lelli2016}. {\sc pynbody} computes magnitudes of star particles in the $K$ band using the Padova simple stellar populations~\citep{Marigo2008}, converts magnitudes to luminosities for all star particles, constructs a galaxy's total luminosity profile, and identifies the radius that encloses half of that luminosity. To verify our effective radii measurements, we generate mock $K$-band images of all our galaxies using {\sc pynbody} and measure their Petrosian profiles~\citep[$\eta(r)$,][]{Petrosian1976} with circular apertures using {\sc PetroFit}~\citep{Geda2022}. The Petrosian radius ($r_{\rm p}$) is defined as the radius at which the profile reaches a predefined Petrosian index ($\eta_{\rm p}$, where $\eta(r_{\rm p})=\eta_{\rm p}$). The total flux radius is then calculated by multiplying the Petrosian radius by a constant  ($r_{\rm total} = \epsilon_{\rm p} \cdot r_{\rm p}$). Measurements using SDSS parameters~\citep[$\eta_p=0.2$ and $\epsilon_{\rm p}=2$,][]{Strauss2002} yield nearly identical values between Petrosian half-light radii (i.e., $L(\leq r_{0.5}) = 0.5L\left(\leq r_{\rm total}\right)$, where $L$ is the projected luminosity) and our {\sc pynbody}-derived effective radii. Furthermore, for galaxies whose light profiles are well described by a single component S\'ersic model~\citep{sersic1963}, we fit 2D S\'ersic profiles using {\sc PetroFit}. The effective radii parameters of the fitted single-component Sérsic models are within $0.3$~kpc of our effective radii measurements.

\begin{figure*}
\epsscale{1.1}
\plotone{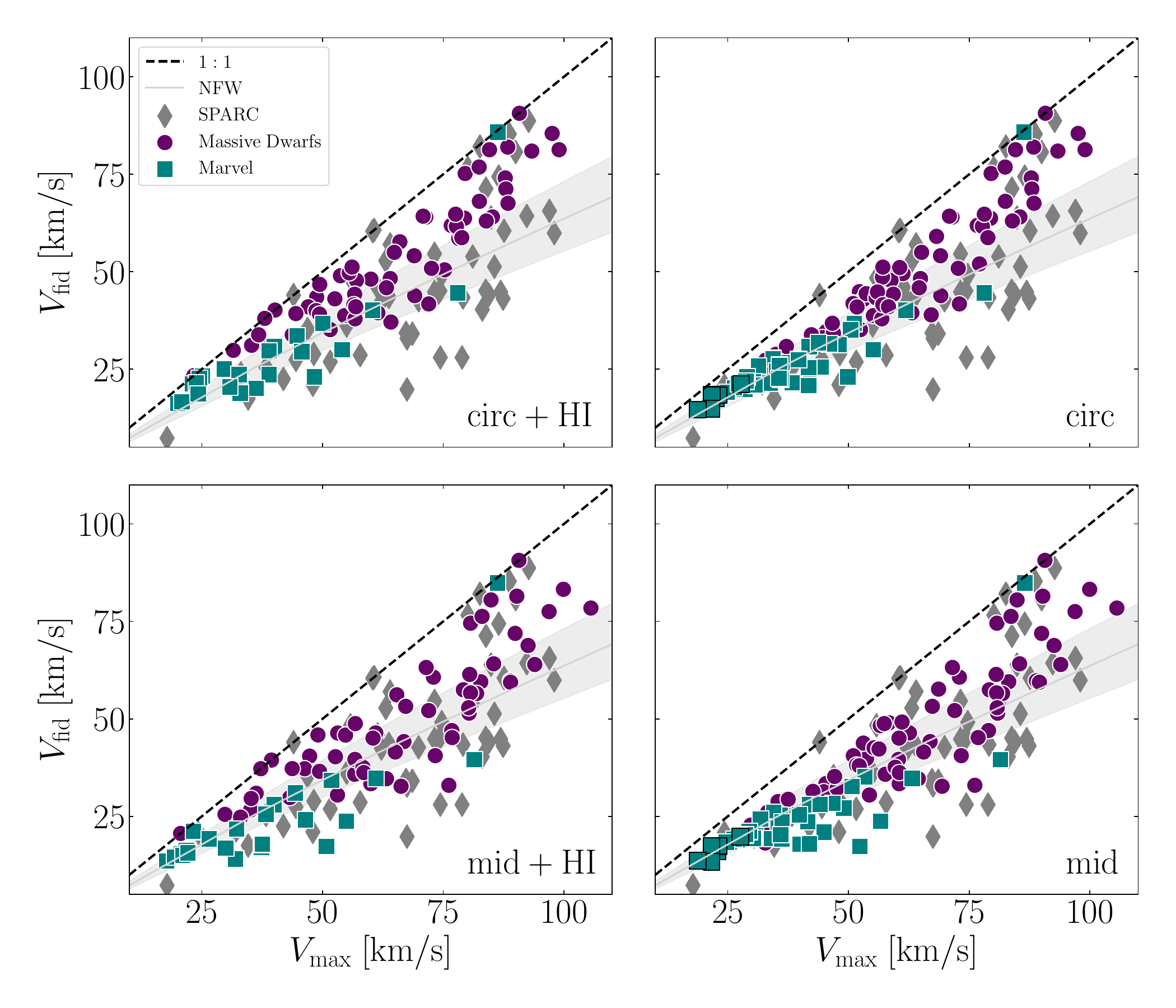}
\caption{{\sc Fiducial velocity versus maximum circular velocity} for SPARC in gray diamonds, the Massive Dwarf zooms in purple circles, and the Marvel dwarfs in teal squares.  The NFW prediction is shown by the gray line with the shaded region representing the root-mean-square scatter in the mass-concentration relation reported in \cite{Ludlow2016}. The dashed black $1:1$ line represents where $V_{\rm fid}$ is equal to $V_{\rm max}$. The results correspond to the following velocities  $V^{\rm circ+HI}$~(upper left), $V^{\rm circ}$~(upper right), $V^{\rm mid+HI}$~(lower left), and $V^{\rm mid}$~(lower right). For some of the lowest-mass Marvel galaxies, there is limited HI and $R_{1} < r_{\rm conv}$; %r_{\rm resol}$, 
we mark these galaxies with black outlines on the teal squares to denote that they are HI-limited and excluded from the left panels. A larger spread in inner velocities at $r_{\rm fid}$ at a fixed $V_{\rm max}$ is predicted for the midplane potential compared to the spherical case. RCs that are more quickly and more slowly rising than NFW are predicted with the midplane method down to $V_{\rm max}^{\rm mid+HI} \sim 50$~km/s in the Massive Dwarf zooms. 
\label{fig:vfid_vmax}}

\end{figure*}

\begin{table*}[ht]
\centering
\renewcommand{\arraystretch}{1.5}
\begin{tabular}{c|c|c|c} 
\hline
 $V_{\rm max}~[{\rm km/s}]$ & SPARC $V_{\rm fid}~[{\rm km/s}]$ & Massive Dwarfs $V_{\rm fid}^{\rm mid+HI}~[{\rm km/s}]$ & Marvel $V_{\rm fid}^{\rm mid+HI}~[{\rm km/s}]$ \\
\hline 
 20--40 & 21.4$_{19.6}^{24.4}$ (5) &  28.1$_{24.9}^{36.5}$ (8) &  17.1$_{15.3}^{22.0}$ (13) \\
 40--60 & 28.8$_{24.4}^{40.6}$ (10) & 37.4$_{36.0}^{45.9}$ (16) & 24.1$_{21.4}^{32.1}$ (5)\\
 60--80 & 44.1$_{33.0}^{54.5}$ (20) &  45.2$_{34.1}^{56.7}$ (17) & 34.7 (1)\\
 80--100 & 63.9$_{44.7}^{80.9}$ (19) &  64.1$_{56.7}^{80.6}$ (19) & 62.2$_{46.8}^{77.6}$ (2)\\
\hline 
\end{tabular}
\caption{Binned RC shape statistics for SPARC and simulated galaxies. $V_{\rm fid}$ statistics are derived for four evenly spaced $V_{\rm max}$ bins between 20--100~km/s. In each column, the 16-50-84th percentiles are provided for the specified sample, with the number of galaxies in that bin indicated in parentheses. For the simulated galaxies, $V_{\rm max}^{\rm mid+HI}$ and $V_{\rm fid}^{\rm mid+HI}$ values are reported. The Massive Dwarfs and SPARC have similar 16-50-84th percentiles between $V_{\rm max} = 60$--$80$~km/s and similar 50-84th percentiles between $V_{\rm max} = 80$--$100$~km/s, but the Massive Dwarfs have a higher 16th percentile in this range, indicating the more slowly rising RCs are missing in this bin. In the two lowest $V_{\rm max}$ bins, neither simulation suite independently matches the SPARC data well.}
\label{tab:vfid_quant}
\end{table*}

\section{Diversity of Rotation Curves} \label{sec:results}

\subsection{Rotation curve shapes} \label{sec:RC_shapes}

We quantify the diversity of the simulated RC shapes by comparing the fiducial velocities, $V_{\rm fid}$, to the maximum circular velocities, $V_{\rm max}$, in Figure~\ref{fig:vfid_vmax}. RCs derived with a spherical enclosed mass~(Equation~\ref{eqn:vcirc}) are plotted in the top row and those using the midplane calculation~(Equation~\ref{eqn:vmid}) are in the bottom row. The left column takes the maximum of the RCs out to $R_1$, while the right column uses the maximum of the total simulated RC. In each panel, the SPARC data points are shown as gray diamonds, Marvel galaxies are the teal squares, and the Massive Dwarf zooms are the purple circles. The $1:1$ line is shown in dashed black, and the NFW prediction is shown by the gray shaded region. For a given $V_{\rm max}$, the associated NFW RC is obtained by sampling the concentration from the \cite{Ludlow2016} relation; the bands in the figure correspond to the root-mean-square scatter from the concentration uncertainty. For an NFW RC with a given $V_{\rm max}$, the associated $r_{\rm fid}$ is derived, and then the $V_{\rm fid}$ is calculated using Equation~\ref{eqn:vfid}. We denote low-mass objects that do not have sufficient HI outside of the resolved region~($r_{\rm conv}$) with black outlines. These objects are not shown in the cases where $V_{\rm max}$ is determined out to $R_1$ (left column). 

To start, we compare the different RC methods against each other, starting with the circ and mid methods. For the case of circ and circ+HI~(top row), most simulated galaxy RCs have a $V_{\rm fid}$ that is above or in agreement with the NFW prediction at a given $V_{\rm max}$. The mid and mid+HI cases~(bottom row) both exhibit a lower median and more extended spread in $V_{\rm fid}$, for given $V_{\rm max}$, than seen in the top row. For example, for halos with $V_{\rm max} = 60$--$80$~km/s, the 16-50-84th percentiles for $V_{\rm fid}$ are $56.3_{48.1}^{61.8}$~km/s and $45.2_{34.1}^{56.7}$~km/s for circ+HI and mid+HI, respectively.

In Appendix \ref{appen:RC_methods}, we examine the shifts for individual galaxies when changing from circ+HI to mid+HI.  The left panel of Figure \ref{fig:RCcomp_vfid_vmax} shows the shift between the upper and lower left panels of Figure \ref{fig:vfid_vmax}. For the more massive galaxies, i.e., with $V_{\rm max} \gtrsim 60$ km/s, the largest shifts occur for systems with disks (see Appendix \ref{appen:RC_methods} for disk definition). These galaxies both shift to higher $V_{\rm max}$ and also lower $V_{\rm fid}$.  As mentioned above, a disk potential causes the RC to peak at larger velocities and at larger radii than a spherical potential \citep{Binney2008}.  Changing from the circ to mid method naturally captures the change in potential in disk-dominated galaxies.  At lower $V_{\rm max}$, shifts can occur in systems without disks.  Previous works have shown that low-mass galaxies have bursty star formation with feedback that drives the galaxies out of equilibrium \citep[e.g.,][]{Read2016, ElBadry2017, Verbeke2017, Downing2023, Sands2024}. If galaxies are undergoing outflows from the midplane, this will reduce the measurement using the midplane potential relative to the mass enclosed in spherical shells. Since outflows are centralized \citep[e.g.,][]{Brook2011}, this will be more likely to reduce $V_{\rm fid}$ than $V_{\rm max}$, consistent with the shifts seen in the lower-mass galaxies in Figure~\ref{fig:RCcomp_vfid_vmax}. In summary, while the specific mechanism that causes the shift differs across $V_{\rm max}$, in both cases the shift is due to deviation from spherical symmetry. 

Next, we examine the shifts in RC shape due to measuring $V_{\rm max}$ within $R_1$ instead of using the full $V_{\rm max}$ determined from the simulations. In the HI-limited case~(left panels of Figure~\ref{fig:vfid_vmax}), there are a number of systems where $V_{\max} \sim V_{\rm fid}$, falling close to the $1:1$ line. These same galaxies do not lie near the $1:1$ line when measuring the full $V_{\rm max}$, instead agreeing well with the NFW prediction.  The right panel of Figure~\ref{fig:RCcomp_vfid_vmax} examines this shift, which occurs predominantly in low-$V_{\rm max}$ galaxies.  In such low-mass systems, the full RCs are still rising at $R_{1}$. RCs that are still rising in the last HI-limited radial bin are also observed~\citep[e.g.,][]{McQuinn2022}, and indeed similar objects exist in the SPARC sample. In these systems, by definition $V_{\rm max}$ is measured at a smaller radius in the HI-limited case than in the full case, resulting in lower $V_{\rm max}$. Because $r_{\rm fid}$ is determined by $V_{\rm max}$, $V_{\rm fid}$ is also measured at a smaller radius.  Figure~\ref{fig:RCcomp_vfid_vmax} shows that the shift tends to be larger in $V_{\rm max}$ than in $V_{\rm fid}$, but the overall effect is that $V_{\rm max}$ and $V_{\rm fid}$ are measured much closer together in radius, and the overall galaxy shifts toward the $1:1$ line.

Throughout the rest of the work, we use the mid+HI method because we expect it to most closely mimic mock observations, i.e., capturing the impact of disk potentials, dynamical disequilibrium, and the observational $R_1$ limit. Hence, it is likely the method that best mimics the SPARC data. However, there are many other effects that can change the observed RCs of galaxies that will not be captured by any of the measurement methods, including mid+HI.  We refer the reader to Section~\ref{subsec:noncirc} for a more in-depth discussion. 

When the mid+HI method is used, we find a broad range of diversity in the $V_{\rm fid}-V_{\rm max}$ plane, with galaxies that both rise more slowly and more quickly than NFW. %especially for $V_{\rm max} \sim 60$--100~km/s. 
To quantify the spread in $V_{\rm fid}$, we split the SPARC and simulation samples into four evenly spaced $V_{\rm max}$ bins between 20--100~km/s and quantify the median $V_{\rm fid}$ and 16-84th percentiles in each. The results are summarized in Table~\ref{tab:vfid_quant}. We note that this method is not perfect as each sample has a different selection function (for the simulation suites the selection criteria is discussed in Section~\ref{sec:sims}), however it aids in comparing the samples. This table helps us explore how well either simulation set matches the observational SPARC data and whether the two simulation sets produce consistent predictions for RC shape at fixed $V_{\rm max}$. Note that we do not comment on bins that are statistically-limited, with fewer than five galaxies in them.

In the upper $V_{\rm max} = 80$--$100$~km/s bin, the median and upper percentile for SPARC and the Massive Dwarf zooms agree well; however, the lower percentile for SPARC is $\sim 10$~km/s smaller. This is visually evident in Figure \ref{fig:vfid_vmax} as we do not reproduce a few of the most slowly rising objects in SPARC between ${V_{\rm max}}\sim80$--$100$~km/s.  As discussed in Section~\ref{subsec:noncirc}, these discrepancies can potentially be explained by effects like non-circular motion in observed RCs, which tend to lower observed rotation velocities. 

In the $V_{\rm max} = 60$--$80$~km/s bin, the Massive Dwarf zooms agree very well with SPARC. In particular, SPARC has $V_{\rm fid} = 44.1_{-33.0}^{+54.5}$, while the mid+HI case has a slightly higher median but similar percentile range with $V_{\rm fid}^{\rm mid+HI}  = 45.2_{-34.1}^{+56.7}$~km/s. The Marvel zoom volumes only have a single object in this $V_{\rm max}$ range, so we cannot derive statistical results. 

In the two lowest $V_{\rm max}$ bins, neither simulation suite independently matches the SPARC data well. The Massive Dwarf zooms have higher medians than SPARC, with the lower quartile not encompassing the SPARC median. The Marvel zoom volumes have lower median compared to SPARC, but the upper quartile does a better job of encompassing the median SPARC $V_{\rm fid}$ values. 

Differences between the two simulation suites could arise from their different feedback prescriptions. The Massive Dwarf zooms are run with superbubble feedback, whereas the Marvel zoom volumes are run with blastwave feedback. In Figure~\ref{fig:storm_vfid_vmax} of Appendix~\ref{appen:feedback_RC}, we compare the diversity of RCs for galaxies in the Marvel Storm zoom volume, run with both blastwave and superbubble feedback, using $V^{\rm mid+HI}$. There is a preference towards larger $V_{\rm max}^{\rm mid+HI}$ for a given galaxy in the runs with superbubble feedback. \cite{Azartash2024} compared galaxy and halo properties in superbubble versus blastwave feedback. They found no change in stellar size between the two feedback models, but they did find that the simulations with superbubble feedback have preferentially higher halo masses compared to their blastwave counterparts (see their Figure~1). We postulate that this preference for higher halo masses is translated into larger $V_{\rm max}^{\rm mid+HI}$ in superbubble compared to blastwave feedback in the Storm simulations.

The impact of the higher $V_{\rm max}^{\rm mid+HI}$ in the superbubble runs on the diversity of RCs is not immediately obvious.  Most of the galaxies in Figure \ref{fig:storm_vfid_vmax} move mostly along the NFW expectation when changing from blastwave to superbubble feedback. This is because $V_{\rm fid}^{\rm mid+HI}$ tends to be slightly higher as well.  However, the most massive galaxy in that set makes it clear that an increase in $V_{\rm max}$ can introduce more diversity if $V_{\rm fid}$ remains approximately constant. At the moment, it is not conclusive whether the choice of feedback model impacts the diversity of RCs, or in what direction.

For a fixed $V_{\rm max}$ in the midplane case, we find values of $V_{\rm fid}$ both above and below the NFW prediction. This is in contrast to other simulation suites such as NIHAO, which only show slowly rising RCs when using the midplane method in the $V_{\rm max}$ range considered in this work~\citep{SantosSantos2018}. Similarly, EAGLE simulations run with a high density threshold for star formation also struggled to produce the most quickly rising RCs \citep{SantosSantos2020}. In both the NIHAO and EAGLE simulations that preferentially made slowly rising RCs, baryonic feedback also induced dark matter density cores in the central galaxy.  In Appendix~\ref{appen:den_profs}, we show that the Marvel and Massive Dwarf simulations can produce cored dark matter matter density profiles, i.e., these are the first simulations to simultaneously make dark matter cores while also producing quickly rising RCs.  In the next section we explore the relation between baryonic distribution and RCs in these new simulations.

\subsection{The influence of baryons on diversity of rotation curves} \label{subsec:baryons_RC}

\begin{figure*}[]
\epsscale{1.17}
\plotone{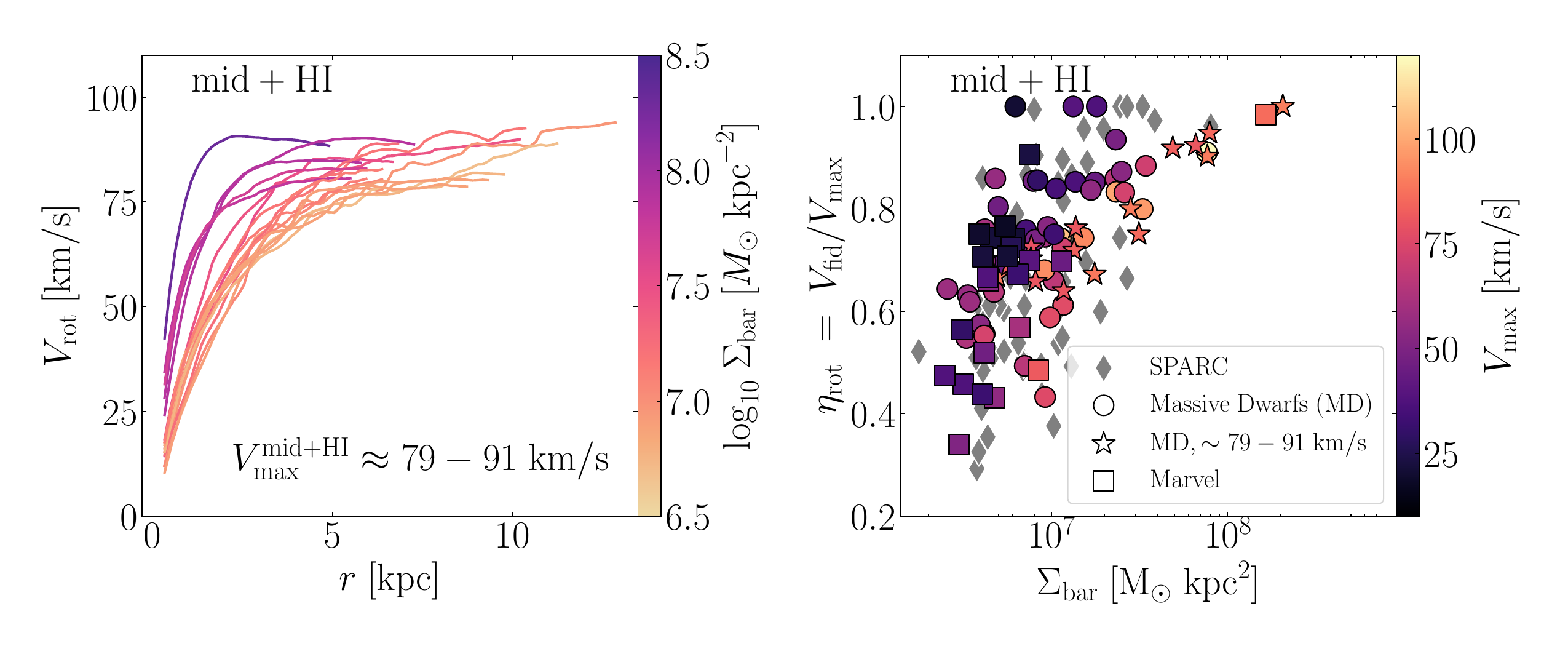} 
\caption{{\bf Left:} {\sc Midplane RCs colored by baryonic surface density}. The RCs are cut off at $R_1$ and are chosen in a $V_{\rm max}$ range that matches with \cite{Ren2019}. The HI-limited midplane method is used to calculate $V_{\rm fid}$ and $V_{\rm max}$. The most quickly rising RCs (violet) have higher surface densities, whereas the more slowly rotating RCs (orange) have lower surface densities. This trend is consistent with observed RCs in the same $V_{\rm max}$ range \citep{Lelli2016}. {\bf Right:} $\eta_{\rm rot}$ versus $\Sigma_{\rm bar}$, with simulated galaxies colored by $V_{\rm max}$ and SPARC data in gray. In the simulated and observed galaxies, higher-mass systems have higher baryonic surface densities and have RCs that tend to rise more quickly. There is large scatter in $\eta_{\rm rot}$ at lower $\Sigma_{\rm bar}$. Systems with $V_{\rm max} \sim 79$--91~km/s that appear in the left panel are highlighted with star markers.} \label{fig:sb_etarot}
\end{figure*}

How baryons contribute to RC shape, and whether the shapes of RCs are correlated with baryonic surface densities, are debated questions.  \cite{Ren2019} showed that the shape of SPARC RCs correlates with surface brightness, at least for galaxies in their highest $V_{\rm max}$ bin, 79--91~km/s.  However, \cite{SantosSantos2020} demonstrated that 
dwarf galaxies show a broad range in RC shape at a given baryonic surface density, concluding that RC shape and surface brightness are not correlated, in contrast to \cite{Ren2019}.  In this subsection, we investigate the relation between baryonic distribution and RC shape in the Marvel and Massive Dwarf galaxy suites.

To quantify the connection between RC shape and baryonic distribution in our simulations, we define baryonic surface densities using the baryonic half-mass radius. Following Section~\ref{subsec:rh}, we calculate the cumulative luminosity profile for the stars and then convert to mass using a mass-to-light ratio of $\Upsilon_* = M_*/L_* = 0.5$~M$_{\odot}/L_{\odot}$~\citep{McGaugh2014, Lelli2016, Lelli2016a}. The value of $\Upsilon_* \approx  0.5$~M$_{\odot}/L_{\odot}$ was found to minimize the scatter around the baryonic Tully-Fisher relationship \citep{Lelli2016a} and to be consistent with stellar population synthesis models with standard initial mass functions~\citep{McGaugh2014}. Mass profiles are calculated this way instead of taking the stellar mass directly from the simulations to mimic the observational procedure. The enclosed HI gas mass profile is calculated out to $R_1$ and multiplied by 1.4 to account for helium and metals in the gas phase~\citep{Arnett1996, Bradford2016}. Consistent with \cite{Lelli2016}, we do not account for the contribution of molecular gas because such data is not available for most SPARC galaxies and molecules are expected to be subdominant compared to stars and atomic gas \citep{Simon2003, Frank2016}. We find that $R_1 > r_{1/2}$, in agreement with SPARC~\citep{Lelli2016}. The simulated stellar and gas enclosed mass profiles are added, and we find the radius $r_{\rm 1/2, bar}$ where half of the total baryonic mass, $M_{\rm bar} = M_* + 1.4 M_{\rm HI}$, is enclosed. Then, we calculate the baryonic surface density as $\Sigma_{\rm bar}~=~M_{\rm bar} / 2 \pi r_{\rm 1/2, bar}^2$, consistent with \cite{SantosSantos2020}.  

To calculate the baryonic surface densities for SPARC galaxies, we derive gravitational baryonic contributions to the RCs, $V_{\rm bar}^2(r) = V_{\rm gas}^2(r) + \Upsilon_* V_{\rm *}^2(r)$ from published mass models~\citep{Lelli2016}. We note that for the SPARC galaxies considered in this study, there are no stellar bulge contributions to $V_*$.  The baryonic half-mass radii are derived using spherical symmetry with $M_{\rm bar}(<r) = rV^2_{\rm bar}(r)/G$. 

To probe the connection between baryonic surface density and RC shape, the left panel of Figure~\ref{fig:sb_etarot} examines the set of galaxies with $V_{\rm max}^{\rm mid+HI} = 79$--91~km/s to match \cite{Ren2019}.\footnote{We do not color by surface brightness as in \cite{Ren2019}, but the baryonic surface density and brightness should be consistent up to a scaling factor for a fixed mass-to-light ratio, as advocated for in \cite{Lelli2016}.} The $R_1$ cut causes each RC to end at a different radius, as for observed galaxies.  For this range of $V_{\rm max}^{\rm mid+HI}$, the most quickly rising RCs have high baryonic surface densities; baryons dominate the central potentials in these systems (see for example density profiles for r488, r502, r468, r555 in Figure \ref{fig:density}). On the other hand, the more slowly rising RCs have low baryonic surface densities. However, this trend does not exist for lower $V_{\rm max}$ values than those shown here, in agreement with \cite{SantosSantos2020}. 
 
\begin{figure*}
\epsscale{1.1}
\vspace{-1mm}
\plotone{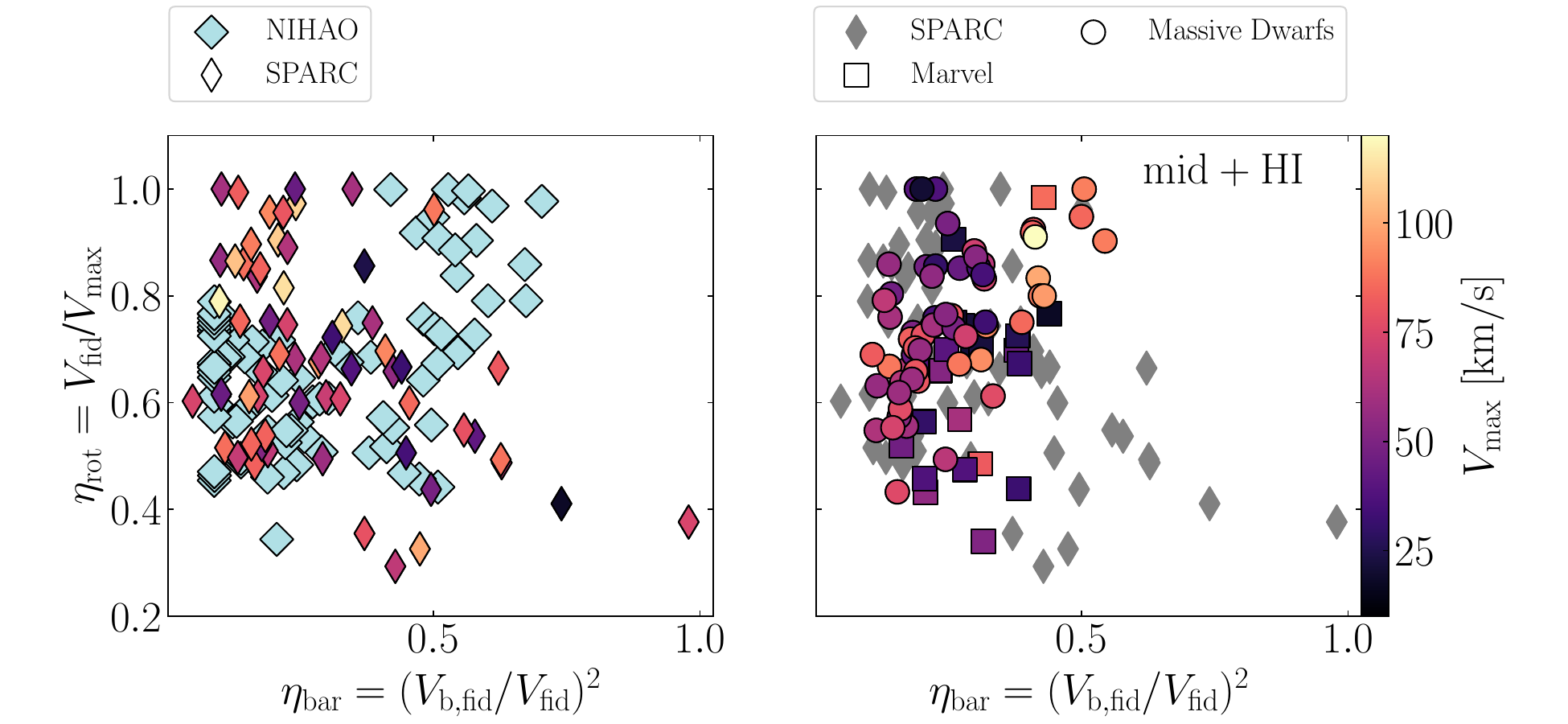}
\caption{{\sc $\eta_{\rm rot}$ versus baryon mass fraction within $r_{\rm fid}$}, colored by $V_{\rm max}$, for {\bf Left:} SPARC data in thin colored diamonds with mass modeling from \cite{Lelli2016}. Objects in the upper-left region have low central baryon mass fractions, but are still more quickly rising in their central regions than the NFW prediction. Previous CDM and baryon simulations have struggled to reproduce these objects. This is demonstrated for NIHAO data \citep[from][]{SantosSantos2020} in light blue diamonds, including galaxies with $V_{\rm max} > 100$~km/s and measured using enclosed mass. {\bf Right:} The Massive Dwarf zooms in circles and the Marvel sample in squares, using the midplane potential to calculate velocities. Most of the simulated galaxies have $\eta_{\rm bar} \lesssim 0.5$. There is a significant spread in $\eta_{\rm rot}$ in this region, including some systems with $\eta_{\rm rot} \gtrsim 0.75$, higher than the NFW prediction.}
\label{fig:etabar_etarot}
\end{figure*}

The parameter 
\begin{equation}
    \eta_{\rm rot} = V_{\rm fid}/V_{\rm max}
\end{equation}
is a proxy for the shape of a RC, as it compares the central velocity with its maximum~\citep{SantosSantos2020}. For the NFW model, $\eta_{\rm rot}$ increases slightly with decreasing $V_{\rm max}$. As a benchmark, note that galaxies with $\eta_{\rm rot} \gtrsim 0.65$--$0.75$ are more quickly rising in their central regions than the corresponding NFW prediction (with scatter from the mass-concentration relation), and galaxies with $\eta_{\rm rot} \lesssim 0.55$ are more slowly rising.  
%To explore how RC shapes and surface densities are related at the population level, t
The right panel of Figure~\ref{fig:sb_etarot} shows the relationship between $\eta_{\rm rot}$, the baryonic surface density $\Sigma_{\rm bar}$, and $V_{\rm max}^{\rm HI+mid}$ for the full simulated and observed samples. The galaxies shown in the left panel are indicated as stars, making it obvious that there is a clear trend between $\eta_{\rm rot}$ and baryonic surface density for galaxies with $V_{\rm max}^{\rm mid+HI} = 79$--91~km/s. In the simulations, galaxies with higher baryonic surface density tend to be more quickly rising, i.e., have larger $\eta_{\rm rot}$ values. However, there is significant scatter for galaxies with $\Sigma_{\rm bar} < 10^7$~M$_{\odot}$~kpc$^{-2}$, showing that the correlation with baryonic surface density evident in the left panel does not hold at lower $V_{\rm max}$ values.    

The simulated galaxies with the most quickly rising RCs (i.e., those with high $\eta_{\rm rot}$) bracket the range of surface densities in the observations, although the (two) highest-mass simulated galaxies have higher surface densities than SPARC.  On the other hand, at every $\Sigma_{\rm bar}~<~3 \times 10^7$~M$_\odot$ kpc$^2$, there is a subset of SPARC galaxies that are more slowly rising than the simulations. These SPARC systems may be influenced by non-circular motions~\citep{Read2016b, Downing2023} or inclination~\citep{Roper2023}, making them appear more slowly rising. We discuss these effects further in Section~\ref{subsec:noncirc}.

In addition to surface densities, we examine the relationship between central baryonic mass fraction and RC shape using mass models for the SPARC data \citep{Lelli2016}. Assuming spherical symmetry, the central baryonic mass fraction within $r_{\rm fid}$ can be quantified using RCs with  
\begin{equation}
    \eta_{\rm bar} \equiv (V_{\rm b, fid}/V_{\rm fid})^2 \, ,
\end{equation}
 where $V_{\rm b, fid}$ is the fiducial velocity of the baryonic RCs. \cite{SantosSantos2020} used this definition and $\eta_{\rm rot}$ to explore how the central baryon mass fraction relates to the shapes of RCs, as in Figure~\ref{fig:etabar_etarot}. The SPARC data are indicated by the diamonds in both panels, colored by $V_{\rm max}$ for the left and indicated in gray for the right. The NIHAO galaxies extracted from \cite{SantosSantos2020} are shown in light blue on the left. The Massive Dwarf galaxies (circles) and Marvel dwarfs (squares) are colored by their $V_{\rm max}^{\rm mid+HI}$ on the right panel.

For galaxies with lower baryonic mass fractions ($\eta_{\rm bar} \lesssim 0.5$), we find simulated galaxies with a spread in RC shape $\eta_{\rm rot}$ that includes objects that rise more quickly than the NFW prediction. This is in contrast to the NIHAO results shown in Figure~\ref{fig:etabar_etarot}---and also for EAGLE and APOSTLE~\citep{SantosSantos2020}, which are not shown. In particular, observed dwarfs in the upper-left of each panel in Figure~\ref{fig:etabar_etarot}, which are more quickly rising than NFW~($\eta_{\rm rot} \gtrsim 0.75$) and have low $\eta_{\rm bar}$, are absent in NIHAO. However, the Massive Dwarf zoom galaxies occupy this space.  These simulated systems are influenced by the HI-limited velocity measurement method, 
pushing them closer the $1:1$ line, as discussed in Section~\ref{sec:RC_shapes} and shown in Appendix~\ref{appen:RC_methods}. 
Although there are simulated galaxies in the upper left region, the higher $V_{\rm max}$ SPARC systems in this region are still missing from our sample. 
\cite{Roper2023} examined two simulated galaxies and showed that systems with high $\eta_{\rm rot}$ and low $\eta_{\rm bar}$ could be found in mock HI RCs along special viewing angles, an effect that we neglect here. 

Other SPARC galaxies of interest have RCs that are slowly rising (low $\eta_{\rm rot}$), but have large central baryonic mass fractions (high $\eta_{\rm bar}$). These are the SPARC galaxies in the lower-right corner of the $\eta_{\rm rot}-\eta_{\rm bar}$ plane in Figure~\ref{fig:etabar_etarot}. They are the same SPARC objects that are missing from the Massive Dwarf and Marvel simulations below the NFW prediction in Figure \ref{fig:vfid_vmax} and the missing slowly rising objects in Figure \ref{fig:sb_etarot}. NIHAO and EAGLE simulations that create dark matter cores do a better job producing  slowly rising objects with larger baryon mass fractions. However, even in NIHAO, a fraction of these SPARC systems are still missing, as shown in the left panel of Figure~\ref{fig:etabar_etarot}. It is possible that all of the simulated galaxies, including those presented here, have mass-loading factors too high relative to real galaxies \citep[e.g.,][]{Collins2022}. Indeed, as we explore below, we believe the NIHAO simulations have stronger feedback than our fiducial runs, preventing them from making both the steeply rising galaxies and compact galaxies. Alternatively, \cite{SantosSantos2018}, \cite{Roper2023}, and \cite{Oman2019} suggest that the observed galaxies in the lower-right corner may be affected by non-circular motion. 

In this subsection, we explored how and if baryons impact the diversity of RC shapes. Our simulations are able to produce quickly rising RCs with low central baryon mass fractions that other simulations have had difficulty producing.  We think the reason is partially due to the inclusion of the $R_1$ cut on RCs, used to mimic observations. Alternatively, $\eta_{\rm bar}$ may be biased by assuming spherical symmetry rather than accounting for a disk geometry. However, there is a subset of slowly-rising SPARC galaxies that we do not reproduce, and we see this in a variety of different galaxy property planes.  Moreover, these galaxies tend to have non-negligible contributions of baryons. While these objects may be better produced for stronger feedback models as in EAGLE and NIHAO \citep{SantosSantos2020}, we suspect they may instead be influenced by non-circular motions, as suggested by \cite{SantosSantos2018, Oman2019, Roper2023}, because the strong feedback scenario prevents the formation of compact galaxies found in SPARC \citep[e.g.,][]{Jiang2019}.  We examine the role of feedback in Section \ref{subsec:subgrid_phys}.

\subsection{Diversity in the size--$M_*$ plane} \label{sec:size_mass}

\begin{figure}
\epsscale{1.17}
\plotone{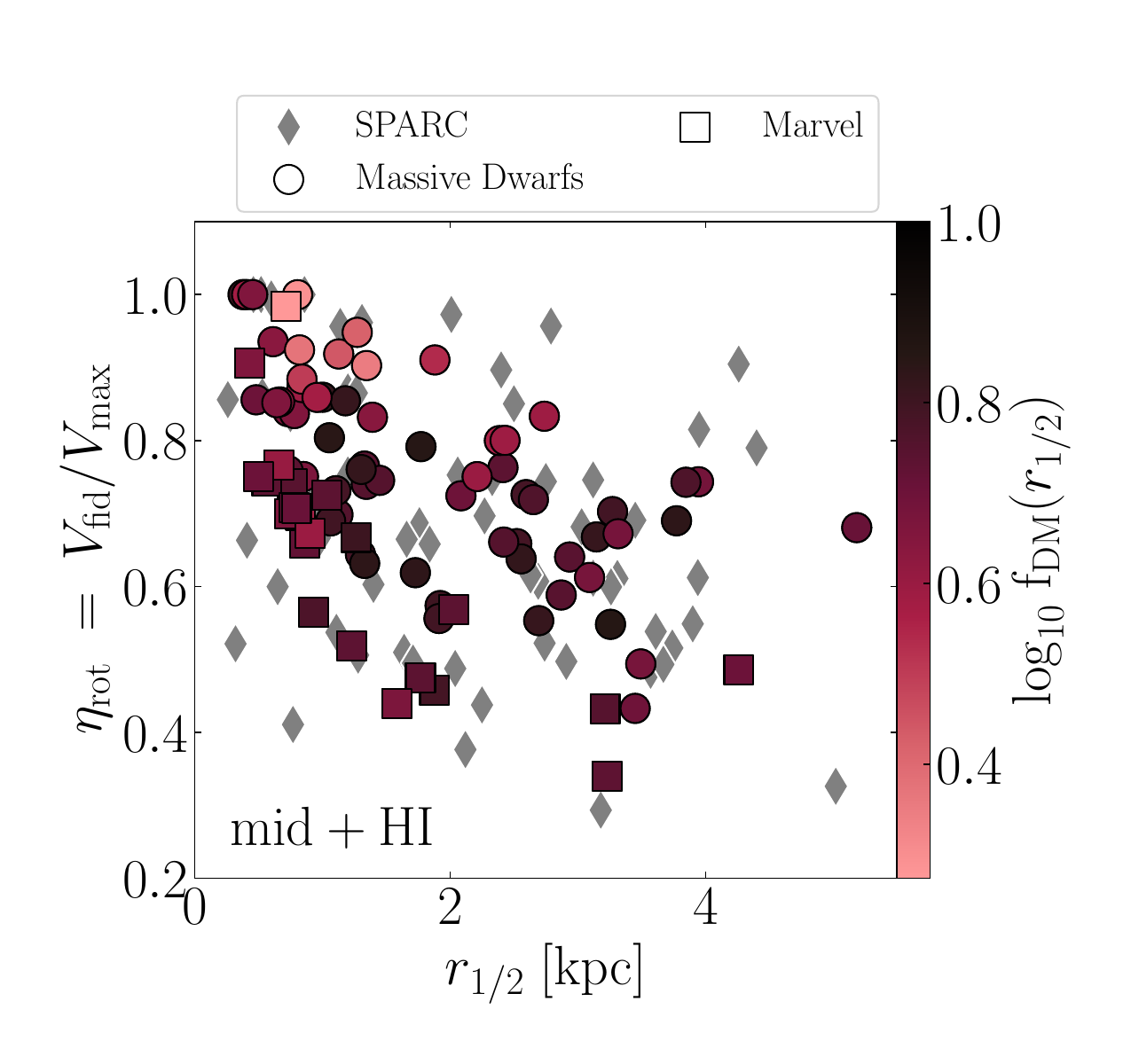}
\caption{{\sc $\eta_{\rm rot}$ versus half-light radius, $r_{1/2}$, colored by dark matter mass fraction within $r_{1/2}$} for our simulated sample at $z = 0$. The objects that are quickly rising have smaller half-light radii and $f_{\rm DM}(r_{1/2})$, meaning they are baryon dominated. However, there are also objects at lower $\eta_{\rm rot}$ that have smaller sizes  because they are less massive. The correlation between $\eta_{\rm rot}$ and $r_{1/2}$ is consistent with most of the SPARC observations.
\label{fig:rh_etarot_fdm}}
\end{figure}

\begin{figure}
\epsscale{1.15}
\plotone{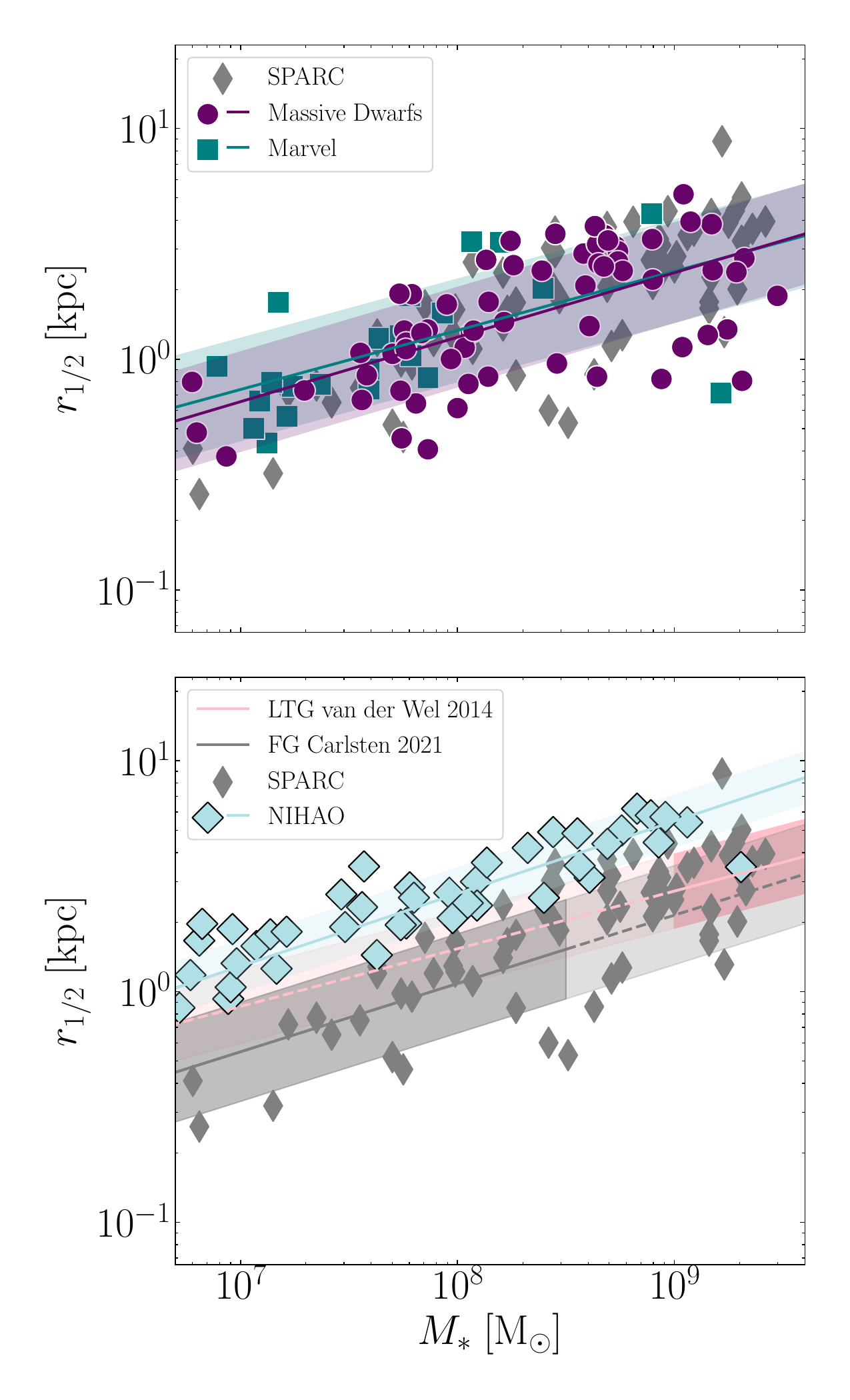}
\caption{{\sc $r_{1/2}$ versus $M_*$} for {\bf Top:} Massive Dwarf zooms in circles, the Marvel dwarfs in squares, and SPARC in gray diamonds. Power-law fits are shown for the Massive Dwarfs and Marvel in purple and teal, respectively. The Massive Dwarfs and the Marvel galaxies cover the same area as SPARC. {\bf Bottom:} The late-type-galaxy~(LTG) size--$M_*$ relation for normal-size galaxies is shown in the pink shaded region for $M_* > 10^9~{\rm M}_{\odot}$~\citep{VanderWel2011}. The field size--$M_*$ relationship is shown by the gray line and shaded region, corresponding to 1--$\sigma$ scatter, for $M_* < 10^{8.5}~{\rm M}_{\odot}$ \citep{Carlsten2021}. The extrapolation of each observed relation is shown by a dashed line. $r_{1/2}$ versus $M_*$ for the NIHAO simulation suite is shown in light blue diamonds \citep{Jiang2019}. The NIHAO simulated suite size--$M_*$ relationship is offset from the SPARC galaxies, as well as the fitted size--$M_*$ relationships, particularly below $M_* \sim 10^9~{\rm M}_{\odot}$.
\label{fig:size_mass}}
\end{figure}

Beyond the diversity in RCs, dwarf galaxies are observed to have a spread in size at a given $M_*$ \citep{Carlsten2021}.  
In Section~\ref{subsec:baryons_RC}, we found that RCs exhibit a diversity of baryonic surface densities, which may be linked to the diversity of the size--$M_*$ relationship. Figure~\ref{fig:rh_etarot_fdm} examines the relationship between $\eta_{\rm rot}$ versus $r_{1/2}$ for the Massive Dwarf and Marvel galaxies. Each simulated galaxy is colored by its dark matter mass fraction within $r_{1/2}$, $f_{\rm DM}$. Although there is scatter, objects with smaller $r_{1/2}$ tend to have higher $\eta_{\rm rot}$ and lower $f_{\rm DM}$, i.e., simulated galaxies with quickly rising RCs (high $\eta_{\rm rot}$) are also the most compact.  However, the converse is not true.  Simulated galaxies with slowly rising RCs (low $\eta_{\rm rot}$) instead show a range of sizes. This is due to the size--$M_*$ relationship and its scatter at a given mass. For fixed $V_{\rm max}$ (e.g., galaxy mass), $\eta_{\rm rot}$ can extend over the full range from $\sim 0.3-1$ (see Figure~\ref{fig:etabar_etarot}) and is also inversely correlated with $r_{1/2}$.  Moreover, given the size--M$_*$ relation,  galaxies with larger stellar masses typically have larger median sizes.  This naturally leads to the result that all galaxies (from small to large $V_{\rm max}$) converge towards  $\eta_{\rm rot}~\sim~1$ as $r_{1/2}$ decreases.  Instead, when $\eta_{\rm rot} \sim 0.4$, the lower-mass~(higher-mass) galaxies will populate the lower~(upper)-end of $r_{1/2}$ values.  The net result is significant scatter in galaxy sizes when  $\eta_{\rm rot} \lesssim 0.6$. Except for a few objects at the extremes, the simulated sample covers a similar area in the $\eta_{\rm rot}-r_{1/2}$ plane to the SPARC sample, indicated by the gray diamonds. 

In addition to struggling to reproduce the diversity of RCs, previous simulation suites have struggled to produce size--$M_*$ relationships consistent with observations. In dwarf galaxies, compact and extended galaxies are observed at a given stellar mass whereas most simulations with strong feedback only produce the most diffuse objects. We examine the size--$M_*$ relationship in the Marvel and Massive Dwarf simulations in Figure~\ref{fig:size_mass}. In the top panel, the SPARC galaxies are shown in gray diamonds, with the Marvel galaxies in teal squares, and the Massive Dwarf zooms in purple circles. The size is given in terms of the half-light radius, as outlined in Section~\ref{subsec:rh}. The bottom panel shows SPARC galaxies in gray diamonds and galaxies from the NIHAO simulation suite extracted from \cite{Jiang2019} as light blue diamonds. 
The observed trend for late-type galaxies from \cite{VanderWel2011} is shown by the pink solid line, which changes to dashed below $M_* \lesssim 10^9$~M$_{\odot}$ where it is extrapolated.
The pink shaded region indicates the 68$\%$ containment interval of scatter in the observed relation. We additionally show the observation from ELVES dwarf galaxies with the field size--$M_*$ relationship in gray for $M_*
< 10^{8.5}$~M$_{\odot}$ from \cite{Carlsten2021}. The dashed gray line denotes the extrapolation of the relationship above $M_* \gtrsim 10^{8.5}$~M$_{\odot}$. \cite{VanderWel2011} examines larger-mass galaxies with $M_* > 10^9$~M$_{\odot}$ whereas \cite{Carlsten2021} focuses on the dwarf regime with $M_* \lesssim 10^{8.5}$~M$_{\odot}$ and each are extrapolated to cover the full range of $M_*$ considered in this work. The \cite{Carlsten2021} size--$M_*$ relationship overlaps with the SPARC galaxies, but the relationship inferred at higher masses is offset below the SPARC data.

At $M_{*} \lesssim 10^9$~M$_{\odot}$, the Massive Dwarf zooms and the Marvel galaxies cover a similar area as the SPARC galaxies in the size--$M_*$ plane (upper panel of Figure~\ref{fig:size_mass}). In contrast, the NIHAO galaxies are offset from the observed relations~(lower panel). A similar offset is noted for FIRE \citep[though see][]{Klein2024} and galaxies from \cite{Lupi2017} in \cite{Jiang2019}.  We fit a power law to the size-$M_*$ relationship as 
\begin{equation} \label{eqn:powlaw}
    {\rm log}_{10}(r_{1/2}/{\rm kpc}) = a + b~{\rm log}_{10}\left(\frac{M_*}{5 \times 10^6~{\rm M}_{\odot}}\right)
\end{equation}
and calculate the scatter $\sigma$ as the standard deviation of the simulated data from the best-fit line. For the Massive Dwarfs,  $a = -0.268$, $b =  0.279$, and $\sigma = 0.218$. Marvel has a slightly shallower slope and larger scatter with $a = -0.208$, $b = 0.256$, and $\sigma = 0.225$. Both Marvel and Massive Dwarfs are in fair agreement with \cite{Carlsten2021}\footnote{We shift the log y-intercept reported in \cite{Carlsten2021} ($a_{c}$) to account for the unit conversion in log space. Additionally, we shift the relation by $M_* = 5 \times 10^6~{\rm M}_{\odot}$. So $a = a_c - 3 ~+b~{\rm log}_{10}(5\times10^6) = 0.347$, where $a_c = 0.667$.} which reports $a = -0.347$, $b = 0.296$, $\sigma = 0.215$. However, the NIHAO relationship has less scatter and is offset with $a = 0.015$, $b =  0.314$, and $ \sigma = 0.116$. For intuition on $a$, note that the $M_* = 5\times 10^6~{\rm M}_{\odot}$ intercept is $0.450$~kpc for the \cite{Carlsten2021} field galaxy relationship, compared to 0.540~kpc for the Massive Dwarfs, and 1.036~kpc for NIHAO.

Previous galaxy formation models in CDM with baryons have failed to produce compact objects below $M_{*} \sim 10^9~{\rm M}_{\odot}$ \citep{Jiang2019, Sales2022}. Given the relation we have found here between steeply rising RCs and compact sizes (Figure \ref{fig:rh_etarot_fdm}), it seems that a failure to make compact objects may be tied to why previous simulations have failed to reproduce the most quickly rising RCs~\citep{SantosSantos2018, SantosSantos2020}. We do find objects as compact and extended as the SPARC galaxies across stellar mass in both the Marvel and Massive Dwarf galaxies. 

In future work, we will examine the origin of the varying sizes in the simulated dwarfs \citep{Geda2025}.  Here, we restrict our analysis to understanding why previous simulations may not have made compact galaxies.  We explore this in the next section.

\section{The role of subgrid physics}\label{subsec:subgrid_phys} % of compact dwarf galaxies} 

The simulations studied in this work include both compact and extended objects, in better agreement with observations than previous numerical studies. For example, the sample of galaxies at $M_{*} \lesssim 10^9~{\rm M}_{\odot}$ in NIHAO and FIRE is offset from the observed size--$M_*$ relationship, with all of the objects being more extended~\citep{Jiang2019}. This section performs a detailed comparison of the Massive Dwarf zoom-in simulations to NIHAO, seeking to identify the source of this difference.
We focus on NIHAO because it was run with {\sc Gasoline2}~\citep{Wadsley2017}, which has a similar hydrodynamics code to {\sc ChaNGA}, allowing us to rerun galaxies with different parameter choices to determine their impact.

To perform this comparison, we run controlled experiments on two galaxies, r556 and r618, that are more compact than the 1--$\sigma$ interval of the observed field galaxy size--$M_*$ relationship found in \cite{Carlsten2021}. We attempt to mitigate numerial butterfly effects by ensuring the two galaxies also have quiescent merger histories after $z \sim 2$~\citep{Genel2019, keller2019}.  \cite{keller2019} showed that major mergers ($1:1$) can significantly alter stellar masses in simulations with the same ICs and subgrid physics, just re-run many times. Because \cite{keller2019} found that sizes are stable to numerical chaos, but morphologies are not, we only focus on differences in size. Beyond this, they showed that galaxies with higher surface densities, like r556 and r618, tended to be more immune to galactic chaos. While we attempt to mitigate these non-linear effects, they may still impact differences in size and mass as the subgrid physics is varied below. 

\begin{table*}
\centering
\begin{tabular}{ c l l l l l }
\hline
\textbf{Test} & \textbf{Resolution} & \textbf{Star Formation} & \textbf{Feedback} & \textbf{Feedback Energy~Budget} & \textbf{Black Holes} \\
\hline
1 (fiducial) & Mint & H$_2$  & Superbubble  & 10$^{49}$~erg/M$_{\odot}$ &  Yes  \\ %\citep{Tremmel2015} \\ 
%& & & & \citep{Tremmel2017} \\
2 & Mint & H$_2$ & Superbubble & 10$^{49}$~erg/M$_{\odot}$ & No \\
3 & Mint & NIHAO  & Superbubble  & 10$^{49}$~erg/M$_{\odot}$ & No \\
4 & Near Mint & H$_2$ & Superbubble & 10$^{49}$~erg/M$_{\odot}$ & No \\
5 & Near Mint & NIHAO & Superbubble & 10$^{49}$~erg/M$_{\odot}$ & No \\
6 & Near Mint & NIHAO &  BW  & 10$^{49}$~erg/M$_{\odot}$ & No \\
7 & Near Mint & NIHAO &  BW + Early feedback & 10$^{49}$~erg/M$_{\odot}$ + $2 \times 10^{50}$~erg/M$_{\odot}$ & No \\
%& & & \citep{Stinson2013} & \\
\hline
\end{tabular}
\caption{Compact galaxy numerical experiments varying subgrid physics. The table lists the resolution of each associated subgrid variation along with the star formation prescription and feedback model used, as well as the feedback energy budget. The details of the H$_2$-based star formation model are found in \citet{Christensen2012}, while the NIHAO star formation parameters are described in \citet{Wang2015}.  Superbubble feedback is described in \citet{Keller2014}, while the blastwave~(BW) feedback and early stellar feedback models are described in \citet{stinson2006} and \citet{Stinson2013}, respectively.  The BH prescription used in our fiducial model is described in \citet{Tremmel2015} and \citet{Tremmel2017}. Each variation is labeled by a test number, given in the first column.}
\label{table:subgrid_var}
\end{table*}

Table~\ref{table:subgrid_var} summarizes the six different subgrid physics variations that are considered. The Tests are performed in steps from the  fiducial model to a subgrid model more closely resembling NIHAO. The fiducal model (see Section~\ref{sec:MassiveDwarfs}) differs from the NIHAO subgrid model (detailed in Section~\ref{sec:nihao}) in star formation, feedback, and resolution. 

\begin{figure*}
\epsscale{1.18}
\plotone{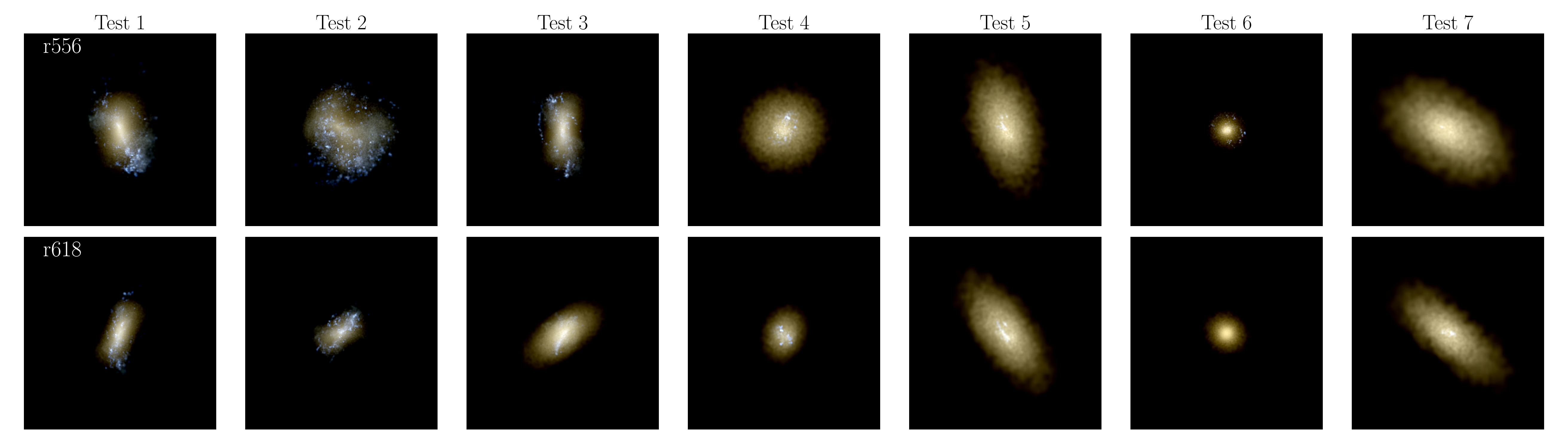}
\caption{{\sc Mock UVI images with different subgrid physics}. Each panel has a side length of 15 kpc. The two different rows are for two simulations, r556 and r618, with the following physics in the columns, {\bf first:} the fiducial H$_2$ star formation and superbubble feedback, {\bf second:} H$_2$ star formation with superbubble feedback and no BHs, {\bf third:} NIHAO star formation and superbubble feedback at the fidicual resolution, {\bf fourth:} H$_2$ star formation, superbubble feedback at Near Mint resolution, {\bf fifth:} NIHAO star formation and superbubble feedback at Near Mint resolution, {\bf sixth:} NIHAO star formation with blastwave feedback, and {\bf seventh:} NIHAO star formation with blastwave and early feedback.  Test~7 is the closest to the original NIHAO simulation model. See text for details. 
\label{fig:subgrid_uvi}}
\end{figure*}

Figure~\ref{fig:subgrid_uvi} shows the differences in UVI images due to the subgrid physics variations. The top row shows variations of simulation r556, and the bottom row shows r618.
The first column of Figure~\ref{fig:subgrid_uvi}  shows the fiducial model, designated Test~1 in Table~\ref{table:subgrid_var}. Relative to the full set of galaxies in  Figure~\ref{fig:uvi_grid}, it is visually apparent that these two are some of the most compact in the Massive Dwarf sample. Figure~\ref{fig:subgrid_size_mass} quantifies the shifts in the size--$M_*$ plane associated with each subgrid variation with r556 in magenta and r618 in black, each numbered with the associated Test listed in Table~\ref{table:subgrid_var}.  Our goal is to understand whether changes that bring the subgrid model in closer alignment to that used by NIHAO also change the compactness of these systems.

\subsection{Black Holes}\label{subsec:subgrid_phys_bh}

First, we rerun the fiducial model without BHs. This is done because the NIHAO simulations we compare to do not include BHs. In Milky Way-mass galaxies, the absence of a BH should increase star formation, altering $M_*$ \citep{Tremmel2017}; however, the expectation for how intermediate-mass BHs impact dwarf galaxies is less well studied. As was discussed in Section~\ref{sec:MassiveDwarfs}, \cite{Bellovary2019} found little BH accretion in the Marvel galaxies, which suggests that there should be little change to $r_{1/2}$ and $M_*$ when BHs are removed. The second column in Figure~\ref{fig:subgrid_uvi} shows the result of Test~2, in which the only change is to remove BHs from the fiducial model. The $r_{1/2}$ remains roughly constant in r618, but increases by roughly 50\% in r556 (see Figure~\ref{fig:subgrid_size_mass}). In addition to the two galaxies shown here, we also explored a larger set of Massive Dwarf zoom galaxies with and without BHs and found no preferential shift in the size--$M_*$ plane. The source of this size increase seen for r556 may thus be random despite attempts to minimize the impact of merger history.  We conclude that there is no clear evidence that BHs are influencing the sizes or stellar masses of these dwarf galaxies. 

\subsection{Star Formation Model}\label{subsec:subgrid_phys_sf}
The third column of Figure~\ref{fig:subgrid_uvi} (Test 3) builds upon Test~2 by altering the star formation model.  As described in Section~\ref{sec:sims}, our fiducial model follows the formation and destruction of H$_2$ and requires that H$_2$ be present when stars form.  NIHAO uses density and temperature thresholds that allow stars to form at lower densities and higher temperatures than our fiducial model. 
However, changing the star formation model at this resolution does not preferentially alter the stellar half-light radius of each galaxy, as r556 becomes smaller (returning to roughly the size of the fiducial run) and r618 becomes larger (though only by $\sim10$\%).  In other words, there is no clear evidence that changing the star formation prescription substantially impacts sizes at the fiducial Mint resolution.  

However, we do find differences when going to lower resolution that we believe are influenced by the star formation model (discussed below).  Hence, while we conclude that the star formation scheme yields similar-size galaxies at the fiducial resolution, we caution the reader that this might not be applicable under all circumstances. 

\subsection{Resolution}\label{subsec:subgrid_phys_res}
Since the mass resolution of the NIHAO galaxies in this halo mass range is lower than our fiducial resolution, we additionally explore tests at lower, Near-Mint resolution (see Section~\ref{sec:nihao} for spatial and mass resolution details). The fourth column of Figure~\ref{fig:subgrid_uvi} shows the results of Test~4, which is identical to Test~2 but at Near-Mint resolution, i.e., it is our fiducial model, minus BHs, but at lower resolution.  Compared to the second column of the figure, the stellar sizes show no preferential change: r556 becomes $\sim$20\% larger, and r618 stays roughly the same. 

Keeping the Near-Mint resolution the same, but changing to the NIHAO star formation model (Test~5)  results in larger stellar sizes compared to Test~4.  The same is true when comparing the results of Test~5 to those of Test~3 (which has the same physics, but at the higher Mint resolution).  We conjecture that the change in size is due to the fact that a star formation model based on a density threshold is inherently resolution dependent. In the NIHAO star formation model, stars start to form right when the gas density crosses the star formation density threshold $n_{*} \geq 10.3$~cm$^{-3}$ at both resolutions. 
As the resolution is lowered, the simulation loses the ability to resolve high-density peaks of gas.  For reasons which are not understood, the lower resolution results in stars at larger radii (as can be seen in Figure \ref{fig:subgrid_uvi}).  While the H$_2$ star formation model is not fully immune to resolution effects, the requirement that H$_2$ be present for star formation may act to keep sizes more stable because it limits star formation to regions with high densities, $n_{*} \geq 100$~cm$^{-3}$.
A full understanding of these effects requires additional work beyond the scope of this paper.

In summary, the NIHAO star formation model does lead to slightly larger galaxies at lower resolution, unlike our fiducial model.  However, the change is not dramatic enough to explain the sizes of dwarfs seen in NIHAO.

\subsection{Feedback}\label{subsec:subgrid_phys_fb}
NIHAO also utilizes different feedback. The remaining two tests switch from superbubble feedback to blastwave.  Test~6 is the same as Test~5, except for this one change.  As is immediately obvious from the sixth column of Figure~\ref{fig:subgrid_uvi}, the galaxies become much more compact after this switch.  The reduction in size is in agreement with \citet{Mina2021}, who found that dwarf galaxies simulated with blastwave feedback result in galaxies that have 2--3$\times$ smaller $r_{1/2}$ than the same systems run with superbubble. They conjecture that superbubble more strongly fluctuates the gravitational potential, which expands the radii of both dark matter (making cores) and stars, since both are collisionless particles \citep[see, e.g.,][]{Graus2019, Riggs2024}.  However, sizes were {\it not} found to be different between the two feedback models in \citet{Azartash2024}.  While \citet{Mina2021} adopts a density threshold for star formation, \citet{Azartash2024} adopts the same H$_2$-based star formation as our fiducial model.  This suggests again that the change in size may be dependent on the star formation model. However, it is clear that a change in feedback can yield dramatic size differences for a given galaxy.

Test~7 is the closest to the NIHAO simulations in \cite{Wang2015}.  Here, we add early stellar feedback to the physics of Test~6 as described in Section~\ref{sec:nihao}. The stellar sizes increase dramatically from Test~6, resulting in the largest galaxies across all the tests. As mentioned in Section \ref{sec:nihao}, the early feedback model is tuned to limit star formation to agree with the abundance matching value at the Milky Way scale, neglecting the impacts on dwarf scales. This choice may be causing $r_{1/2}$ to be large below $M_* \sim 10^9~{\rm M}_{\odot}$ (bottom panel of Figure \ref{fig:size_mass}). Beyond this, early feedback is added as thermal energy, similar to feedback from Type~II SNe, though cooling is not shut off.  This effectively gives a $20\times$ boost (see feedback energy budgets in Table \ref{table:subgrid_var}) to the total SN energy per star particle. Additionally, the energy is injected in a short time period, which likely helps maximize the effect. Given the change from Test~6 to Test~7, we conclude that the early feedback is the strongest contributor to making diffuse galaxies in NIHAO.

\begin{figure}
\epsscale{1.15}
\plotone{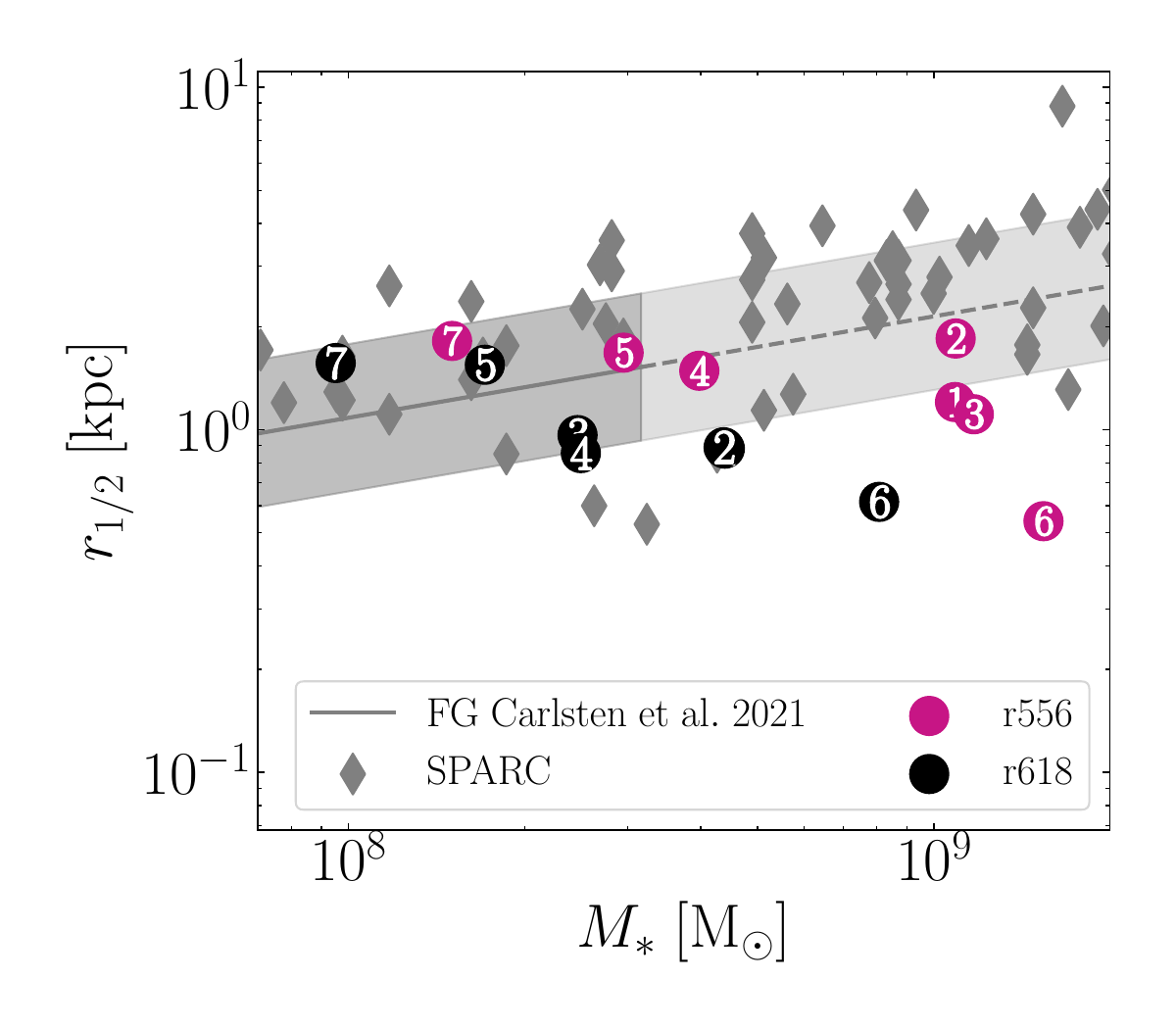}
\caption{{\sc $r_{1/2}$ versus $M_*$} for galaxy r556~(magenta) and r618~(black), run with different subgrid physics.  The marker number denotes the particular sub-grid physics test from Table~\ref{table:subgrid_var}. We note that Test 1 in r618 is sitting directly under Test 2. The SPARC data is shown by the gray diamonds, and the field size--$M_*$ relationship is shown by the gray-shaded region for $M_*
< 10^{8.5}~{\rm M}_{\odot}$ from \cite{Carlsten2021}, with the extrapolation to higher masses indicated by the dashed line. The compact objects in Test 1 (fiducial model) progressively move from compact galaxies in the lower right of the size--$M_*$ plane to the upper left; $r_{1/2}$ increases while stellar masses decrease.
\label{fig:subgrid_size_mass}}
\end{figure}

\subsection{Summary: Changes in the Size--Mass Plane}\label{subsec:sum_size_mass}
Figure~\ref{fig:subgrid_size_mass} presents the size--$M_*$ relationship for galaxy r556~(magenta) and r618~(black) as each undergoes variations in its subgrid physics, following Table~\ref{table:subgrid_var}. Each test's position in the size--$M_*$ plane is plotted with its associated number. (Note that Test~1 in r618 is sitting directly under Test~2 in the figure.) The SPARC data is shown in gray diamonds, and we again present the observed size--$M_*$ relationship with 1--$\sigma$ in the shaded region from \cite{Carlsten2021}, with the line styles as in Figure~\ref{fig:size_mass}.  The Mint resolution runs (Tests~1--3) cluster on the small side of the \cite{Carlsten2021} median relation, with some scatter.  The Near Mint test with superbubble feedback and H$_2$ (Test~4) does not change much in size, though the stellar mass of r556 decreases, moving closer to the mean size--$M_*$ relation.  Switching to the NIHAO star formation scheme increases $r_{1/2}$ and further decreases stellar mass (Test~5), but switching to blastwave feedback (Test~6) makes the most compact galaxies, indicating that resolution alone is not limiting the production of compact galaxies. Including early feedback as in NIHAO makes the galaxies the most diffuse, while also decreasing their stellar mass (Test~7).

We have not reproduced the most diffuse NIHAO galaxies in the size--$M_*$ plane, however, we have reproduced the most compact of the NIHAO galaxies (lower panel of Figure \ref{fig:size_mass}). Additionally, we have not achieved a perfect match to NIHAO with our Near Mint runs, only stepped in that direction (e.g., our particle masses and gas force resolution are slightly larger than theirs, our early feedback is a bit weaker, and we do not adopt a different, lower force resolution for the dark matter as they do). Test 6 makes it clear that resolution is not limiting the formation of compact galaxies. Further, given that the early feedback has the largest impact on size and stellar mass in our tests, we expect that increasing the early feedback efficiency from the value used here ($\epsilon_{\rm EF} = 10\%$) to the value used in \cite{Wang2015} ($\epsilon_{\rm EF} = 13\%$) would further increase sizes and decrease stellar masses. Still, the shift towards the upper left of the size--$M_*$ plane is significant ($M_*$ differs by ~1-dex between Test~1 and Test~7 in r618), and the galaxies are nearly off the median relationship. The larger driver from the bottom right to the upper left of the size--$M_*$ plane seems to be due partially to a resolution-dependent star formation model, in addition to very strong early stellar feedback.  We conclude that the strong early feedback in NIHAO may prevent the formation of the most compact galaxies. 

\section{Discussion} \label{sec:discussion}

Until now, high-resolution simulations of galaxy formation have not been able to reproduce the full diversity of RCs in dwarf galaxies, at least before turning the simulated galaxies into mock observations of RCs~\citep{Roper2023}.   %While they still lack the most slowly rising RCs, 
Our simulations are able to create dwarf galaxies that both rise more quickly and more slowly than predicted by an NFW profile, as measured by the midplane gravitational potential (bottom panels of Figure~\ref{fig:vfid_vmax}). The next subsections will discuss reasons why we might miss the most extreme objects. However, it is clear that the simulations presented here span a broader range of RC diversity than previously seen.   

There has been a debate about whether feedback from baryons can solve the RC diversity problem. For example, \citet{Williams2025} predict dark matter density profiles within CDM using a theoretical model based on statistical mechanics.  Their model shows that maximum entropy solutions for cold collisionless self-gravitating dark matter halos
can have a range of inner density profiles, including flat density cores.  They conclude that CDM density profiles can reproduce the whole range of diversity, without considering the contribution from baryons. On the other hand, \citet{DelPopolo2024} also used a large set of analytical, spherical collapse models to predict dark matter density profiles.  Their models include the impact of baryonic feedback on dark matter density profiles.  They conclude that the RCs follow the baryonic distribution, though we note that they still struggled to produce objects with the most quickly rising RCs in the $V_{\rm max} \sim 70$--$100~{\rm km/s}$ range.  

In galaxy simulations, the consensus to this point has been that baryons are responsible for creating the most slowly rising RCs in CDM, via the creation of dark matter cores from baryonic feedback \citep{SantosSantos2018, SantosSantos2020}.  However, to date, no one has been able to simultaneously create dark matter cores and the most steeply rising RCs. These troubles have led to questions about whether baryonic feedback within CDM can produce the full range of diversity \citep{Kaplinghat2020}.

We conclude that the ability to simultaneously make quickly rising RCs is related to how dense the gas is allowed to get in the centers of the simulated galaxies. To form dwarfs with high surface densities, gas must be allowed to cool and form stars at the centers of halos, without being removed by strong feedback. The ability of the Marvel and Massive Dwarf simulations to simultaneously collect cold gas at their centers while also resolving clustered, bursty star formation allows them to be the first cosmological dwarf simulations to capture a broader range of RC diversity in the absense of mock RCs.  In particular, these are the first simulations that create dark matter cores while still allowing for steep RCs in compact galaxies. 

\subsection{Observational Considerations} 
\label{subsec:noncirc}

While the simulations in this work produce a broader range of RC shapes than previous simulations, we still find a subset of SPARC galaxies in Figures \ref{fig:vfid_vmax}, \ref{fig:sb_etarot}, and \ref{fig:etabar_etarot} that are more slowly rising than produced in the simulations.  This subsection speculates how the inclusion of observational effects might impact these specific cases.

Many previous works have examined the origin of non-circular gas motions and their impact on observed RCs \citep[e.g.,][]{Valenzuela2007}. The feedback produced by bursty star formation histories in dwarf galaxies can drive them out of dynamical equilibrium 
\citep[e.g.,][]{Read2016, ElBadry2017, Verbeke2017, Downing2023, Sands2024}. Clustered SNe can create HI holes~\citep{Read2016b} inside galaxies, which can be hard to detect depending on the inclination angle~\citep[see also][]{
Verbeke2017, Sands2024}, and feedback also drives non-circular motions such as turbulence or outflows.  On top of feedback-induced non-circular motions, \citet{Marasco2018} showed that triaxial halos can induce radial gas flows (bars) in dwarf galaxies \citep[see also][]{Oman2019}. Analytically derived RC speeds in triaxial halos deviate from spherical halos \cite[see][]{Binney2008, DynamicsAstrophysicsGalaxies}. Gas can also experience pressure support that reduces the velocity relative to stars and dark matter \citep{Dalcanton2010, Oman2019, Pineda2017, Sands2024}. Finally, because the relative contribution of rotational velocity is expected to decrease with decreasing galaxy mass, galaxies instead become dominated by dispersion and their disk thickness increases~\citep{Oman2019, ManceraPina2022}.  This effect is likely to matter only in the lowest-mass galaxies, rather than our most-massive dwarfs, where diversity is maximized.  However, the presence of dispersion, at all galaxy masses, is another factor that would lower observed rotation velocities \citep[e.g.,][]{Jahn2023}

Many of the non-circular motions mentioned above tend to lower the velocities, particularly $V_{\rm fid}$, relative to what we have presented here~\citep{Oman2019, Downing2023}.  $V_{\rm fid}$ will be lowered because many of the processes discussed above are more centrally located, e.g., star formation rates are higher at the centers of galaxies, leading to more clustered star formation, feedback, and turbulence at the centers of galaxies. $V_{\rm max}$ can be lowered as well, but the result would be to shift $r_{\rm fid}$ to smaller radii and thus lower $V_{\rm fid}$ again.

If $V_{\rm fid}$ were to be lowered in mock observations, we can conjecture how our results would change (assuming that $V_{\rm max}$ is largely unaffected). First, we would observe a broader range of $V_{\rm fid}$ in Figure~\ref{fig:vfid_vmax} to lower velocities. The most slowly rising RCs are most likely to {\it appear} slowly rising due to non-circular motions \citep{Oman2019, Roper2023}. 
Second, lowering $V_{\rm fid}$ while keeping $V_{\rm max}$ unchanged will lower $\eta_{\rm rot}$.  While $\Sigma_{\rm bar}$ is not impacted (since the half-mass radius does not change), lowering $\eta_{\rm rot}$ might help to fill the space across all $\Sigma_{\rm bar}$ in the right panel of Figure \ref{fig:sb_etarot}.  Finally, \citet{Roper2023} have already shown that lower $V_{\rm fid}$ shifts galaxies to lower $\eta_{\rm rot}$ and higher $\eta_{\rm bar}$, i.e., galaxies shift into the lower-right corner in Figure \ref{fig:etabar_etarot}. 

While the most likely outcome of non-circular motion is a lowering of $V_{\rm fid}$, \citet{Roper2023} also showed some instances where $\eta_{\rm rot}$ could increase. In particular, bisymmetrical non-circular motions (i.e., bars), such as those introduced within a triaxial halo, lead to sinusoidal variations in $\eta_{\rm rot}$ as azimuthal viewing angle changes (see their Figure~8).  In some viewing angles, $\eta_{\rm rot}$ is much larger than the intrinsic value.  We reproduce several galaxies with the steepest $\eta_{\rm rot}$, but they tend to have lower $V_{\rm max}$ values than the SPARC data points with such high $\eta_{\rm rot}$. % (e.g., Figure \ref{fig:etabar_etarot}).  
While this might be simply small number statistics, it could be that such radial motions might bring some of our galaxies as high as the SPARC sample.  The simulated galaxies predominantly reside in triaxial halos below M$_{\star} < 10^8$~M$_{\odot}$ \citep{Keith2025}, and we have witnessed the existence of bars in some of them (to be explored further in future work).  Such radial motions might cause the most extreme steeply rising RCs.

\subsection{The Role of Resolution} 
In this subsection, we explore how the ratios of dark matter to baryon particle mass impact RCs and stellar sizes. Numerical work has been conducted to examine where artificial two-body relaxation impacts central dark matter density profiles \citep{Power2003, LSB2019}. Cosmological simulations often include mixtures of heavier dark matter and lighter baryonic particles. The unequal masses can result in so-called ``mass segregation"\footnote{This nomenclature is taken from \cite{Ludlow2019}.} as the particles try to reach energy equipartition, causing the more-massive particles to lose energy and fall inwards while the lower-mass particles heat up \citep{Ludlow2019}. 

In simulations where the stellar particles are less massive than their dark matter counterparts, stellar sizes can be artificially extended for low-mass systems at $z = 0$~\citep{Ludlow2019}. The simulations presented in this work have large mass ratios between dark matter and stars: $\mu = m_*/m_{\rm DM} \sim 18$ for the  Massive Dwarf zoom ins and $\mu \sim 15.8$ for the Marvel galaxies. This is because stars form with 30\% the mass of their parent gas particles. \cite{Ludlow2019} found that the convergence radius $r_{\rm conv}$ for two-body scattering, rather than the gravitational softening, sets the smallest stellar sizes obtainable. They additionally found that convergence was attained at all $z$ when the stellar sizes exceeded $r_{\rm conv}$. All the galaxy sizes in our simulation suites are larger than $r_{\rm conv}$, as calculated in Section~\ref{subsec:rotcurves}, which indicates that the stellar sizes should be converged.\footnote{\citet{Azartash2024} showed that stellar sizes in Marvel galaxies stop changing at $r_{1/2} \sim 0.34~{\rm kpc}$, smaller than the galaxies we consider here, but consistent with \cite{Ludlow2019}.} 

RCs are also influenced by large mass ratios between dark matter and star particles because of mass segregation. \cite{Ludlow2019} found that dark matter RCs were only marginally impacted for large mass ratios~($\mu > 1$). However, the RCs associated with the stellar components were more significantly affected, with the full RC increasing in velocity with time more than the $\mu =1$ case \citep[see Figure~1 of][]{Ludlow2019}. 

Dark matter dominates the majority of our RCs at all radii. Larger $\mu$ values have little impact on the dark matter contribution to the RC. On the other hand, the most rapidly rising RCs in our sample are dominated by stars (see density profiles in  Appendix~\ref{appen:den_profs}), and the baryonic component of RCs is strongly impacted by mass segregation. In the absence of mass segregation, RCs would rise even more quickly, particularly in the systems dominated by stars in the central regions. Hence, while mass segregation could be lowering $V_{\rm fid}$ in these galaxies, it does not negate the conclusion that we can make steeply rising RCs and hence that we make a broader range of RC diversity than previous hydrodynamical simulations within CDM.

\section{Summary and Conclusions} \label{sec:conclusions} 
This paper explored the diversity of dwarf galaxy properties in 85 simulated galaxies from the Massive Dwarf zoom ins and the Marvel Dwarf zoom volumes. We compared the properties of their simulated RCs and sizes to SPARC data \citep{Lelli2016}. We found RCs that span a broader range of diversity than previously seen in $\Lambda$CDM simulations with baryons. 

We explored the shapes of RCs, considering both the true halo $V_{\rm max}$ and that inferred when HI is resolved out to 1 M$_{\odot}$ pc$^{-2}$. For the Massive Dwarfs, we find RCs that both rise slower and faster than the NFW prediction for $V_{\rm max} \gtrsim 60$~km/s when RCs are calculated using the gravitational midplane potential instead of the mass enclosed. 

We also found for the first time that the simulated galaxies are a better match to the stellar size--$M_*$ relationship, in contrast with previous simulations, which showed an offset in the size--$M_*$ plane by $\gtrsim 1$--2~kpc, with simulated galaxies more diffuse than observed~\citep{Jiang2019}. Instead, we found dwarf galaxies from the Massive Dwarf zoom ins with diverse sizes at a fixed stellar mass, covering a similar area in the size--$M_*$ plane as the SPARC galaxies and having similar power-law slopes, scatters, and offsets to the \cite{Carlsten2021} size--$M_*$ power-law relationship.

We examined whether RC shapes are correlated with baryonic distributions, as previous works have drawn differing conclusions on this topic. We found the correlation varies with $V_{\rm max}$. In a fixed bin of higher $V_{\rm max}$, the most quickly rising RCs belong to more compact, high baryonic surface density galaxies, and the most slowly rising occurred in lower baryonic surface density galaxies. This is in agreement with previous work by \cite{Ren2019}. However, this was not true for lower $V_{\rm max}$ systems, which showed a larger spread in RC shapes (measured by $\eta_{\rm rot}$) at a fixed baryonic surface density, in line with \cite{SantosSantos2020}. We found significant diversity at lower surface densities and low baryon mass fractions within $r_{\rm fid}$.  

By re-running two of our most compact galaxies, we conclude that the early stellar feedback used in the NIHAO simulations likely prevents the formation of the most dense galaxies, leading to galaxies that are more diffuse than those observed \citep{Jiang2019}.  \citet{SantosSantos2020} showed that their EAGLE-CHT10 model, which has clustered, bursty star formation that leads to dark matter cores, also makes diffuse dwarf galaxies and none that are more quickly rising than NFW below $V_{\rm max} \sim$ 100 km s$^{-1}$ \citep[see also][]{Roper2023}. Although we were unable to rerun their models, it seems clear that they likely also have feedback strong enough to prevent the formation of compact galaxies. 

However, there is a subset of the most slowly rising SPARC RCs with large baryon mass fractions that we were unable to reproduce, and this is apparent in multiple galaxy property planes. The existence of these slowly rising RCs with large baryon mass fractions and high surface densities remains physically perplexing. Simulations with stronger feedback do a better job of populating this region, but at the expense of making steeply rising RCs \citep{SantosSantos2020} and the most compact observed dwarf galaxies. We postulate that the RCs of these systems may be impacted by non-circular motions, as found using mock HI RCs in \cite{Roper2023}. We also note that works that generate mock RCs should ensure and report that they are also able to properly produce the size--$M_*$ relationship in dwarf galaxies to fully address the diversity of dwarfs problem. We do not perform a similar analysis in this paper, but hope to in future work. 

We conclude that the ability to simultaneously make quickly rising RCs is related to how dense the gas is allowed to get in the centers of the simulated galaxies. To form dwarfs with high surface densities, gas must be allowed to cool and form stars at the centers of halos, without being removed by strong feedback.  The ability of these ChaNGa dwarfs to simultaneously collect cold gas at their centers while also resolving clustered, bursty star formation allows them to be the first cosmological dwarf simulations in CDM to capture a broader range of size and RC diversity. In particular, these are the first CDM simulations that create dark matter cores while still allowing for steep RCs in compact galaxies.

It has also been argued that the diversity of RC shapes is better reproduced in a self-interacting dark matter~(SIDM) model \citep{Spergel2000} than in CDM \citep{Kaplinghat2014, Ren2019, Kaplinghat2020}. 
Assuming a low SIDM cross section allows dwarfs in this mass range to produce thermalized dark matter density cores that react to the potential of the baryonic distribution. In this model, RC shapes correlate with high and low surface brightness galaxies. However, the absence of the most compact objects in previous simulations will make producing diversity impossible for galaxies simulated in this SIDM case, as well as the CDM case. We will explore this in a forthcoming work.

\section{Acknowledgements}
AC thanks Julianne Dalcanton, Nico Garavito Camargo, Adrian Price-Whelan, and Isabel Santos-Santos for helpful discussions on this work. AMB acknowledges support from NSF grant AST-2306340 and from grant FI-CCA-Research-00011826 from the Simons Foundation. ML is supported by the Department of Energy~(DOE) under Award Number DE-SC0007968 and the Simons Investigator in Physics Award. AHGP is supported by National Science Foundation Grant No. AST-2008110 and AST-2510899. FDM acknowledges support from NSF grant PHY2013909. This research was supported in part by grant NSF PHY-2309135 to the Kavli Institute for Theoretical Physics (KITP). This work was performed in part at Aspen Center for Physics, which is supported by National Science Foundation grant PHY-2210452. BWK acknowledges support provided by NASA through a grant from the Space Telescope Science Institute, through grant HST AR-17547. This work used Stampede2 at the Texas Advanced Computing Center (TACC) through allocation MCA94P018 from
the Advanced Cyberinfrastructure Coordination Ecosystem: Services $\&$ Support (ACCESS) program, which is supported by U.S. National Science Foundation grants 2138259, 2138286, 2138307, 2137603, and 2138296. Resources supporting this work were provided by the NASA High-End Computing (HEC) Program through the NASA Advanced Supercomputing (NAS) Division at Ames Research Center. Some of the simulations were performed using resources made available by the Flatiron Institute. The Flatiron Institute is a division of the Simons Foundation.

\vspace{5mm}

\appendix

\section{Comparison of rotation curve shape measurement methods}\label{appen:RC_methods}
\setcounter{figure}{0}

The main text presents four different ways of measuring maximum and central velocities, $V_{\rm max}$ and $V_{\rm fid}$. Here, we present a direct comparison between methods for each of the Massive Dwarf zoom galaxies. We first compare the midplane method for deriving rotation curves~(RCs) to the spherical mass-enclosed method (Equation~\ref{eqn:vmid} versus~\ref{eqn:vcirc}). The left panel of Figure~\ref{fig:RCcomp_vfid_vmax} compares the mid+HI in dark purple to the circ+HI method in light purple, with a black line connecting the same galaxy measured with the two different approaches. If the galaxy has a stellar disk, it is denoted with an X marker; if it does not have a disk, it is marked with a circle. We use the specific angular momentum $j$ of the stars to define disks. After aligning the galaxy in the x-y plane using the angular momentum vector of the gas in the central 5 kpc, galaxies with a median stellar z-component $j_z/j > 0.5$ are defined as disks. We find that galaxies with disks have larger $V_{\rm max}$ and lower $V_{\rm fid}$ values. This is in agreement with the analytic prediction that RCs derived for disks using gravitational potentials peak at larger velocity and galactocentric radius compared to the spherical mass-enclosed case. Thus, disk galaxies shift to the lower right of the $V_{\rm fid}-V_{\rm max}$ plane, sometimes falling below the NFW prediction when the mid+HI method is used.  Non-disks also shift, with lower $V_{\rm max}$ systems shifting more in $V_{\rm fid}$ than $V_{\rm max}$.  As we discuss in the main text, this is likely due to the fact that the feedback in lower-mass galaxies can easily throw them out of dynamical equilibrium.

The right panel of Figure~\ref{fig:RCcomp_vfid_vmax} compares the case where RCs are measured out to the true $V_{\rm max}$ versus that of truncating the RCs at $R_1$ to mimic observational limits. Most velocity measurements are unaffected above $V_{\rm max} \sim 60$~km/s. For the lower-mass systems, galaxies shift to the lower left of the $V_{\rm fid}-V_{\rm max}$ plane, closer to or at the $1:1$ line when the mid+HI is used instead of the mid method. In the lower-mass systems, there is less HI and $R_1$ decreases \citep[see][]{Ruan2025}. The smaller $R_1$ leads to a decrease in the measured $V_{\rm max}$.  Thus, $r_{\rm fid}$ decreases since it is proportional to $V_{\rm max}$, and $V_{\rm fid}$ also decreases. 

\renewcommand{\thefigure}{\Alph{section}\arabic{figure}}
\setcounter{figure}{0}

\begin{figure}[]
\epsscale{1.15}
\plotone{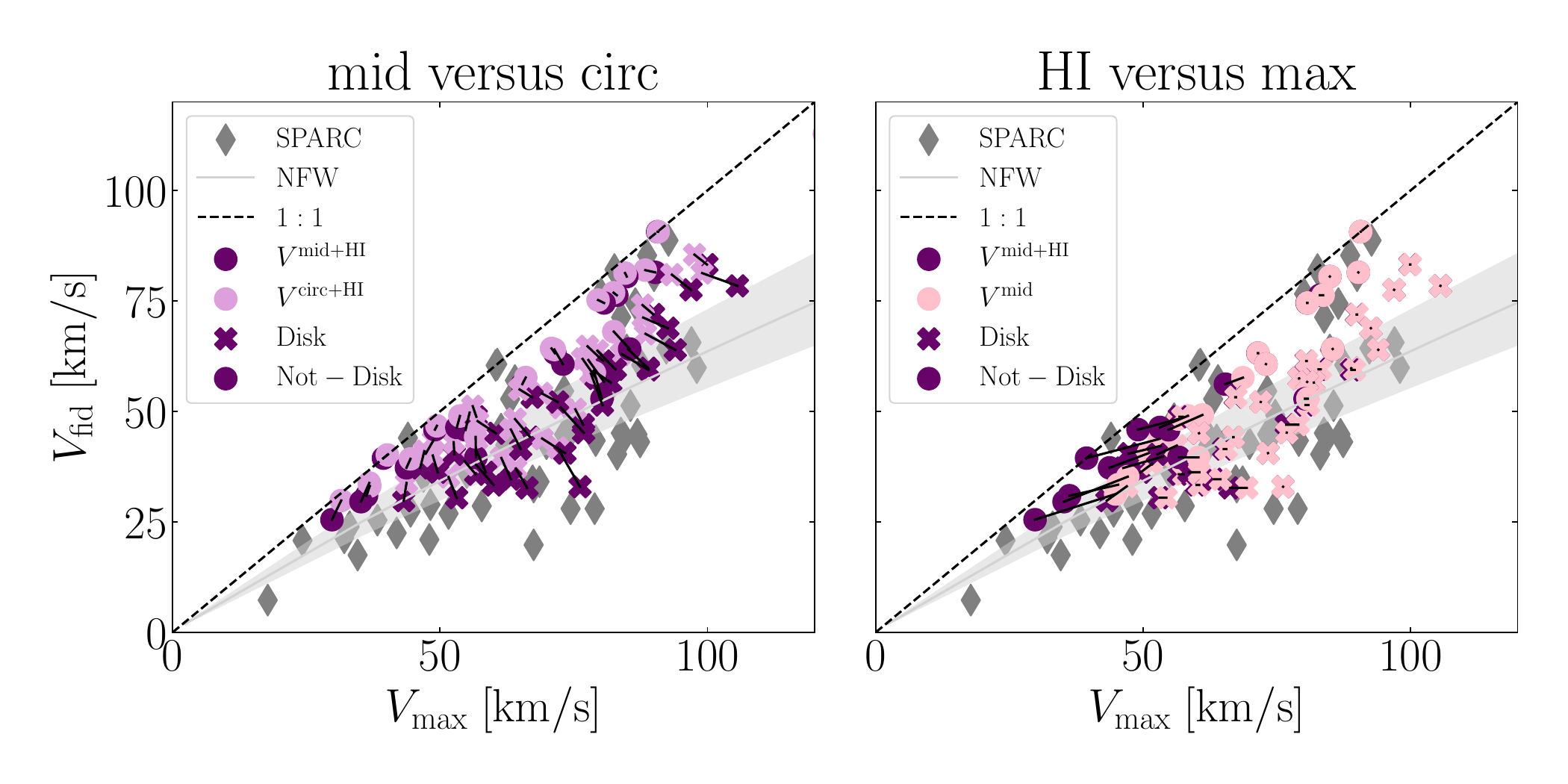}
\caption{{\sc RC shape measurement comparison} for the Massive Dwarf galaxies. In both panels, the SPARC galaxies are shown in gray diamonds, the NFW prediction is the gray shaded region corresponding to the scatter in the mass-concentration relationship, and the black dashed line is the $1:1$ line. Additionally, disk galaxies are denoted as X markers and the galaxies which are not disk are shown as circles. {\bf Left:} A comparison between the mid+HI method (dark purple) and the circ+HI method (light purple) is shown. When disks are present the Massive Dwarfs shift to lower $V_{\rm fid}$ and larger $V_{\rm max}$ for the mid+HI compared to circ+HI. {\bf Right:} A comparison between the mid+HI (dark purple) and the mid method (pink) is shown. For the higher-mass systems, there is little change between the two methods, and it is only at the low $V_{\rm max}$ end where galaxies start to have lower $V_{\rm max}$ and $V_{\rm fid}$ for the mid+HI compared to circ+HI case. This is because in the HI case, the RC is still rising at $R_1$, causing a shift to lower $V_{\rm max}$. This shift causes $r_{\rm fid}$ to be measured at a lower radius along the same RC curve, decreasing $V_{\rm fid}$.  
\label{fig:RCcomp_vfid_vmax}}
\end{figure}

\section{The impact of superbubble versus blastwave feedback}\label{appen:feedback_RC}

\cite{Azartash2024} explored differences in burstiness of star formation and impact on dark matter density profiles in superbubble compared to blastwave feedback for the Marvel volume Storm galaxies. They also explored additional galaxy scaling relations. We select the galaxies in both Storm zoom volumes that match our stellar mass cut, and we match each halo in the two different runs using dark matter particle IDs as detailed in \cite{Azartash2024}. In Figure~\ref{fig:storm_vfid_vmax}, we show the pairs with the matches connected with black lines. The superbubble runs are shown in maroon and the blastwave runs in black.  The mid+HI method is used in both cases. There is a preference towards larger $V_{\rm max}$ for a given galaxy in the runs with superbubble feedback. \cite{Azartash2024} noted a systematic increase in $M_{\rm vir}$ in the superbubble runs compared to the blastwave runs, likely translating to the preference for larger $V_{\rm max}$ in the superbubble compared to blastwave runs. For the lower-mass systems, the RCs are pushed towards or above the NFW prediction shown by the gray line. In the most massive system however, only $V_{\rm max}$ increases. 

\renewcommand{\thefigure}{\Alph{section}\arabic{figure}}
\setcounter{figure}{0}
 
\begin{figure}[]
\epsscale{0.65}
\plotone{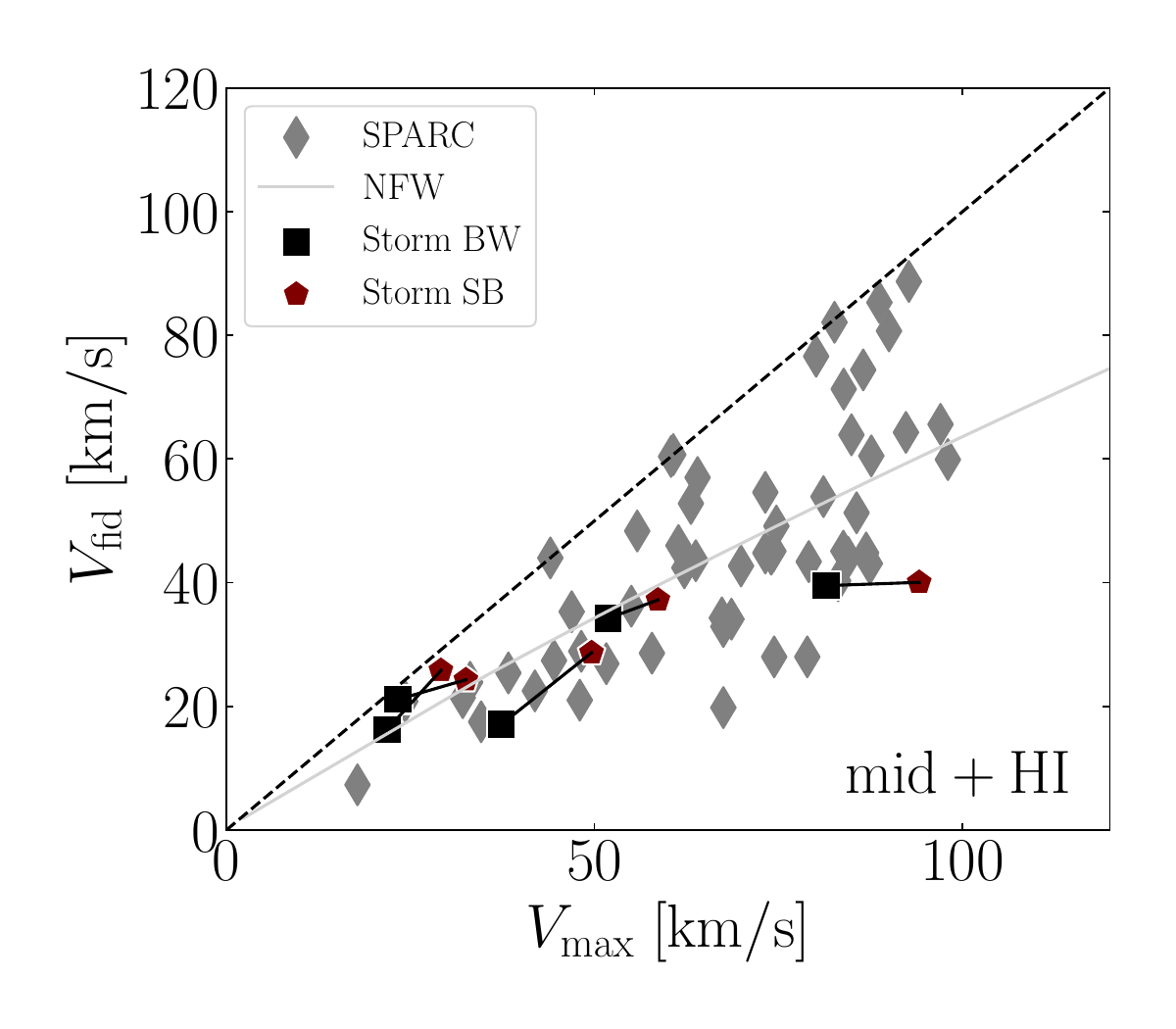}
\caption{{\sc $V_{\rm fid}$ versus $V_{\rm max}$ comparison} in galaxies run with superbubble (maroon pentagons) versus blastwave feedback (black squares), with SPARC data shown by the gray diamonds, and the NFW prediction is the gray. The black lines connect the same halo in the blastwave versus superbubble cases. There is a preference for higher $V_{\rm max}$ for galaxies with superbubble compared to those in blastwave.
\label{fig:storm_vfid_vmax}}
\end{figure}

\section{Density profiles}\label{appen:den_profs}
\setcounter{figure}{0}

\begin{figure}
\epsscale{1.15}
\plotone{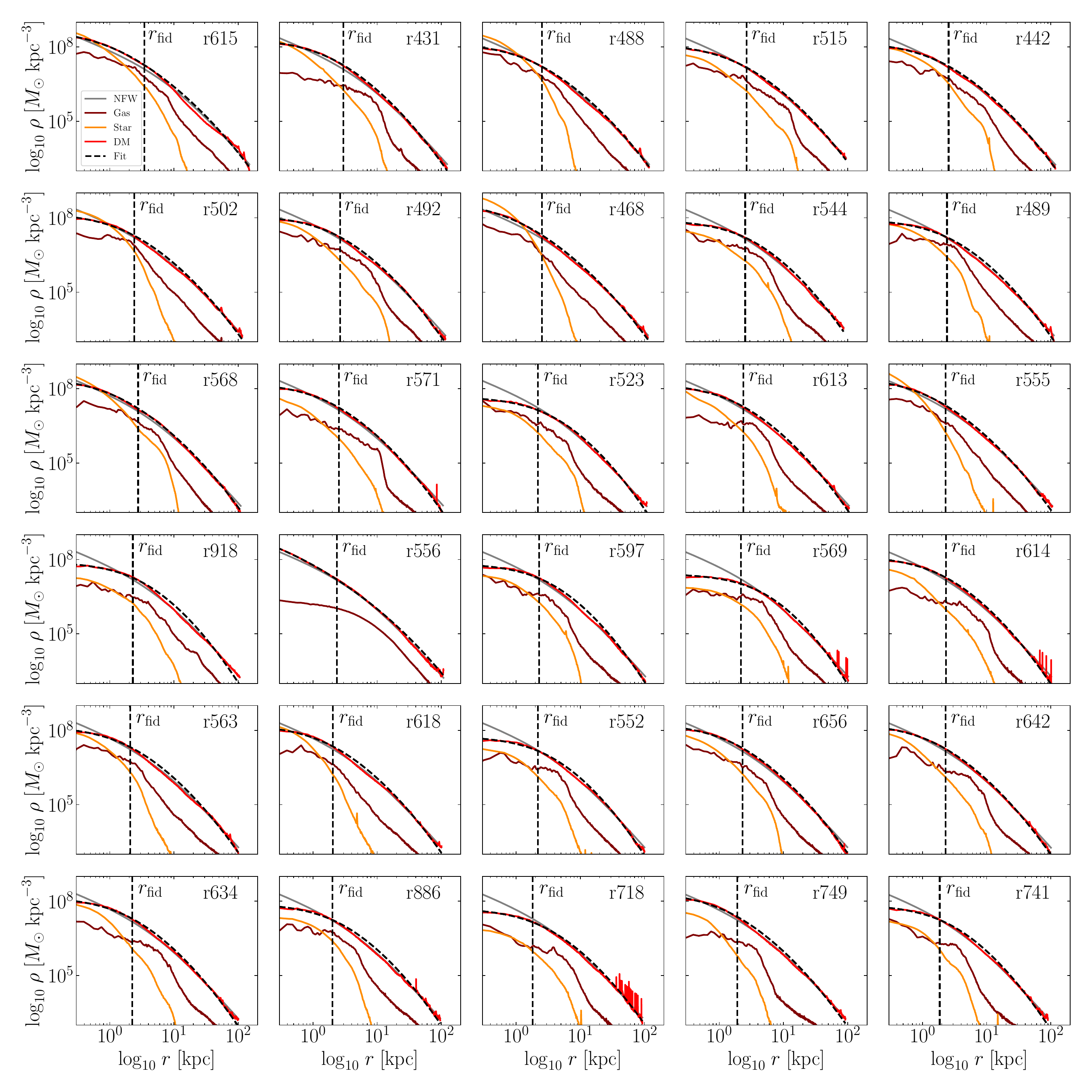}
\caption{{\sc Density profiles for the 30 most-massive Massive Dwarf zooms} in maroon for gas, orange for stars, and red for dark matter. Cored Einasto fits are shown in black dashed lines and NFW in gray. The dashed black vertical line in each sub-panel indicates $r_{\rm fid}$, the point where we measure the density slope $\alpha$. 
\label{fig:density}}
\end{figure}

Figure~\ref{fig:density} shows the density profiles  (outside the convergence radius $r_{\rm conv}$) for the 30 most-massive galaxies in the Massive Dwarf set.  Using the simulation particle data, we show the spherically averaged densities for dark matter, gas, and stars in red, maroon, and orange, respectively.  The sub-panels are ordered by $M_{*}$ from  top to bottom and left to right in a given row, with the simulation name in the upper right-hand corner of each sub-panel. The value of $r_{\rm fid}$ is indicated by the dashed black vertical line in each sub-panel.

We fit all $z = 0$ dark matter density profiles to a core-Einasto profile \citep{Lazar2020}, 

\begin{equation} \label{eqn:ce_fit}
    {\rm log} \Bigg[\frac{\rho_{\rm cEin}(r)}{ \rho_s}\Bigg]  = - \frac{2}{\alpha_{\epsilon}} \Bigg[ \Bigg( \frac{r + r_c}{r_s} \Bigg)^{\alpha_{\epsilon}} - 1\Bigg] \, ,
\end{equation} 
where $\rho_s$ is a normalization parameter, $r_c$ is the core radius, $r_s$ is the scale radius, and  $\alpha_{\epsilon}=0.16$.  
We restrict the density profile fits to the radial range of r$_{\rm conv}$ to $R_{200}$. The lower bound is intended to limit artificial two-body heating at low radii \citep{Power2003, LSB2019}.

Following \cite{Lazar2020}, we quantify the goodness of fit using the quality factor $Q$, defined as 
\begin{equation} \label{eqn:Q}
    Q^2 = \frac{1}{N} \sum_i^N ~[{\rm log}_{10} \rho_i - {\rm log}_{10} \rho_i^{\rm fit}]^2 \, ,
\end{equation} 
where $N$ is the number of radial bins, and $\rho^{\rm fit}$ is the fit defined using Equation~\ref{eqn:ce_fit} with the relevant parameters chosen to minimize $Q$. The best-fit core-Einasto curve is shown as the dashed~black curve in each sub-panel of Figure~\ref{fig:density}.

The slope of the dark matter density profile is 
\begin{equation} \label{eqn:ce_fit_slope}
    \frac{{\rm dlog}~\rho_{\rm cEin}(r)}{{\rm dlog}~r}  = - 2 \Bigg(\frac{r}{r_s}\Bigg)^{\alpha_\epsilon} \Bigg( 1 +  \frac{r_c}{r} \Bigg)^{\alpha_{\epsilon} -1} 
\end{equation} 
and is related to the inner slope of the RC, i.e., the shape of the inner RC. More explicitly, halos with central densities with slope $\alpha$ ($\rho \sim r^{\alpha}$) have central mass profiles  
\begin{equation}
    M(r) \sim \int_{0}^r dr' r'^{\alpha} r'^2  \sim r^{3 + \alpha} 
\end{equation}
and the associated spherical mass-enclosed circular velocity profiles rise as 
\begin{equation}
    V_{\rm circ} \sim \sqrt{r^{3 + \alpha} / r} \sim r^{1 + \alpha/2} \, .
\end{equation}

When a disk is present, it will dominate the midplane potential, further altering the shape of the RCs~\citep{Binney2008}.  As noted in \cite{Oman2015}, the size of the core also influences the shape of RCs since $V_{\rm circ}$ depends on mass enclosed.

\begin{figure}[]
\epsscale{0.75}
\plotone{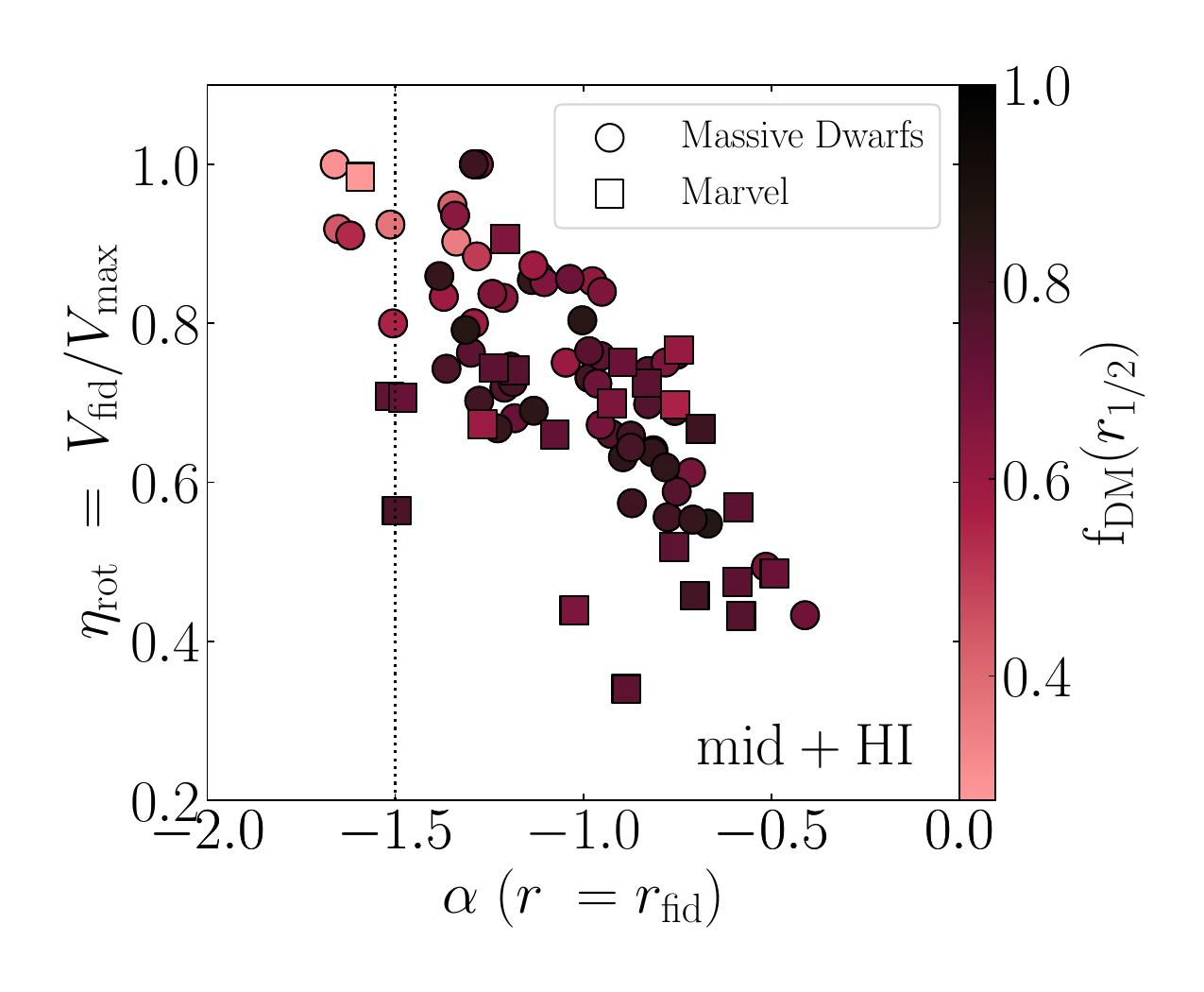}
\caption{{\sc $\eta_{\rm rot}$ versus inner slope $\alpha$ colored by dark matter mass fraction within $r_{1/2}$, $f_{\rm DM}(r_{1/2})$}, for the Massive Dwarf~(circles) and Marvel~(squares) galaxies at $z = 0$. The inner slope $\alpha$ is measured at $r_{\rm fid}$. More quickly~(slowly) rising RCs have steeper~(shallower) dark matter slopes. 
\label{fig:alpha_etarot}}

\end{figure}

Figure~\ref{fig:alpha_etarot} explores how the  dark matter slope $\alpha(r_{\rm fid})$ is related to the RC shape parameter $\eta_{\rm rot}$, with points colored by the dark matter mass fraction within $r_{1/2}$. There is a clear correlation between $\eta_{\rm rot}$ and $\alpha(r_{\rm fid})$.   The most slowly rising objects  have  $\alpha(r_{\rm fid}) \sim -0.5$ while the most quickly rising objects have $\alpha(r_{\rm fid}) \lesssim -1.5$.  The latter also have the smallest dark matter fractions; in these cases, prominent central stellar distributions dominate the total inner densities, setting the central velocities (Figure \ref{fig:density}).  

We also check the slopes in the corresponding dark-matter-only~(DMO) Massive Dwarf zooms using the same fitting methods. These simulations have the same cosmology, gravitational softening, and dark matter particle-mass resolution as the hydro Massive Dwarf zooms. Measuring the slope at $r_{\rm fid}$ of the corresponding hydro run, we find $\alpha(r_{\rm fid}) \sim-1.5$ in all cases. Thus, while the hydro simulations do not produce halos with central slopes $\alpha(r_{\rm fid}) \sim 0$, most of the sample is still more cored than the corresponding DMO case. Moreover, $r_{\rm fid}$ is well outside of the dark matter core size. Only the most quickly rising systems have slopes slightly higher than DMO runs, likely because of the presence of a dense stellar component driving adiabatic contraction \citep{Blumenthal1986, Dutton2005, Gnedin2011}. 

\bibliography{main}{}

\begin{thebibliography}{}
\expandafter\ifx\csname natexlab\endcsname\relax\def\natexlab#1{#1}\fi
\providecommand{\url}[1]{\href{#1}{#1}}
\providecommand{\dodoi}[1]{doi:~\href{http://doi.org/#1}{\nolinkurl{#1}}}
\providecommand{\doeprint}[1]{\href{http://ascl.net/#1}{\nolinkurl{http://ascl.net/#1}}}
\providecommand{\doarXiv}[1]{\href{https://arxiv.org/abs/#1}{\nolinkurl{https://arxiv.org/abs/#1}}}

\bibitem[{{Arnett}(1996)}]{Arnett1996}
{Arnett}, D. 1996, {Supernovae and Nucleosynthesis: An Investigation of the History of Matter from the Big Bang to the Present}

\bibitem[{{Azartash-Namin} {et~al.}(2024){Azartash-Namin}, {Engelhardt}, {Munshi}, {Keller}, {Brooks}, {Van Nest}, {Christensen}, {Quinn}, \& {Wadsley}}]{Azartash2024}
{Azartash-Namin}, B., {Engelhardt}, A., {Munshi}, F., {et~al.} 2024, \apj, 970, 40, \dodoi{10.3847/1538-4357/ad49a5}

\bibitem[{{Bellovary} {et~al.}(2019){Bellovary}, {Cleary}, {Munshi}, {Tremmel}, {Christensen}, {Brooks}, \& {Quinn}}]{Bellovary2019}
{Bellovary}, J.~M., {Cleary}, C.~E., {Munshi}, F., {et~al.} 2019, \mnras, 482, 2913, \dodoi{10.1093/mnras/sty2842}

\bibitem[{{Benavides} {et~al.}(2025){Benavides}, {Sales}, {Wetzel}, {Moreno}, {Feldmann}, {Mercado}, {Bullock}, {Hopkins}, {Faucher-Guig{\`e}re}, {Stern}, {Wheeler}, \& {Kere{\v{s}}}}]{Benavides2025}
{Benavides}, J.~A., {Sales}, L.~V., {Wetzel}, A., {et~al.} 2025, arXiv e-prints, arXiv:2508.00991, \dodoi{10.48550/arXiv.2508.00991}

\bibitem[{{Ben{\'\i}tez-Llambay} {et~al.}(2019){Ben{\'\i}tez-Llambay}, {Frenk}, {Ludlow}, \& {Navarro}}]{Llambay2019}
{Ben{\'\i}tez-Llambay}, A., {Frenk}, C.~S., {Ludlow}, A.~D., \& {Navarro}, J.~F. 2019, \mnras, 488, 2387, \dodoi{10.1093/mnras/stz1890}

\bibitem[{{Binney} \& {Tremaine}(2008)}]{Binney2008}
{Binney}, J., \& {Tremaine}, S. 2008, {Galactic Dynamics: Second Edition}

\bibitem[{{Blumenthal} {et~al.}(1986){Blumenthal}, {Faber}, {Flores}, \& {Primack}}]{Blumenthal1986}
{Blumenthal}, G.~R., {Faber}, S.~M., {Flores}, R., \& {Primack}, J.~R. 1986, \apj, 301, 27, \dodoi{10.1086/163867}

\bibitem[{{Booth} \& {Schaye}(2009)}]{Booth2009}
{Booth}, C.~M., \& {Schaye}, J. 2009, \mnras, 398, 53, \dodoi{10.1111/j.1365-2966.2009.15043.x}

\bibitem[{{Bosma}(1981)}]{Bosma1981}
{Bosma}, A. 1981, \aj, 86, 1825, \dodoi{10.1086/113063}

\bibitem[{{Bovy}(in press 2026)}]{DynamicsAstrophysicsGalaxies}
{Bovy}, J. in press 2026, Dynamics and Astrophysics of Galaxies (Princeton, NJ: Princeton University Press)

\bibitem[{{Bradford} {et~al.}(2016){Bradford}, {Geha}, \& {van den Bosch}}]{Bradford2016}
{Bradford}, J.~D., {Geha}, M.~C., \& {van den Bosch}, F.~C. 2016, \apj, 832, 11, \dodoi{10.3847/0004-637X/832/1/11}

\bibitem[{{Broeils} \& {Rhee}(1997)}]{Broeils1997}
{Broeils}, A.~H., \& {Rhee}, M.~H. 1997, \aap, 324, 877

\bibitem[{{Brook} {et~al.}(2011){Brook}, {Governato}, {Ro{\v{s}}kar}, {Stinson}, {Brooks}, {Wadsley}, {Quinn}, {Gibson}, {Snaith}, {Pilkington}, {House}, \& {Pontzen}}]{Brook2011}
{Brook}, C.~B., {Governato}, F., {Ro{\v{s}}kar}, R., {et~al.} 2011, \mnras, 415, 1051, \dodoi{10.1111/j.1365-2966.2011.18545.x}

\bibitem[{{Carlsten} {et~al.}(2021){Carlsten}, {Greene}, {Greco}, {Beaton}, \& {Kado-Fong}}]{Carlsten2021}
{Carlsten}, S.~G., {Greene}, J.~E., {Greco}, J.~P., {Beaton}, R.~L., \& {Kado-Fong}, E. 2021, \apj, 922, 267, \dodoi{10.3847/1538-4357/ac2581}

\bibitem[{{Catinella} {et~al.}(2006){Catinella}, {Giovanelli}, \& {Haynes}}]{Catinella2006}
{Catinella}, B., {Giovanelli}, R., \& {Haynes}, M.~P. 2006, \apj, 640, 751, \dodoi{10.1086/500171}

\bibitem[{{Celiz} {et~al.}(2025){Celiz}, {Navarro}, {Abadi}, \& {Springel}}]{Celiz2025}
{Celiz}, B.~M., {Navarro}, J.~F., {Abadi}, M.~G., \& {Springel}, V. 2025, \aap, 699, A12, \dodoi{10.1051/0004-6361/202554847}

\bibitem[{{Christensen} {et~al.}(2012){Christensen}, {Quinn}, {Governato}, {Stilp}, {Shen}, \& {Wadsley}}]{Christensen2012}
{Christensen}, C., {Quinn}, T., {Governato}, F., {et~al.} 2012, \mnras, 425, 3058, \dodoi{10.1111/j.1365-2966.2012.21628.x}

\bibitem[{{Christensen} {et~al.}(2024){Christensen}, {Brooks}, {Munshi}, {Riggs}, {Van Nest}, {Akins}, {Quinn}, \& {Chamberland}}]{Christensen2024}
{Christensen}, C.~R., {Brooks}, A.~M., {Munshi}, F., {et~al.} 2024, \apj, 961, 236, \dodoi{10.3847/1538-4357/ad0c5a}

\bibitem[{{Christensen} {et~al.}(2014){Christensen}, {Governato}, {Quinn}, {Brooks}, {Shen}, {McCleary}, {Fisher}, \& {Wadsley}}]{Christensen2014}
{Christensen}, C.~R., {Governato}, F., {Quinn}, T., {et~al.} 2014, \mnras, 440, 2843, \dodoi{10.1093/mnras/stu399}

\bibitem[{{Collins} \& {Read}(2022)}]{Collins2022}
{Collins}, M. L.~M., \& {Read}, J.~I. 2022, Nature Astronomy, 6, 647, \dodoi{10.1038/s41550-022-01657-4}

\bibitem[{{Crain} {et~al.}(2015){Crain}, {Schaye}, {Bower}, {Furlong}, {Schaller}, {Theuns}, {Dalla Vecchia}, {Frenk}, {McCarthy}, {Helly}, {Jenkins}, {Rosas-Guevara}, {White}, \& {Trayford}}]{Crain2015}
{Crain}, R.~A., {Schaye}, J., {Bower}, R.~G., {et~al.} 2015, \mnras, 450, 1937, \dodoi{10.1093/mnras/stv725}

\bibitem[{{Dalcanton} \& {Stilp}(2010)}]{Dalcanton2010}
{Dalcanton}, J.~J., \& {Stilp}, A.~M. 2010, \apj, 721, 547, \dodoi{10.1088/0004-637X/721/1/547}

\bibitem[{{de Blok}(2010)}]{deBlok2010}
{de Blok}, W.~J.~G. 2010, Advances in Astronomy, 2010, 789293, \dodoi{10.1155/2010/789293}

\bibitem[{{de Blok} {et~al.}(2001){de Blok}, {McGaugh}, {Bosma}, \& {Rubin}}]{deBlok2001}
{de Blok}, W.~J.~G., {McGaugh}, S.~S., {Bosma}, A., \& {Rubin}, V.~C. 2001, \apjl, 552, L23, \dodoi{10.1086/320262}

\bibitem[{{Del Popolo} {et~al.}(2024){Del Popolo}, {Le Delliou}, \& {Lee}}]{DelPopolo2024}
{Del Popolo}, A., {Le Delliou}, M., \& {Lee}, X. 2024, arXiv e-prints, arXiv:2410.05948, \dodoi{10.48550/arXiv.2410.05948}

\bibitem[{{Di Cintio} {et~al.}(2014{\natexlab{a}}){Di Cintio}, {Brook}, {Dutton}, {Macci{\`o}}, {Stinson}, \& {Knebe}}]{Dicinto2014a}
{Di Cintio}, A., {Brook}, C.~B., {Dutton}, A.~A., {et~al.} 2014{\natexlab{a}}, \mnras, 441, 2986, \dodoi{10.1093/mnras/stu729}

\bibitem[{{Di Cintio} {et~al.}(2014{\natexlab{b}}){Di Cintio}, {Brook}, {Macci{\`o}}, {Stinson}, {Knebe}, {Dutton}, \& {Wadsley}}]{Dicintio2014b}
{Di Cintio}, A., {Brook}, C.~B., {Macci{\`o}}, A.~V., {et~al.} 2014{\natexlab{b}}, \mnras, 437, 415, \dodoi{10.1093/mnras/stt1891}

\bibitem[{{Downing} \& {Oman}(2023)}]{Downing2023}
{Downing}, E.~R., \& {Oman}, K.~A. 2023, \mnras, 522, 3318, \dodoi{10.1093/mnras/stad868}

\bibitem[{{Dutton} {et~al.}(2020){Dutton}, {Buck}, {Macci{\`o}}, {Dixon}, {Blank}, \& {Obreja}}]{Dutton2020}
{Dutton}, A.~A., {Buck}, T., {Macci{\`o}}, A.~V., {et~al.} 2020, \mnras, 499, 2648, \dodoi{10.1093/mnras/staa3028}

\bibitem[{{Dutton} {et~al.}(2005){Dutton}, {Courteau}, {de Jong}, \& {Carignan}}]{Dutton2005}
{Dutton}, A.~A., {Courteau}, S., {de Jong}, R., \& {Carignan}, C. 2005, \apj, 619, 218, \dodoi{10.1086/426375}

\bibitem[{{Dutton} {et~al.}(2019){Dutton}, {Macci{\`o}}, {Buck}, {Dixon}, {Blank}, \& {Obreja}}]{Dutton2019}
{Dutton}, A.~A., {Macci{\`o}}, A.~V., {Buck}, T., {et~al.} 2019, \mnras, 486, 655, \dodoi{10.1093/mnras/stz889}

\bibitem[{{El-Badry} {et~al.}(2017){El-Badry}, {Wetzel}, {Geha}, {Quataert}, {Hopkins}, {Kere{\v{s}}}, {Chan}, \& {Faucher-Gigu{\`e}re}}]{ElBadry2017}
{El-Badry}, K., {Wetzel}, A.~R., {Geha}, M., {et~al.} 2017, \apj, 835, 193, \dodoi{10.3847/1538-4357/835/2/193}

\bibitem[{{El-Badry} {et~al.}(2018){El-Badry}, {Quataert}, {Wetzel}, {Hopkins}, {Weisz}, {Chan}, {Fitts}, {Boylan-Kolchin}, {Kere{\v{s}}}, {Faucher-Gigu{\`e}re}, \& {Garrison-Kimmel}}]{ElBadry2018a}
{El-Badry}, K., {Quataert}, E., {Wetzel}, A., {et~al.} 2018, \mnras, 473, 1930, \dodoi{10.1093/mnras/stx2482}

\bibitem[{{Flores} \& {Primack}(1994)}]{Flores1994}
{Flores}, R.~A., \& {Primack}, J.~R. 1994, \apjl, 427, L1, \dodoi{10.1086/187350}

\bibitem[{{Frank} {et~al.}(2016){Frank}, {de Blok}, {Walter}, {Leroy}, \& {Carignan}}]{Frank2016}
{Frank}, B.~S., {de Blok}, W.~J.~G., {Walter}, F., {Leroy}, A., \& {Carignan}, C. 2016, \aj, 151, 94, \dodoi{10.3847/0004-6256/151/4/94}

\bibitem[{{Frosst} {et~al.}(2022){Frosst}, {Courteau}, {Arora}, {Stone}, {Macci{\`o}}, \& {Blank}}]{Frosst2022}
{Frosst}, M., {Courteau}, S., {Arora}, N., {et~al.} 2022, \mnras, 514, 3510, \dodoi{10.1093/mnras/stac1497}

\bibitem[{{Geda} {et~al.}(2022){Geda}, {Crawford}, {Hunt}, {Bershady}, {Tollerud}, \& {Randriamampandry}}]{Geda2022}
{Geda}, R., {Crawford}, S.~M., {Hunt}, L., {et~al.} 2022, \aj, 163, 202, \dodoi{10.3847/1538-3881/ac5908}

\bibitem[{Geda {et~al.}(2025)Geda, Cruz, Wright, Greene, Brooks, Quinn, Wadsley, \& Keller}]{Geda2025}
Geda, R., Cruz, A., Wright, A.~C., {et~al.} 2025

\bibitem[{{Genel} {et~al.}(2019){Genel}, {Bryan}, {Springel}, {Hernquist}, {Nelson}, {Pillepich}, {Weinberger}, {Pakmor}, {Marinacci}, \& {Vogelsberger}}]{Genel2019}
{Genel}, S., {Bryan}, G.~L., {Springel}, V., {et~al.} 2019, \apj, 871, 21, \dodoi{10.3847/1538-4357/aaf4bb}

\bibitem[{{Gnedin} {et~al.}(2011){Gnedin}, {Ceverino}, {Gnedin}, {Klypin}, {Kravtsov}, {Levine}, {Nagai}, \& {Yepes}}]{Gnedin2011}
{Gnedin}, O.~Y., {Ceverino}, D., {Gnedin}, N.~Y., {et~al.} 2011, arXiv e-prints, arXiv:1108.5736, \dodoi{10.48550/arXiv.1108.5736}

\bibitem[{{Governato} {et~al.}(2010){Governato}, {Brook}, {Mayer}, {Brooks}, {Rhee}, {Wadsley}, {Jonsson}, {Willman}, {Stinson}, {Quinn}, \& {Madau}}]{Governato2010}
{Governato}, F., {Brook}, C., {Mayer}, L., {et~al.} 2010, \nat, 463, 203, \dodoi{10.1038/nature08640}

\bibitem[{{Governato} {et~al.}(2012){Governato}, {Zolotov}, {Pontzen}, {Christensen}, {Oh}, {Brooks}, {Quinn}, {Shen}, \& {Wadsley}}]{Governato2012}
{Governato}, F., {Zolotov}, A., {Pontzen}, A., {et~al.} 2012, \mnras, 422, 1231, \dodoi{10.1111/j.1365-2966.2012.20696.x}

\bibitem[{{Graus} {et~al.}(2019){Graus}, {Bullock}, {Fitts}, {Cooper}, {Boylan-Kolchin}, {Weisz}, {Wetzel}, {Feldmann}, {Faucher-Gigu{\`e}re}, {Quataert}, {Hopkins}, \& {Kere{\v{s}}}}]{Graus2019}
{Graus}, A.~S., {Bullock}, J.~S., {Fitts}, A., {et~al.} 2019, \mnras, 490, 1186, \dodoi{10.1093/mnras/stz2649}

\bibitem[{{Haardt} \& {Madau}(2012)}]{Haardt2012}
{Haardt}, F., \& {Madau}, P. 2012, \apj, 746, 125, \dodoi{10.1088/0004-637X/746/2/125}

\bibitem[{{Hopkins} {et~al.}(2018){Hopkins}, {Wetzel}, {Kere{\v{s}}}, {Faucher-Gigu{\`e}re}, {Quataert}, {Boylan-Kolchin}, {Murray}, {Hayward}, {Garrison-Kimmel}, {Hummels}, {Feldmann}, {Torrey}, {Ma}, {Angl{\'e}s-Alc{\'a}zar}, {Su}, {Orr}, {Schmitz}, {Escala}, {Sanderson}, {Grudi{\'c}}, {Hafen}, {Kim}, {Fitts}, {Bullock}, {Wheeler}, {Chan}, {Elbert}, \& {Narayanan}}]{Hopkins2018}
{Hopkins}, P.~F., {Wetzel}, A., {Kere{\v{s}}}, D., {et~al.} 2018, \mnras, 480, 800, \dodoi{10.1093/mnras/sty1690}

\bibitem[{{Jahn} {et~al.}(2023){Jahn}, {Sales}, {Marinacci}, {Vogelsberger}, {Torrey}, {Qi}, {Smith}, {Li}, {Kannan}, {Burger}, \& {Zavala}}]{Jahn2023}
{Jahn}, E.~D., {Sales}, L.~V., {Marinacci}, F., {et~al.} 2023, \mnras, 520, 461, \dodoi{10.1093/mnras/stad109}

\bibitem[{{Jiang} {et~al.}(2019){Jiang}, {Dekel}, {Freundlich}, {Romanowsky}, {Dutton}, {Macci{\`o}}, \& {Di Cintio}}]{Jiang2019}
{Jiang}, F., {Dekel}, A., {Freundlich}, J., {et~al.} 2019, \mnras, 487, 5272, \dodoi{10.1093/mnras/stz1499}

\bibitem[{{Kaplinghat} {et~al.}(2014){Kaplinghat}, {Keeley}, {Linden}, \& {Yu}}]{Kaplinghat2014}
{Kaplinghat}, M., {Keeley}, R.~E., {Linden}, T., \& {Yu}, H.-B. 2014, \prl, 113, 021302, \dodoi{10.1103/PhysRevLett.113.021302}

\bibitem[{{Kaplinghat} {et~al.}(2020){Kaplinghat}, {Ren}, \& {Yu}}]{Kaplinghat2020}
{Kaplinghat}, M., {Ren}, T., \& {Yu}, H.-B. 2020, \jcap, 2020, 027, \dodoi{10.1088/1475-7516/2020/06/027}

\bibitem[{{Keith} {et~al.}(2025){Keith}, {Munshi}, {Brooks}, {Van Nest}, {Engelhardt}, {Cruz}, {Keller}, {Quinn}, \& {Wadsley}}]{Keith2025}
{Keith}, B., {Munshi}, F., {Brooks}, A.~M., {et~al.} 2025, \apj, 986, 138, \dodoi{10.3847/1538-4357/add40d}

\bibitem[{{Keller} {et~al.}(2014){Keller}, {Wadsley}, {Benincasa}, \& {Couchman}}]{Keller2014}
{Keller}, B.~W., {Wadsley}, J., {Benincasa}, S.~M., \& {Couchman}, H.~M.~P. 2014, \mnras, 442, 3013, \dodoi{10.1093/mnras/stu1058}

\bibitem[{{Keller} {et~al.}(2019){Keller}, {Wadsley}, {Wang}, \& {Kruijssen}}]{keller2019}
{Keller}, B.~W., {Wadsley}, J.~W., {Wang}, L., \& {Kruijssen}, J.~M.~D. 2019, \mnras, 482, 2244, \dodoi{10.1093/mnras/sty2859}

\bibitem[{{Klein} {et~al.}(2024){Klein}, {Bullock}, {Moreno}, {Mercado}, {Hopkins}, {Cochrane}, \& {Benavides}}]{Klein2024}
{Klein}, C., {Bullock}, J.~S., {Moreno}, J., {et~al.} 2024, \mnras, 532, 538, \dodoi{10.1093/mnras/stae1505}

\bibitem[{{Knollmann} \& {Knebe}(2009)}]{Knollmann2009}
{Knollmann}, S.~R., \& {Knebe}, A. 2009, \apjs, 182, 608, \dodoi{10.1088/0067-0049/182/2/608}

\bibitem[{{Krause} {et~al.}(2020){Krause}, {Offner}, {Charbonnel}, {Gieles}, {Klessen}, {V{\'a}zquez-Semadeni}, {Ballesteros-Paredes}, {Girichidis}, {Kruijssen}, {Ward}, \& {Zinnecker}}]{Krause2020}
{Krause}, M. G.~H., {Offner}, S. S.~R., {Charbonnel}, C., {et~al.} 2020, \ssr, 216, 64, \dodoi{10.1007/s11214-020-00689-4}

\bibitem[{{Kroupa}(2001)}]{Kroupa2001}
{Kroupa}, P. 2001, \mnras, 322, 231, \dodoi{10.1046/j.1365-8711.2001.04022.x}

\bibitem[{{Kuzio de Naray} {et~al.}(2006){Kuzio de Naray}, {McGaugh}, {de Blok}, \& {Bosma}}]{Kuzio2006}
{Kuzio de Naray}, R., {McGaugh}, S.~S., {de Blok}, W.~J.~G., \& {Bosma}, A. 2006, \apjs, 165, 461, \dodoi{10.1086/505345}

\bibitem[{{Lazar} {et~al.}(2020){Lazar}, {Bullock}, {Boylan-Kolchin}, {Chan}, {Hopkins}, {Graus}, {Wetzel}, {El-Badry}, {Wheeler}, {Straight}, {Kere{\v{s}}}, {Faucher-Gigu{\`e}re}, {Fitts}, \& {Garrison-Kimmel}}]{Lazar2020}
{Lazar}, A., {Bullock}, J.~S., {Boylan-Kolchin}, M., {et~al.} 2020, \mnras, 497, 2393, \dodoi{10.1093/mnras/staa2101}

\bibitem[{{Lelli} {et~al.}(2016{\natexlab{a}}){Lelli}, {McGaugh}, \& {Schombert}}]{Lelli2016}
{Lelli}, F., {McGaugh}, S.~S., \& {Schombert}, J.~M. 2016{\natexlab{a}}, \aj, 152, 157, \dodoi{10.3847/0004-6256/152/6/157}

\bibitem[{{Lelli} {et~al.}(2016{\natexlab{b}}){Lelli}, {McGaugh}, \& {Schombert}}]{Lelli2016a}
---. 2016{\natexlab{b}}, \apjl, 816, L14, \dodoi{10.3847/2041-8205/816/1/L14}

\bibitem[{{Lelli} {et~al.}(2012){Lelli}, {Verheijen}, {Fraternali}, \& {Sancisi}}]{Lelli2012b}
{Lelli}, F., {Verheijen}, M., {Fraternali}, F., \& {Sancisi}, R. 2012, \aap, 544, A145, \dodoi{10.1051/0004-6361/201219457}

\bibitem[{{Ludlow} {et~al.}(2016){Ludlow}, {Bose}, {Angulo}, {Wang}, {Hellwing}, {Navarro}, {Cole}, \& {Frenk}}]{Ludlow2016}
{Ludlow}, A.~D., {Bose}, S., {Angulo}, R.~E., {et~al.} 2016, \mnras, 460, 1214, \dodoi{10.1093/mnras/stw1046}

\bibitem[{{Ludlow} {et~al.}(2019{\natexlab{a}}){Ludlow}, {Schaye}, \& {Bower}}]{LSB2019}
{Ludlow}, A.~D., {Schaye}, J., \& {Bower}, R. 2019{\natexlab{a}}, \mnras, 488, 3663, \dodoi{10.1093/mnras/stz1821}

\bibitem[{{Ludlow} {et~al.}(2019{\natexlab{b}}){Ludlow}, {Schaye}, {Schaller}, \& {Richings}}]{Ludlow2019}
{Ludlow}, A.~D., {Schaye}, J., {Schaller}, M., \& {Richings}, J. 2019{\natexlab{b}}, \mnras, 488, L123, \dodoi{10.1093/mnrasl/slz110}

\bibitem[{{Lupi} {et~al.}(2017){Lupi}, {Volonteri}, \& {Silk}}]{Lupi2017}
{Lupi}, A., {Volonteri}, M., \& {Silk}, J. 2017, \mnras, 470, 1673, \dodoi{10.1093/mnras/stx1313}

\bibitem[{{Mancera Pi{\~n}a} {et~al.}(2022){Mancera Pi{\~n}a}, {Fraternali}, {Oosterloo}, {Adams}, {di Teodoro}, {Bacchini}, \& {Iorio}}]{ManceraPina2022}
{Mancera Pi{\~n}a}, P.~E., {Fraternali}, F., {Oosterloo}, T., {et~al.} 2022, \mnras, 514, 3329, \dodoi{10.1093/mnras/stac1508}

\bibitem[{{Marasco} {et~al.}(2018){Marasco}, {Oman}, {Navarro}, {Frenk}, \& {Oosterloo}}]{Marasco2018}
{Marasco}, A., {Oman}, K.~A., {Navarro}, J.~F., {Frenk}, C.~S., \& {Oosterloo}, T. 2018, \mnras, 476, 2168, \dodoi{10.1093/mnras/sty354}

\bibitem[{{Marigo} {et~al.}(2008){Marigo}, {Girardi}, {Bressan}, {Groenewegen}, {Silva}, \& {Granato}}]{Marigo2008}
{Marigo}, P., {Girardi}, L., {Bressan}, A., {et~al.} 2008, \aap, 482, 883, \dodoi{10.1051/0004-6361:20078467}

\bibitem[{{Mashchenko} {et~al.}(2008){Mashchenko}, {Wadsley}, \& {Couchman}}]{Mashchenko2008}
{Mashchenko}, S., {Wadsley}, J., \& {Couchman}, H.~M.~P. 2008, Science, 319, 174, \dodoi{10.1126/science.1148666}

\bibitem[{{McGaugh} \& {Schombert}(2014)}]{McGaugh2014}
{McGaugh}, S.~S., \& {Schombert}, J.~M. 2014, \aj, 148, 77, \dodoi{10.1088/0004-6256/148/5/77}

\bibitem[{{McQuinn} {et~al.}(2022){McQuinn}, {Adams}, {Cannon}, {Fuson}, {Skillman}, {Brooks}, {Rhode}, {Haynes}, {Inoue}, {Marine}, {Salzer}, \& {Talluri}}]{McQuinn2022}
{McQuinn}, K. B.~W., {Adams}, E. A.~K., {Cannon}, J.~M., {et~al.} 2022, \apj, 940, 8, \dodoi{10.3847/1538-4357/ac9285}

\bibitem[{{Menon} {et~al.}(2015){Menon}, {Wesolowski}, {Zheng}, {Jetley}, {Kale}, {Quinn}, \& {Governato}}]{Menon2015}
{Menon}, H., {Wesolowski}, L., {Zheng}, G., {et~al.} 2015, Computational Astrophysics and Cosmology, 2, 1, \dodoi{10.1186/s40668-015-0007-9}

\bibitem[{{Mina} {et~al.}(2021){Mina}, {Shen}, {Keller}, {Mayer}, {Madau}, \& {Wadsley}}]{Mina2021}
{Mina}, M., {Shen}, S., {Keller}, B.~W., {et~al.} 2021, \aap, 655, A22, \dodoi{10.1051/0004-6361/202039420}

\bibitem[{{Moore}(1994)}]{Moore1994}
{Moore}, B. 1994, \nat, 370, 629, \dodoi{10.1038/370629a0}

\bibitem[{{Munshi} {et~al.}(2021){Munshi}, {Brooks}, {Applebaum}, {Christensen}, {Quinn}, \& {Sligh}}]{Munshi2021}
{Munshi}, F., {Brooks}, A.~M., {Applebaum}, E., {et~al.} 2021, \apj, 923, 35, \dodoi{10.3847/1538-4357/ac0db6}

\bibitem[{{Navarro} {et~al.}(1996){Navarro}, {Frenk}, \& {White}}]{NFW1996}
{Navarro}, J.~F., {Frenk}, C.~S., \& {White}, S. D.~M. 1996, \apj, 462, 563, \dodoi{10.1086/177173}

\bibitem[{{Navarro} {et~al.}(1997){Navarro}, {Frenk}, \& {White}}]{NFW1997}
---. 1997, \apj, 490, 493, \dodoi{10.1086/304888}

\bibitem[{{Oh} {et~al.}(2011){Oh}, {de Blok}, {Brinks}, {Walter}, \& {Kennicutt}}]{Oh2011}
{Oh}, S.-H., {de Blok}, W.~J.~G., {Brinks}, E., {Walter}, F., \& {Kennicutt}, Robert~C., J. 2011, \aj, 141, 193, \dodoi{10.1088/0004-6256/141/6/193}

\bibitem[{{Oman} {et~al.}(2019){Oman}, {Marasco}, {Navarro}, {Frenk}, {Schaye}, \& {Ben{\'\i}tez-Llambay}}]{Oman2019}
{Oman}, K.~A., {Marasco}, A., {Navarro}, J.~F., {et~al.} 2019, \mnras, 482, 821, \dodoi{10.1093/mnras/sty2687}

\bibitem[{{Oman} {et~al.}(2015){Oman}, {Navarro}, {Fattahi}, {Frenk}, {Sawala}, {White}, {Bower}, {Crain}, {Furlong}, {Schaller}, {Schaye}, \& {Theuns}}]{Oman2015}
{Oman}, K.~A., {Navarro}, J.~F., {Fattahi}, A., {et~al.} 2015, \mnras, 452, 3650, \dodoi{10.1093/mnras/stv1504}

\bibitem[{{Petrosian}(1976)}]{Petrosian1976}
{Petrosian}, V. 1976, \apjl, 210, L53, \dodoi{10.1086/18230110.1086/182253}

\bibitem[{{Pfalzner} {et~al.}(2012){Pfalzner}, {Kaczmarek}, \& {Olczak}}]{Pfalzner2012}
{Pfalzner}, S., {Kaczmarek}, T., \& {Olczak}, C. 2012, \aap, 545, A122, \dodoi{10.1051/0004-6361/201219881}

\bibitem[{{Piacitelli} {et~al.}(2025){Piacitelli}, {Brooks}, {Christensen}, {Sanchez}, {Faerman}, {Shen}, {Cruz}, {Keller}, {Quinn}, \& {Wadsley}}]{Piacitelli2025}
{Piacitelli}, D.~R., {Brooks}, A.~M., {Christensen}, C., {et~al.} 2025, arXiv e-prints, arXiv:2505.08861, \dodoi{10.48550/arXiv.2505.08861}

\bibitem[{{Pineda} {et~al.}(2017){Pineda}, {Hayward}, {Springel}, \& {Mendes de Oliveira}}]{Pineda2017}
{Pineda}, J. C.~B., {Hayward}, C.~C., {Springel}, V., \& {Mendes de Oliveira}, C. 2017, \mnras, 466, 63, \dodoi{10.1093/mnras/stw3004}

\bibitem[{{Planck Collaboration} {et~al.}(2014){Planck Collaboration}, {Ade}, {Aghanim}, {Armitage-Caplan}, {Arnaud}, {Ashdown}, {Atrio-Barand ela}, {Aumont}, {Baccigalupi}, {Banday}, \& et~al.}]{Plank2014}
{Planck Collaboration}, {Ade}, P.~A.~R., {Aghanim}, N., {et~al.} 2014, \aap, 571, A16, \dodoi{10.1051/0004-6361/201321591}

\bibitem[{{Planck Collaboration} {et~al.}(2020){Planck Collaboration}, {Aghanim}, {Akrami}, {Ashdown}, {Aumont}, {Baccigalupi}, {Ballardini}, {Banday}, {Barreiro}, {Bartolo}, {Basak}, {Battye}, {Benabed}, {Bernard}, {Bersanelli}, {Bielewicz}, {Bock}, {Bond}, {Borrill}, {Bouchet}, {Boulanger}, {Bucher}, {Burigana}, {Butler}, {Calabrese}, {Cardoso}, {Carron}, {Challinor}, {Chiang}, {Chluba}, {Colombo}, {Combet}, {Contreras}, {Crill}, {Cuttaia}, {de Bernardis}, {de Zotti}, {Delabrouille}, {Delouis}, {Di Valentino}, {Diego}, {Dor{\'e}}, {Douspis}, {Ducout}, {Dupac}, {Dusini}, {Efstathiou}, {Elsner}, {En{\ss}lin}, {Eriksen}, {Fantaye}, {Farhang}, {Fergusson}, {Fernandez-Cobos}, {Finelli}, {Forastieri}, {Frailis}, {Fraisse}, {Franceschi}, {Frolov}, {Galeotta}, {Galli}, {Ganga}, {G{\'e}nova-Santos}, {Gerbino}, {Ghosh}, {Gonz{\'a}lez-Nuevo}, {G{\'o}rski}, {Gratton}, {Gruppuso}, {Gudmundsson}, {Hamann}, {Handley}, {Hansen}, {Herranz}, {Hildebrandt}, {Hivon}, {Huang}, {Jaffe}, {Jones}, {Karakci}, {Keih{\"a}nen},
  {Keskitalo}, {Kiiveri}, {Kim}, {Kisner}, {Knox}, {Krachmalnicoff}, {Kunz}, {Kurki-Suonio}, {Lagache}, {Lamarre}, {Lasenby}, {Lattanzi}, {Lawrence}, {Le Jeune}, {Lemos}, {Lesgourgues}, {Levrier}, {Lewis}, {Liguori}, {Lilje}, {Lilley}, {Lindholm}, {L{\'o}pez-Caniego}, {Lubin}, {Ma}, {Mac{\'\i}as-P{\'e}rez}, {Maggio}, {Maino}, {Mandolesi}, {Mangilli}, {Marcos-Caballero}, {Maris}, {Martin}, {Martinelli}, {Mart{\'\i}nez-Gonz{\'a}lez}, {Matarrese}, {Mauri}, {McEwen}, {Meinhold}, {Melchiorri}, {Mennella}, {Migliaccio}, {Millea}, {Mitra}, {Miville-Desch{\^e}nes}, {Molinari}, {Montier}, {Morgante}, {Moss}, {Natoli}, {N{\o}rgaard-Nielsen}, {Pagano}, {Paoletti}, {Partridge}, {Patanchon}, {Peiris}, {Perrotta}, {Pettorino}, {Piacentini}, {Polastri}, {Polenta}, {Puget}, {Rachen}, {Reinecke}, {Remazeilles}, {Renzi}, {Rocha}, {Rosset}, {Roudier}, {Rubi{\~n}o-Mart{\'\i}n}, {Ruiz-Granados}, {Salvati}, {Sandri}, {Savelainen}, {Scott}, {Shellard}, {Sirignano}, {Sirri}, {Spencer}, {Sunyaev}, {Suur-Uski}, {Tauber}, {Tavagnacco},
  {Tenti}, {Toffolatti}, {Tomasi}, {Trombetti}, {Valenziano}, {Valiviita}, {Van Tent}, {Vibert}, {Vielva}, {Villa}, {Vittorio}, {Wandelt}, {Wehus}, {White}, {White}, {Zacchei}, \& {Zonca}}]{Planck2020}
{Planck Collaboration}, {Aghanim}, N., {Akrami}, Y., {et~al.} 2020, \aap, 641, A6, \dodoi{10.1051/0004-6361/201833910}

\bibitem[{{Pontzen} \& {Governato}(2012)}]{Pontzen2012}
{Pontzen}, A., \& {Governato}, F. 2012, \mnras, 421, 3464, \dodoi{10.1111/j.1365-2966.2012.20571.x}

\bibitem[{{Pontzen} {et~al.}(2013){Pontzen}, {Ro{\v s}kar}, {Stinson}, {Woods}, {Reed}, {Coles}, \& {Quinn}}]{pynbody}
{Pontzen}, A., {Ro{\v s}kar}, R., {Stinson}, G.~S., {et~al.} 2013, {pynbody: Astrophysics Simulation Analysis for Python}

\bibitem[{{Pontzen} {et~al.}(2008){Pontzen}, {Governato}, {Pettini}, {Booth}, {Stinson}, {Wadsley}, {Brooks}, {Quinn}, \& {Haehnelt}}]{Pontzen2008}
{Pontzen}, A., {Governato}, F., {Pettini}, M., {et~al.} 2008, \mnras, 390, 1349, \dodoi{10.1111/j.1365-2966.2008.13782.x}

\bibitem[{{Power} {et~al.}(2003){Power}, {Navarro}, {Jenkins}, {Frenk}, {White}, {Springel}, {Stadel}, \& {Quinn}}]{Power2003}
{Power}, C., {Navarro}, J.~F., {Jenkins}, A., {et~al.} 2003, \mnras, 338, 14, \dodoi{10.1046/j.1365-8711.2003.05925.x}

\bibitem[{{Read} {et~al.}(2016{\natexlab{a}}){Read}, {Agertz}, \& {Collins}}]{Read2016}
{Read}, J.~I., {Agertz}, O., \& {Collins}, M.~L.~M. 2016{\natexlab{a}}, \mnras, 459, 2573, \dodoi{10.1093/mnras/stw713}

\bibitem[{{Read} \& {Gilmore}(2005)}]{Read2005}
{Read}, J.~I., \& {Gilmore}, G. 2005, \mnras, 356, 107, \dodoi{10.1111/j.1365-2966.2004.08424.x}

\bibitem[{{Read} {et~al.}(2016{\natexlab{b}}){Read}, {Iorio}, {Agertz}, \& {Fraternali}}]{Read2016b}
{Read}, J.~I., {Iorio}, G., {Agertz}, O., \& {Fraternali}, F. 2016{\natexlab{b}}, \mnras, 462, 3628, \dodoi{10.1093/mnras/stw1876}

\bibitem[{{Regan}(2023)}]{Regan2023}
{Regan}, J. 2023, The Open Journal of Astrophysics, 6, 12, \dodoi{10.21105/astro.2210.04899}

\bibitem[{{Regan} {et~al.}(2020){Regan}, {Wise}, {Woods}, {Downes}, {O'Shea}, \& {Norman}}]{Regan2020}
{Regan}, J.~A., {Wise}, J.~H., {Woods}, T.~E., {et~al.} 2020, The Open Journal of Astrophysics, 3, 15, \dodoi{10.21105/astro.2008.08090}

\bibitem[{{Relatores} {et~al.}(2019{\natexlab{a}}){Relatores}, {Newman}, {Simon}, {Ellis}, {Truong}, {Blitz}, {Bolatto}, {Martin}, \& {Morrissey}}]{Relatores2019a}
{Relatores}, N.~C., {Newman}, A.~B., {Simon}, J.~D., {et~al.} 2019{\natexlab{a}}, \apj, 873, 5, \dodoi{10.3847/1538-4357/ab0382}

\bibitem[{{Relatores} {et~al.}(2019{\natexlab{b}}){Relatores}, {Newman}, {Simon}, {Ellis}, {Truong}, {Blitz}, {Bolatto}, {Martin}, {Matuszewski}, {Morrissey}, \& {Neill}}]{Relatores2019b}
---. 2019{\natexlab{b}}, \apj, 887, 94, \dodoi{10.3847/1538-4357/ab5305}

\bibitem[{{Ren} {et~al.}(2019){Ren}, {Kwa}, {Kaplinghat}, \& {Yu}}]{Ren2019}
{Ren}, T., {Kwa}, A., {Kaplinghat}, M., \& {Yu}, H.-B. 2019, Physical Review X, 9, 031020, \dodoi{10.1103/PhysRevX.9.031020}

\bibitem[{{Riggs} {et~al.}(2024){Riggs}, {Brooks}, {Munshi}, {Christensen}, {Cohen}, {Quinn}, \& {Wadsley}}]{Riggs2024}
{Riggs}, C.~L., {Brooks}, A.~M., {Munshi}, F., {et~al.} 2024, \apj, 977, 20, \dodoi{10.3847/1538-4357/ad8b1e}

\bibitem[{{Roper} {et~al.}(2023){Roper}, {Oman}, {Frenk}, {Ben{\'\i}tez-Llambay}, {Navarro}, \& {Santos-Santos}}]{Roper2023}
{Roper}, F.~A., {Oman}, K.~A., {Frenk}, C.~S., {et~al.} 2023, \mnras, 521, 1316, \dodoi{10.1093/mnras/stad549}

\bibitem[{{Ruan} {et~al.}(2025){Ruan}, {Brooks}, {Cruz}, {Peter}, {Keller}, {Quinn}, {Wadsley}, \& {Adams}}]{Ruan2025}
{Ruan}, D., {Brooks}, A.~M., {Cruz}, A., {et~al.} 2025, arXiv e-prints, arXiv:2503.16607, \dodoi{10.48550/arXiv.2503.16607}

\bibitem[{{Rubin} {et~al.}(1980){Rubin}, {Ford}, \& {Thonnard}}]{Rubin1980}
{Rubin}, V.~C., {Ford}, W.~K., J., \& {Thonnard}, N. 1980, \apj, 238, 471, \dodoi{10.1086/158003}

\bibitem[{{Sales} {et~al.}(2022){Sales}, {Wetzel}, \& {Fattahi}}]{Sales2022}
{Sales}, L.~V., {Wetzel}, A., \& {Fattahi}, A. 2022, Nature Astronomy, 6, 897, \dodoi{10.1038/s41550-022-01689-w}

\bibitem[{{Sales} {et~al.}(2017){Sales}, {Navarro}, {Oman}, {Fattahi}, {Ferrero}, {Abadi}, {Bower}, {Crain}, {Frenk}, {Sawala}, {Schaller}, {Schaye}, {Theuns}, \& {White}}]{Salas2017}
{Sales}, L.~V., {Navarro}, J.~F., {Oman}, K., {et~al.} 2017, \mnras, 464, 2419, \dodoi{10.1093/mnras/stw2461}

\bibitem[{{Sands} {et~al.}(2024){Sands}, {Hopkins}, {Shen}, {Boylan-Kolchin}, {Bullock}, {Faucher-Giguere}, {Mercado}, {Moreno}, {Necib}, {Ou}, {Wellons}, \& {Wetzel}}]{Sands2024}
{Sands}, I.~S., {Hopkins}, P.~F., {Shen}, X., {et~al.} 2024, arXiv e-prints, arXiv:2404.16247, \dodoi{10.48550/arXiv.2404.16247}

\bibitem[{{Santos-Santos} {et~al.}(2018){Santos-Santos}, {Di Cintio}, {Brook}, {Macci{\`o}}, {Dutton}, \& {Dom{\'\i}nguez-Tenreiro}}]{SantosSantos2018}
{Santos-Santos}, I.~M., {Di Cintio}, A., {Brook}, C.~B., {et~al.} 2018, \mnras, 473, 4392, \dodoi{10.1093/mnras/stx2660}

\bibitem[{{Santos-Santos} {et~al.}(2020){Santos-Santos}, {Navarro}, {Robertson}, {Ben{\'\i}tez-Llambay}, {Oman}, {Lovell}, {Frenk}, {Ludlow}, {Fattahi}, \& {Ritz}}]{SantosSantos2020}
{Santos-Santos}, I. M.~E., {Navarro}, J.~F., {Robertson}, A., {et~al.} 2020, \mnras, 495, 58, \dodoi{10.1093/mnras/staa1072}

\bibitem[{{Schaye} {et~al.}(2015){Schaye}, {Crain}, {Bower}, {Furlong}, {Schaller}, {Theuns}, {Dalla Vecchia}, {Frenk}, {McCarthy}, {Helly}, {Jenkins}, {Rosas-Guevara}, {White}, {Baes}, {Booth}, {Camps}, {Navarro}, {Qu}, {Rahmati}, {Sawala}, {Thomas}, \& {Trayford}}]{Schaye2015}
{Schaye}, J., {Crain}, R.~A., {Bower}, R.~G., {et~al.} 2015, \mnras, 446, 521, \dodoi{10.1093/mnras/stu2058}

\bibitem[{{S{\'e}rsic}(1963)}]{sersic1963}
{S{\'e}rsic}, J.~L. 1963, Boletin de la Asociacion Argentina de Astronomia La Plata Argentina, 6, 41

\bibitem[{{Simon} {et~al.}(2003){Simon}, {Bolatto}, {Leroy}, \& {Blitz}}]{Simon2003}
{Simon}, J.~D., {Bolatto}, A.~D., {Leroy}, A., \& {Blitz}, L. 2003, \apj, 596, 957, \dodoi{10.1086/378200}

\bibitem[{{Spergel} \& {Steinhardt}(2000)}]{Spergel2000}
{Spergel}, D.~N., \& {Steinhardt}, P.~J. 2000, \prl, 84, 3760, \dodoi{10.1103/PhysRevLett.84.3760}

\bibitem[{{Spergel} {et~al.}(2007){Spergel}, {Bean}, {Dor{\'e}}, {Nolta}, {Bennett}, {Dunkley}, {Hinshaw}, {Jarosik}, {Komatsu}, {Page}, {Peiris}, {Verde}, {Halpern}, {Hill}, {Kogut}, {Limon}, {Meyer}, {Odegard}, {Tucker}, {Weiland}, {Wollack}, \& {Wright}}]{Spergel2007}
{Spergel}, D.~N., {Bean}, R., {Dor{\'e}}, O., {et~al.} 2007, \apjs, 170, 377, \dodoi{10.1086/513700}

\bibitem[{{Springel} {et~al.}(2006){Springel}, {Frenk}, \& {White}}]{Springel2006}
{Springel}, V., {Frenk}, C.~S., \& {White}, S. D.~M. 2006, \nat, 440, 1137, \dodoi{10.1038/nature04805}

\bibitem[{{Stinson} {et~al.}(2006){Stinson}, {Seth}, {Katz}, {Wadsley}, {Governato}, \& {Quinn}}]{stinson2006}
{Stinson}, G., {Seth}, A., {Katz}, N., {et~al.} 2006, \mnras, 373, 1074, \dodoi{10.1111/j.1365-2966.2006.11097.x}

\bibitem[{{Stinson} {et~al.}(2013){Stinson}, {Brook}, {Macci{\`o}}, {Wadsley}, {Quinn}, \& {Couchman}}]{Stinson2013}
{Stinson}, G.~S., {Brook}, C., {Macci{\`o}}, A.~V., {et~al.} 2013, \mnras, 428, 129, \dodoi{10.1093/mnras/sts028}

\bibitem[{{Strauss} {et~al.}(2002){Strauss}, {Weinberg}, {Lupton}, {Narayanan}, {Annis}, {Bernardi}, {Blanton}, {Burles}, {Connolly}, {Dalcanton}, {Doi}, {Eisenstein}, {Frieman}, {Fukugita}, {Gunn}, {Ivezi{\'c}}, {Kent}, {Kim}, {Knapp}, {Kron}, {Munn}, {Newberg}, {Nichol}, {Okamura}, {Quinn}, {Richmond}, {Schlegel}, {Shimasaku}, {SubbaRao}, {Szalay}, {Vanden Berk}, {Vogeley}, {Yanny}, {Yasuda}, {York}, \& {Zehavi}}]{Strauss2002}
{Strauss}, M.~A., {Weinberg}, D.~H., {Lupton}, R.~H., {et~al.} 2002, \aj, 124, 1810, \dodoi{10.1086/342343}

\bibitem[{{Teyssier} {et~al.}(2013){Teyssier}, {Pontzen}, {Dubois}, \& {Read}}]{Teyssier2013}
{Teyssier}, R., {Pontzen}, A., {Dubois}, Y., \& {Read}, J.~I. 2013, \mnras, 429, 3068, \dodoi{10.1093/mnras/sts563}

\bibitem[{{Tollet} {et~al.}(2016){Tollet}, {Macci{\`o}}, {Dutton}, {Stinson}, {Wang}, {Penzo}, {Gutcke}, {Buck}, {Kang}, {Brook}, {Di Cintio}, {Keller}, \& {Wadsley}}]{Tollet2015}
{Tollet}, E., {Macci{\`o}}, A.~V., {Dutton}, A.~A., {et~al.} 2016, \mnras, 456, 3542, \dodoi{10.1093/mnras/stv2856}

\bibitem[{Tremmel {et~al.}(2015)Tremmel, Governato, Volonteri, \& Quinn}]{Tremmel2015}
Tremmel, M., Governato, F., Volonteri, M., \& Quinn, T.~R. 2015, Monthly Notices of the Royal Astronomical Society, 451, 1868, \dodoi{10.1093/mnras/stv1060}

\bibitem[{{Tremmel} {et~al.}(2018){Tremmel}, {Governato}, {Volonteri}, {Quinn}, \& {Pontzen}}]{Tremmel2018a}
{Tremmel}, M., {Governato}, F., {Volonteri}, M., {Quinn}, T.~R., \& {Pontzen}, A. 2018, \mnras, 475, 4967, \dodoi{10.1093/mnras/sty139}

\bibitem[{{Tremmel} {et~al.}(2017){Tremmel}, {Karcher}, {Governato}, {Volonteri}, {Quinn}, {Pontzen}, {Anderson}, \& {Bellovary}}]{Tremmel2017}
{Tremmel}, M., {Karcher}, M., {Governato}, F., {et~al.} 2017, \mnras, 470, 1121, \dodoi{10.1093/mnras/stx1160}

\bibitem[{{Valenzuela} {et~al.}(2007){Valenzuela}, {Rhee}, {Klypin}, {Governato}, {Stinson}, {Quinn}, \& {Wadsley}}]{Valenzuela2007}
{Valenzuela}, O., {Rhee}, G., {Klypin}, A., {et~al.} 2007, \apj, 657, 773, \dodoi{10.1086/508674}

\bibitem[{{van der Wel} {et~al.}(2011){van der Wel}, {Straughn}, {Rix}, {Finkelstein}, {Koekemoer}, {Weiner}, {Wuyts}, {Bell}, {Faber}, {Trump}, {Koo}, {Ferguson}, {Scarlata}, {Hathi}, {Dunlop}, {Newman}, {Dickinson}, {Jahnke}, {Salmon}, {de Mello}, {Kocevski}, {Lai}, {Grogin}, {Rodney}, {Guo}, {McGrath}, {Lee}, {Barro}, {Huang}, {Riess}, {Ashby}, \& {Willner}}]{VanderWel2011}
{van der Wel}, A., {Straughn}, A.~N., {Rix}, H.~W., {et~al.} 2011, \apj, 742, 111, \dodoi{10.1088/0004-637X/742/2/111}

\bibitem[{{Verbeke} {et~al.}(2017){Verbeke}, {Papastergis}, {Ponomareva}, {Rathi}, \& {De Rijcke}}]{Verbeke2017}
{Verbeke}, R., {Papastergis}, E., {Ponomareva}, A.~A., {Rathi}, S., \& {De Rijcke}, S. 2017, \aap, 607, A13, \dodoi{10.1051/0004-6361/201730758}

\bibitem[{{Wadsley} {et~al.}(2017){Wadsley}, {Keller}, \& {Quinn}}]{Wadsley2017}
{Wadsley}, J.~W., {Keller}, B.~W., \& {Quinn}, T.~R. 2017, \mnras, 471, 2357, \dodoi{10.1093/mnras/stx1643}

\bibitem[{{Wadsley} {et~al.}(2004){Wadsley}, {Stadel}, \& {Quinn}}]{Wadsley2004}
{Wadsley}, J.~W., {Stadel}, J., \& {Quinn}, T. 2004, \na, 9, 137, \dodoi{10.1016/j.newast.2003.08.004}

\bibitem[{{Wang} {et~al.}(2016){Wang}, {Koribalski}, {Serra}, {van der Hulst}, {Roychowdhury}, {Kamphuis}, \& {Chengalur}}]{WangJ2016}
{Wang}, J., {Koribalski}, B.~S., {Serra}, P., {et~al.} 2016, \mnras, 460, 2143, \dodoi{10.1093/mnras/stw1099}

\bibitem[{{Wang} {et~al.}(2015){Wang}, {Dutton}, {Stinson}, {Macci{\`o}}, {Penzo}, {Kang}, {Keller}, \& {Wadsley}}]{Wang2015}
{Wang}, L., {Dutton}, A.~A., {Stinson}, G.~S., {et~al.} 2015, \mnras, 454, 83, \dodoi{10.1093/mnras/stv1937}

\bibitem[{{Williams} {et~al.}(2025){Williams}, {Hjorth}, \& {Skillman}}]{Williams2025}
{Williams}, L. L.~R., {Hjorth}, J., \& {Skillman}, E.~D. 2025, arXiv e-prints, arXiv:2502.15879, \dodoi{10.48550/arXiv.2502.15879}

\bibitem[{{Wise} {et~al.}(2019){Wise}, {Regan}, {O'Shea}, {Norman}, {Downes}, \& {Xu}}]{Wise2019}
{Wise}, J.~H., {Regan}, J.~A., {O'Shea}, B.~W., {et~al.} 2019, \nat, 566, 85, \dodoi{10.1038/s41586-019-0873-4}

\bibitem[{{Zentner} {et~al.}(2022){Zentner}, {Dandavate}, {Slone}, \& {Lisanti}}]{Zentner2022}
{Zentner}, A., {Dandavate}, S., {Slone}, O., \& {Lisanti}, M. 2022, \jcap, 2022, 031, \dodoi{10.1088/1475-7516/2022/07/031}

\end{thebibliography}
\bibliographystyle{aasjournal}

\end{document}